\documentclass[12pt]{iopart}
\usepackage{graphicx}
\usepackage{float}
\usepackage{dsfont}
\usepackage{subcaption}
\usepackage{cite} 
\usepackage{nicematrix}
\usepackage{amsfonts}
\usepackage{xcolor}

\begin{document}

\title[Continuous-variable quantum key distribution with noisy squeezed states]{Continuous-variable quantum key distribution with noisy squeezed states}

\author{Akash nag Oruganti$^\dagger$, Ivan Derkach, Radim Filip, Vladyslav C. Usenko$^\ddagger$}

\address{Department of Optics, Palacky University, 17. listopadu 12, 77146 Olomouc, Czech Republic}
\ead{$^\dagger$akashnag.oruganti@upol.cz, $^\ddagger$usenko@optics.upol.cz}
\vspace{10pt}

\begin{abstract}

We address the crucial role of noisy squeezing in security and performance of continuous-variable (CV) quantum key distribution (QKD) protocols. Squeezing has long been recognized for its numerous advantages in CV QKD, such as enhanced robustness against channel noise and loss, and improved secret key rates. However, the excess noise of the squeezed states, that unavoidably originates already from optical loss as well as other imperfections in the source, raises concerns about its potential exploitation by an eavesdropper. For the widely adopted trust assumption on the excess noise in the signal states, we confirm the stability of the protocol against the noisy squeezing in both purely attenuating as well as noisy channels in the asymptotic limit, which implies perfect parameter estimation. In the finite-size regime we show that this stability largely holds at up to $10^7$ data points using optimal biased homodyne detection for key distribution and parameter estimation. Untrusted assumption on the noisy squeezing, on the other hand, introduces additional security bounds on the squeezing excess noise already in the asymptotic regime, which is further enforced by the finite-size effects. Additionally, we show the critical negative role of noisy squeezing in the case of atmospheric free-space channels, already in the trusted-noise and asymptotic assumptions, which emphasizes the importance of squeezing purity in the free-space quantum channels. Our results pave the way towards practical realization of squeezed-state CV QKD protocols in both fibre and free-space channels. 
\end{abstract}

\section{Introduction}
Quantum key distribution (QKD) \cite{Gisin2002, Scarani2009, Diamanti2016,xu2020secure, pirandolaadvances} is arguably the most mature of the quantum technologies, which has its goal in the development of protocols for provably secure distribution of symmetrical cryptographic keys. Recent advances in the field are particularly concerned with the continuous-variable (CV) \cite{Braunstein2005}  approach to QKD, focused on realizations using off-the-shelf high-speed and efficient coherent homodyne detection. CV QKD was intensively developed, studied, and tested using coherent signal states \cite{GG02, grosshans2003quantum,  jouguet2013experimental}, typically using Gaussian quadrature modulation, compatible with the most advanced security proofs \cite{PhysRevA.81.062343, PhysRevLett.118.200501}. In recent years, significant progress has been made in enhancing the performance and the security of CV-QKD protocols under realistic conditions, including improved noise tolerance \cite{PRXQuantum.2.040334,su2023experimental}, higher key generation speed \cite{wang2020high, hajomer2024highspeed, pietri2024experimental}, and longer secure distance \cite{zhang2020long, hajomer2023long}. All of which can be further refined by integrating automated machine learning algorithms \cite{chin2021machine, liu2022automated}.

Alternatively, CV QKD can be realized using quadrature modulation of squeezed states, which was the initial proposal for Gaussian CV QKD \cite{PhysRevA.63.052311}. Such states are known to have quantum fluctuations of a quadrature observable of light field (referred to as a squeezed quadrature) to be suppressed under the level of vacuum fluctuations (being also known as the shot-noise level) \cite{lvovsky2015squeezed}. To comply with the Heisenberg uncertainty principle, a complementary quadrature observable then has the fluctuations above the shot-noise level, and is referred to as the anti-squeezed (AS) quadrature. In the ideal case of a pure squeezed state, the AS quadrature variance is inverse to the squeezed one (putting $\hbar=2$).

It was shown that squeezing can bring numerous advantages to CV QKD. In particular, it can provide higher robustness against channel noise and loss \cite{PhysRevLett.102.130501, naturecom2012} or make protocols more tolerant against imperfect data processing \cite{Usenko_2011}, which is particularly demanding for Gaussian-distributed data, and even completely eliminate information leakage from the lossy quantum channels \cite{jacobsen2018complete}. In free-space channels, where atmospheric turbulence leads to transmittance fluctuations, limiting CV QKD \cite{usenko2012entanglement}, squeezing can offer advantages as well \cite{Derkach_2020}, and can therefore be particularly useful for satellite-based CV QKD \cite{e23010055}. 

Moreover, squeezed-state CV QKD provides certain advantages over alternative QKD protocols. Studies have demonstrated that in low phase noise but high thermal noise environments, squeezed-state CV QKD can tolerate more loss compared to DV QKD protocols like the six-state protocol \cite{Kish2024comparisonof}. Additionally, non-Gaussian CV QKD protocols, which integrate operations like photon subtraction, can extend the communication range, but their implementation complexity makes squeezed-state CV QKD a favorable choice for practical applications \cite{PhysRevA.102.012608}. These distinctions underline the importance of assessing the security and performance of squeezed states under realistic conditions.

Importantly, squeezed states are never perfectly pure in practice and contain some excess noise in the AS quadrature above the level of inverse of the squeezing variance (currently the highest achieved purity of strongly squeezed state is 89\% \cite{vahlbruch2016detection}). This is concerned with the fact that optical loss, phase noise, parasitic nonlinearities etc. are unavoidably present in a device that produces squeezed states. This leads to effective emergence of the AS excess noise, which is larger, the stronger are the initial squeezing and the imperfections in the source. The typical values of AS noise observed in the experiment can range from 0.8dB \cite{mondain2019}, to 4.3dB and 8.4dB \cite{sun2019 ,kashiwazaki2020}. Such noise was recently studied in the context of phase sensitive amplification in optical communication in Ref.\cite{PSA10068233}.

The AS noise is typically assumed trusted in CV QKD security analysis, being a part of a trusted sender station \cite{naturecom2012, gehring2015implementation, jacobsen2018complete,PhysRevApplied.10.064028} and hence does not directly affect the security of CV QKD, at least in the idealized case of infinitely large data ensemble (also referred to as asymptotic case), which implies perfect parameter estimation in CV QKD. Furthermore, the outcomes of the measurement of the AS quadrature do not directly contribute to the key data. This is the main reason why the role of AS noise was not previously studied in the context of CV QKD. 

Trusted noise refers to noise that arises from known, well-characterized imperfections within the sender’s and receiver’s own devices, located in their isolated stations. Because these imperfections are closely monitored and their effects are understood, this noise is assumed not to assist an eavesdropper in gaining information about the key. In contrast, untrusted noise is any noise that cannot be confidently attributed to such internal, controlled sources. It includes noise originating from external or poorly characterized components, and thus cannot be ruled out as a potential avenue for an eavesdropper’s attack.

However, an eavesdropper may at least partially control the preparation device and hold the purification of the AS noise, which undermines the trust assumption on this type of noise. Also, the trusted parties may not be able to distinguish the AS noise from the noise introduced in the channel, and hence assume the noise to be untrusted. Furthermore, in the finite-size setting of a limited data ensemble, the parameter estimation, which is an essential part of a CV QKD protocol, may probabilistically rely on the measurements of the AS quadrature and be corrupt even by the trusted AS noise. Finally, the AS noise, even if assumed trusted, leads to untrusted excess noise in the fluctuating free-space channels and imposes limitations on the levels of squeezing already for the pure states \cite{Derkach_2020}.

Hence it is necessary to analyse the role of AS noise in CV QKD, which we do in the current paper. We develop a purification model for incorporating the AS noise in security analysis of CV QKD. Using the model, we demonstrate that trusted AS noise in CV-QKD has no negative impact and can even have a slight positive effect in the asymptotic regime of infinitely large data ensembles. However, when the assumption of trusted AS noise is waived, the protocol's security is substantially and adversely impacted. We also present the negative effects on the parameter estimation and subsequent reduction of security range if the AS quadrature measurements are used for parameter estimation, which holds true already for the trusted AS and is worsen if the AS excess noise is untrusted. Finally, we analyse the negative role of AS noise in fluctuating free-space channels in both trusted and untrusted scenarios. The obtained results are essential for realization of the squeezed-state CV QKD protocols in practical scenarios and further development of the protocols with realistic squeezed states.

Our paper is structured as follows: we describe the protocol and adopt the most conservative and general approach in Sec. \ref{sec:protocol} where we describe the protocol, build the model for purification of trusted AS noise, and derive the relevant entropic quantities; in Sec. \ref{sec:asymptotic} we discuss the effects of trusted and untrusted AS in the asymptotic regime; in Sec. \ref{sec:finite} we address the role of AS in the parameter estimation and obtain the implications on the key rate and security range of the protocol as well as suggest the optimal protocol implementations in the presence of AS noise; in Sec \ref{sec:fluctuating channels} we study the role of AS in fluctuating free space channels, and then finish the paper with concluding remarks.

\section{CV QKD protocol with AS excess noise}\label{sec:protocol}
\begin{figure}
    \centering
    \includegraphics[width=0.99\textwidth]{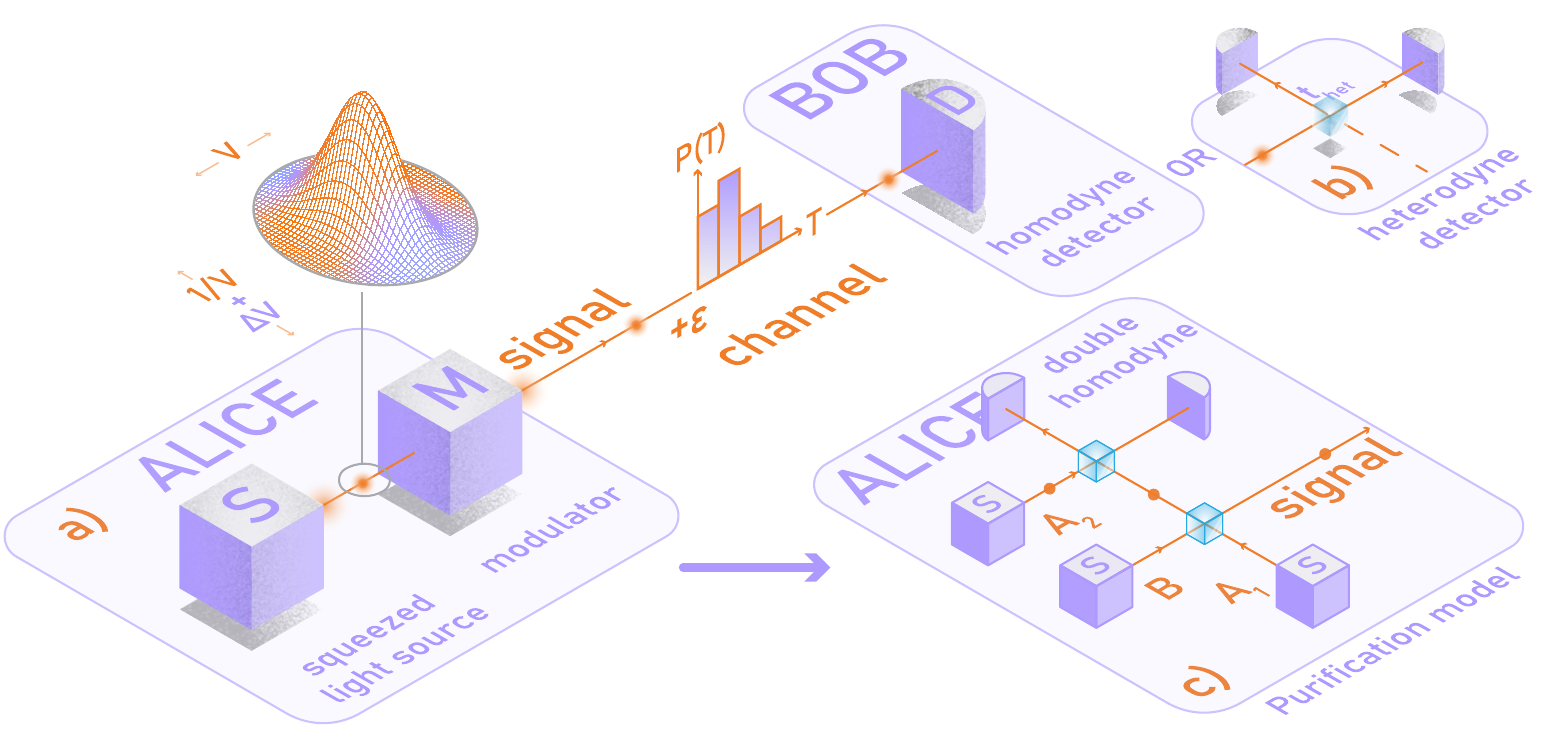}
    \caption{a) Prepare and measure scheme of a CV QKD protocol with noisy squeezed states. Alice modulates the squeezed state, characterized by the squeezed variance $V$ and AS excess noise $\Delta V$, according to normally distributed random variables and sends the Gaussian modulated squeezed state through a quantum channel characterized by transmittance distribution $P(T)$ and excess noise $\epsilon$. All modulator imperfections are attributed to the source. Bob at his end performs homodyne detection or b) unbalanced heterodyne detection with $t_{het}$. c) Equivalent general source model, used for purification and security analysis, based on the balanced coupling of two oppositely squeezed states in modes $A_1$ and $B$ with different squeezed variances $V_1$, $V_2$ and a balanced heterodyne measurement at the Alice's side, fed by a squeezed state with variance $V_m$ in mode $A_2$.}
    \label{fig:model}
\end{figure}

To investigate the role of AS noise in the squeezed-state CV-QKD, we consider the prepare-and-measure (P\&M) squeezed-state protocol, as shown in Fig. \ref{fig:model}, in which a trusted party (hereafter referred to as Alice) generates a squeezed signal state with squeezed quadrature variance V and AS quadrature variance $1/V+\Delta V$, $\Delta V$ being the excess AS noise. Alice then modulates the squeezed quadrature and AS quadrature (here and further with no loss of generality we assume the $x$-quadrature to be squeezed) according to Gaussian distributed random variables $x_m$ and $p_m$ with variances $V_x$ and $V_p$, respectively. Note that while the secret key is typically obtained from the measurements of the squeezed quadrature, we assume that both quadratures are modulated in the general case, particularly for estimating the channel parameters. State preparation imperfections (including modulation) are attributed to the squeezed light source. The modulated squeezed state is sent to a remote trusted party (hereafter referred to as Bob) via a quantum channel characterized by transmittance $T$, and quadrature excess noise $\epsilon$ (which we, without loss of generality, quantify with respect to the channel input). 
Bob then performs a homodyne detection to measure either x- or p-quadrature of the quantum state he receives. Alternatively, Bob can measure both quadratures simultaneously with a generally unbalanced heterodyne detector (double homodyne) with the balancing $t_{het}$ as shown in Fig. \ref{fig:model}b. After a sufficient number of transmitted pulses are measured, trusted parties perform error correction and privacy amplification on their data to obtain the secret key. The lower bound on the secret key, secure against collective attacks (directly extendable to security against general attacks in the asymptotic regime \cite{PhysRevLett.118.200501}) is given by: 

\begin{equation}\label{eq:key_general}
    K_\infty^{DR/RR}=\text{Max}\{0,\beta I_{AB}-\chi_{AE/BE}\}
\end{equation} 
for direct reconciliation (DR) and reverse reconciliation (RR) respectively, when either Alice or Bob is the reference side for the error correction of the key data. Here $\beta \in (0,1)$ stands for the post-processing (error-correction) efficiency, which scales down the achievable classical mutual information $I_{AB}$ between Alice and Bob, $\chi_{AE}$ and $\chi_{BE}$ are the upper bounds on the accessible information of an eavesdropper (hereafter referred to as Eve) on Alice's or Bob's data in DR and RR scenarios respectively, referred to as the Holevo bound \cite{Holevo}.
Mutual information between Alice and Bob's data sets $I_{AB}$ can be calculated from the variances and covariance of Alice and Bob's data (with variances $V_A$ and $V_B$ respectively) and reads
\begin{equation}
    I_{AB}=\frac{1}{2}\log_2\left[\frac{V_A}{V_{A|B}}\right]=\frac{1}{2}\log_2\left[\frac{V_B}{V_{B|A}}\right].
    \label{eq:IAB}
\end{equation}
The conditional variance $V_{i|j}=V_i-C^2_{ij}/V_j$ (for $i=B,A$ and $j=A,B$ respectively) is obtained from the variances $V_{i,j}$ and covariance $C_{ij}$ of Alice and Bob data. 
The Holevo bounds are obtained from the von Neumann entropy of the state available to Eve $S(E)$ before and $S(E|A/B)$ after conditioning on the measurements by Alice or Bob for DR or RR respectively, $\chi_{AE/BE}=S(E)-S(E|A/B)$. \\
To assess the amount of information on the raw key available to Eve, we model this protocol using the general equivalent entanglement-based (EB) scheme presented in \cite{Usenko_2011}, as shown in Fig. \ref{fig:model}b. This scheme involves coupling two oppositely squeezed states with $x$-quadrature variances $V_1$ and $V_2$ so that Alice then performs heterodyne detection on her mode \textit{A} by coupling it on a balanced beamsplitter to a mode containing a pure squeezed state with $V_m$ $x$-quadrature variance. \par
We analyze security of the protocol against Gaussian collective attacks, which were shown optimal against Gaussian CV QKD \cite{PhysRevLett.97.190502,PhysRevLett.97.190503}, by obtaining the Holevo bound from the covariance matrices of the respective states shared between the trusted parties and an eavesdropper. The three-mode covariance matrix of the general EB scheme, taking into account the effect of the quantum channel in mode B, is

\begin{center}
\begin{equation}\label{eq:3-mode-cm}
\gamma_{A_1A_2B}=\begin{bNiceMatrix}[margin]
\gamma_{A_1}&\sigma_A&\sqrt{T}\sigma_{AB}&  \\
\sigma_A&\gamma_{A_2}&\sqrt{T}\sigma_{AB}&  \\
\sqrt{T}\sigma_{AB}&\sqrt{T}\sigma_{AB}& \gamma_{B}
\end{bNiceMatrix},
\end{equation}
    \end{center}
    where the submatrices are  
    \begin{center}
    \begin{gather*}
        \gamma_{A_1}=\gamma_{A_2}=\left[\begin{array}{cc}
\frac{V_1+V_2+2V_m}{4}& 0  \\
      0 &\frac{V_m(V_1+V_2)+2V_1 V_2}{4V_1V_2V_m }  \\
        \end{array}\right], \sigma_A=\left[\begin{array}{cc}
\frac{V_1+V_2-2V_m}{4}& 0  \\
      0 &\frac{V_m(V_1+V_2)-2V_1 V_2}{4V_1V_2V_m }  \\
        \end{array}\right],
        \end{gather*}
         \end{center}
         \begin{center}
        \begin{gather*}\sigma_{AB}=\left[\begin{array}{cc}
\frac{V_2-V_1}{2\sqrt{2}}& 0  \\
      0 &\frac{V_1-V_2}{2\sqrt{2}V_1 V_2}  \\
        \end{array}\right],\hspace{0.1cm} \gamma_{B}=\left[\begin{array}{cc}
        T\left(\frac{V_1+V_2}{2}+\epsilon\right) +1-T & 0  \\
       0  & T\left(\frac{V_1+V_2}{2V_1V_2}+\epsilon+\Delta V_u\right)+1-T 
    \end{array}\right].\hspace{0.1cm}
    \end{gather*}
    \end{center}
When Bob performs imbalanced heterodyne measurement, the covariance matrix $\gamma_{A_1A_2BB'}$ reads
\begin{center}
\begin{equation}\label{eq:3-mode-het-cm}
\begin{bNiceMatrix}[margin]
\gamma_{A_1}&\sigma_A&\sqrt{t_{het} T}\sigma_{AB}& -\sqrt{(1-t_{het}) T}\sigma_{AB} \\
\sigma_A&\gamma_{A_2}&\sqrt{t_{het} T}\sigma_{AB}& -\sqrt{(1-t_{het}) T}\sigma_{AB} \\
\sqrt{t_{het} T}\sigma_{AB}&\sqrt{t_{het} T}\sigma_{AB}& t_{het}\gamma_{B}+1-t_{het}&\sqrt{1-t_{het}}\sqrt{t_{het}}(\mathds{1}-\gamma_B)\\
-\sqrt{(1-t_{het}) T}\sigma_{AB}&-\sqrt{(1-t_{het}) T}\sigma_{AB}&\sqrt{1-t_{het}}\sqrt{t_{het}}(\mathds{1}-\gamma_B)&(1-t_{het})\gamma_{B}+t_{het}
\end{bNiceMatrix}.
\end{equation}
    \end{center}

The squeezed variances of the prepare-and-measure and entanglement-based \\ schemes are related by the equivalence conditions \cite{Usenko_2011}:

\begin{equation}
    V_{1,2}=V+V_x\pm\sqrt{\frac{(V+V_x)[V_x+V V_p(V+V_x)]}{1+V V_p}},
\end{equation}
\begin{equation}
    V_m=\frac{V^2 V_p (V+V_x)}{V_x(1+V V_p)}.
\end{equation}

The trusted parties may assume the AS noise $\Delta V$ to be either trusted (hence, being out of control by an eavesdropper) or untrusted (fully controlled by Eve). We further denote the trusted and untrusted AS noise as $\Delta V_t$ and $\Delta V_u$, respectively. We model the trusted AS noise by redefining the modulation variance of the AS quadrature $V_p\rightarrow V_p+\Delta V_t$ ($\Delta V_t$ does not contribute to the mutual information between Alice and Bob). On the other hand, the untrusted AS noise $\Delta V_u$ is added to the AS quadrature similarly to the channel noise $\epsilon$ (for a detailed explanation of the assumption that all noise is concentrated in the AS quadrature, see Sec.\ref{sec: noisy squeezed state}).

In the worst-case scenario, we assume that Eve can purify the state shared between Alice and Bob; thus $S(E)=S(AB)$, where $S(AB)$ is the Von Neumann entropy of the state shared between trusted parties, it can be calculated from the bosonic entropic function \cite{Serafini_2005}, as $S(AB)=\sum^3_{i=1} G[(\lambda_i-1)/2]$, where $\lambda_{1,2,3}$ are the symplectic eigenvalues of the covariance matrix $\gamma_{AB}$ and $G(x)=(x+1)\log(x+1)-x\log x$. Similarly, conditional entropies $S(E|A)$ and $S(E|B)$ can be obtained from the corresponding conditional covariance matrices:
\begin{equation}
    \gamma_{i|j}=\gamma_i - \sigma_{ij}(X.\gamma_j .X)^{MP}\sigma^T_{ij},
    \label{eq:condmat}
\end{equation}
where $i=A,B$ and $j=B,A$ respectively, with matrices $X=Diag(1,0)$ for x-quadrature measurement and $X=Diag(0,1)$ for p-quadrature measurement, while for the heterodyne measurement the conditioning (\ref{eq:condmat}) has to be applied twice on both x- and p-quadrature measurements on the two modes of the heterodyne detector. \\ Although for the two-mode conditional matrices, the symplectic eigenvalues can be found analytically, the eigenvalues of the 3-mode covariance matrix (\ref{eq:3-mode-cm}) and the larger matrices are found numerically to be used further in the calculation of the Holevo bound and the key rates.

\begin{figure}
    \centering
    \includegraphics[width=1\linewidth]{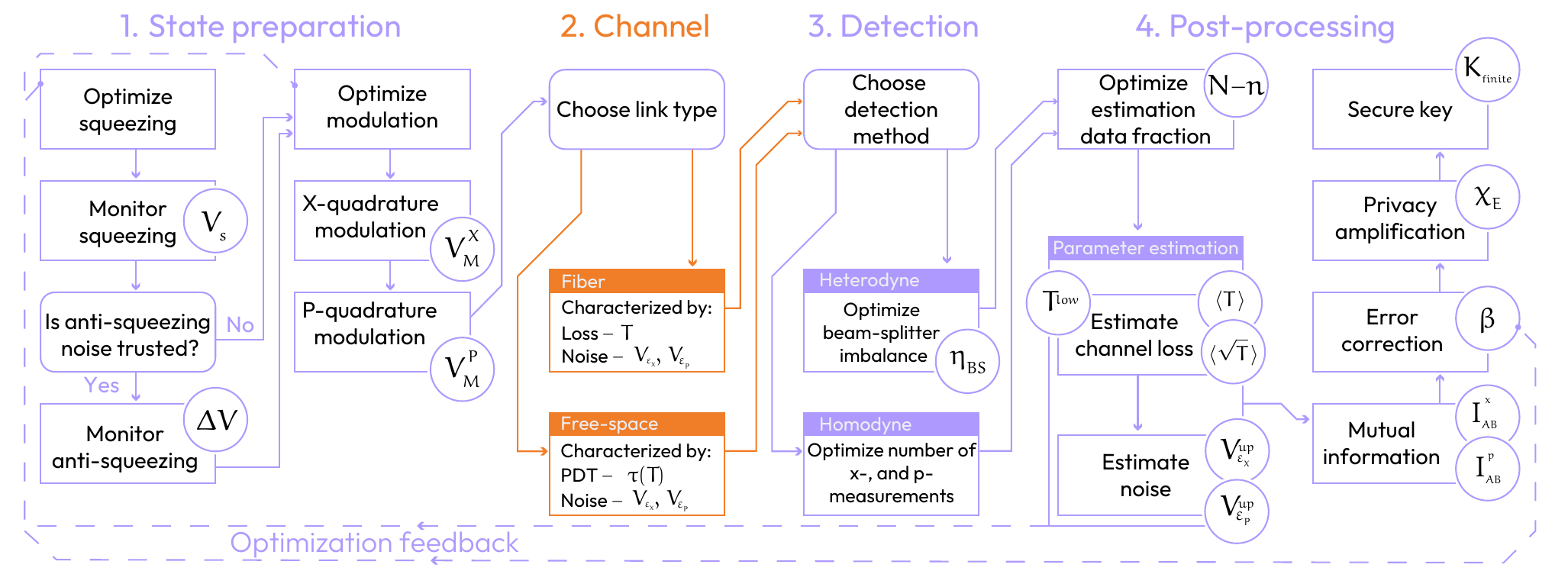}
    \caption{Diagram of the protocol operation indicating relevant parameters and estimated quantities. \textit{State preparation} optimization is based on estimated channel parameters and error correction efficiency $\beta$. Not trusting anti-squeezing noise will translate into significant increase of channel noise in respective quadrature. Monitoring is required to trust the anti-squeezing noise $\Delta V$. In a fiber link channel loss remains fixed and (aside from channel noise) only lower bound on transmittance $T^{low}$ needs to be estimated. In a free-space link statistical properties of probability distribution of transmittance, namely $\langle T \rangle$ and $\langle \sqrt{T} \rangle$, are required to correctly assess the secure key. Note that only optimization of data fraction can be done after the block measurement has already been commenced, other parameters are optimized before the protocol round begins, i.e. prior to the generation and transmission of quantum states. }
    \label{fig:steps}
\end{figure}

The mutual information and the Holevo bound depend on the type of measurement Bob performs. We consider two major scenarios as follows:

\begin{description}
   \item[Bob's homodyne measurement approach:] Bob performs homodyne detection on mode $B$ and switches between the measurement of squeezed and AS quadratures with a certain ratio. Equivalently, Alice can choose different quadratures (x or p) to be squeezed to achieve the same result. Since Bob measures both the quadratures, albeit at different times, when all the measurements are utilized for key generation, the asymptotic key rate $K_{\infty}$ from Eq. (\ref{eq:key_general}) is given by 
   \begin{equation}
       K_{\infty}=\underbrace{\text{Max}\{0,I^x_{AB}-[S(AB)-S(A|B^x)]\}}_{K^x_{\infty}}+\underbrace{\text{Max}\{0,I^p_{AB}-[S(AB)-S(A|B^p)]\}}_{K^p_{\infty}}
   \end{equation}
In this expression, the switching ratio between squeezed and AS quadrature measurements does not explicitly appear. This is because in the asymptotic regime, where the number of measurements approaches infinity, statistical fluctuations and biases due to finite sample sizes can be ignored. Consequently, the estimations of parameters like noise and channel transmission are exact, and no bias correction is necessary. Therefore, biases do not appear in the key rate expression for the asymptotic case.\\
However, in practical implementations with a finite number of signals, the switching ratio plays a crucial role. It affects the precision of parameter estimations and thus impacts the finite-size key rate. We will elaborate on the finite-size effects and the optimization of the switching ratio in subsequent sections (Sec.\ref{sec:finite}).\\
   For simplicity, we present all the key rate expressions hereafter in the more relevant RR scenario (allowing for much longer secure distances \cite{grosshans2003quantum}). The key rate expressions for DR can be obtained by interchanging modes $A$ and $B$.
    $K^x_{\infty}$ and $K^p_{\infty}$ are the asymptotic key rates between Alice and Bob, when Bob performs homodyne measurement of squeezed and AS quadrature, respectively. $I^i_{AB}$ and $S(AB)-S(A|B^i)$ are the mutual information and Holevo bound when Bob performs the homodyne measurement of squeezed ($i=x$) and AS quadrature ($i=p$).  \\

   \item[Bob's heterodyne measurement approach:] Bob performs imbalanced heterodyne measurement, as shown in Fig. \ref{fig:model}b. Similarly to the previous case, the asymptotic key rate expression is different depending on whether the AS quadrature measurements are used for key generation. The asymptotic key rate when AS quadrature is used for key generation is (where $B^x$ and $B^p$ means that Bob measures x-quadrature in mode B and p-quadrature in mode $B'$ of the heterodyne detector):
   \begin{equation}
       K_{\infty}=
       \text{Max}\{0,I^x_{AB}+I^p_{AB}-\left[S(AB)-S(A|B^xB^p)\right]\}
   \end{equation}
   When the AS quadrature is not used for key generation, the asymptotic key rate is:
   \begin{equation}
       K_{\infty}=
       \text{Max}\{0,I^x_{AB}-\left[S(AB^xB^p)-S(AB^p|B^x)\right]\}
   \end{equation}
    Since the trusted parties do not use the AS quadrature for key generation, they can also disclose (publicly announce) the AS quadrature measurement outcomes, which can provide a practical advantage in parameter estimation and choosing error correction codes. The key rate when AS quadrature measurements are disclosed is:
   \begin{equation}
       K_{\infty}=
       \text{Max}\{0,I^x_{AB}-\left[S(AB^x|B^p)-S(A|B^xB^p)\right]\}
   \end{equation}
\end{description}
 To study the optimal performance of CV QKD with untrusted AS noise (on top of symmetric noise $\epsilon$) in the asymptotic regime, we only consider the measurement of squeezed quadrature by Bob. A brief summary of all stages of the protocol implementation and relevant quantities are shown in Fig. \ref{fig:steps}. 

\section{Performance of CV QKD with noisy squeezed states}
\subsection{Asymptotic key rate}\label{sec:asymptotic}
For the case when the AS noise can be trusted and the channel is purely attenuating ($\epsilon=0$), there is no need to build an equivalent EPR state, as the information available to Eve can be obtained directly from the other output of the beamsplitter, which explicitly models the channel attenuation \cite{PhysRevLett.94.020504}. The covariance matrix of the state available to Eve is
\begin{equation}
    \gamma_{E}=\left[\begin{array}{cc}
        (1-T)(V_x+V)+T & 0 \\
        0 & (1-T)(\frac{1}{V}+\Delta V_t)+T 
    \end{array}\right],
\end{equation}
where $\Delta V_t$ is the trusted AS excess noise.
Conditional state of Eve on the homodyne measurement of $x$-quadrature by Bob (the most basic scenario, as the AS quadrature measurements are not required in the asymptotic regime) is
\begin{equation}
    \gamma_{E|B}=\left[\begin{array}{cc}
        (V + V_x)/[1 + T(-1 + V + V_x)] & 0 \\
        0 & T + (1 - T) (1/V + \Delta V_t) 
    \end{array}\right].
\end{equation}
Conditional state of Eve on measurement of $x$-quadrature by Alice is
\begin{equation}
    \gamma_{E|A}=\left[\begin{array}{cc}
        (1-T)V+T & 0 \\
        0 & T + (1 - T) (1/V + \Delta V_t) 
    \end{array}\right].
\end{equation}

The symplectic eigenvalues of the single mode matrices $\gamma_y$ given above (where $y$ is $E,E|A$ and $E|B$) are obtained by the relation $\lambda_y=\sqrt{Det(\gamma_y)}$.
Mutual information between trusted parties is not influenced by AS noise, as the AS quadrature is not used for key generation. On the other hand it reduces the Holevo bound and we obtain the following analytical expressions for the Holevo bound in the limit of infinitely strong AS noise:
\begin{equation}
    \lim_{\Delta V_t \to \infty} \chi_{EB}=\frac{1}{2}\log_2 \left[\frac{[V+V_x-T(V+V_x-1)][1+T(V+V_x-1)]}{V+V_x}\right]=I_{EB}
\end{equation}

\begin{equation}
    \lim_{\Delta V_t \to \infty} \chi_{EA}=\frac{1}{2}\log_2 \left[1+\frac{V_x( 1- T)}{T + V(1 - T)}\right]=I_{EA},
\end{equation}
where $I_{EB}$ and $I_{EA}$ are the classical mutual information quantities between Eve and Bob, and Eve and Alice respectively. This implies that for a purely lossy channel and infinite trusted AS noise, collective attacks are equivalent to individual attacks. Since $\lim_{\Delta V_t \to \infty} \chi_{EB}$ is a function of $(V+V_x)$, where $0<V\leq 1$, for large modulation variance $V_x$, the Holevo bound for RR barely depends on the degree of squeezing. In the
limit of infinitely strong signal modulation $V_x \to \infty$ and
infinitely strong signal squeezing $V \to 0$ the improvement to the
Holevo bound due to the strong trusted AS noise vanishes and
the key rate analytically hits the known bounds for the squeezed-state
protocol, $\log_2\left[T/(1-T)\right]$ for DR and $\log_2\left[1/(1-T)\right]$ for RR \cite{e18010020}. \\ 

\begin{figure}
    \centering
    \begin{subfigure}{.45\linewidth}
        \includegraphics[width=\linewidth]{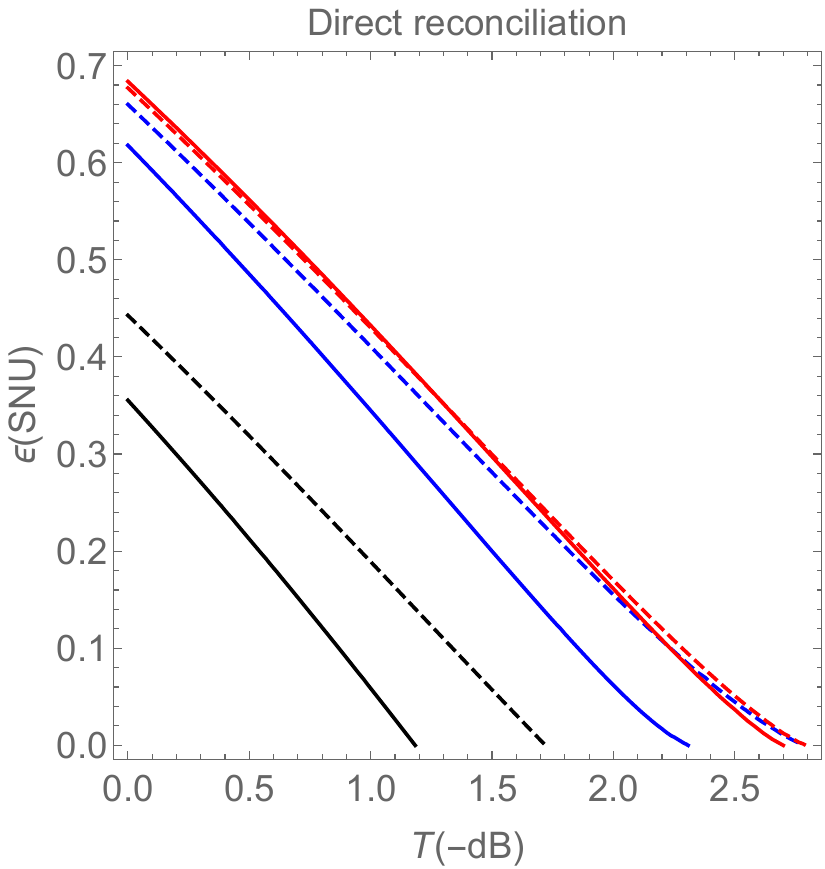}
    \end{subfigure}\hfill
    \begin{subfigure}{.45\linewidth}
        \includegraphics[width=\linewidth]{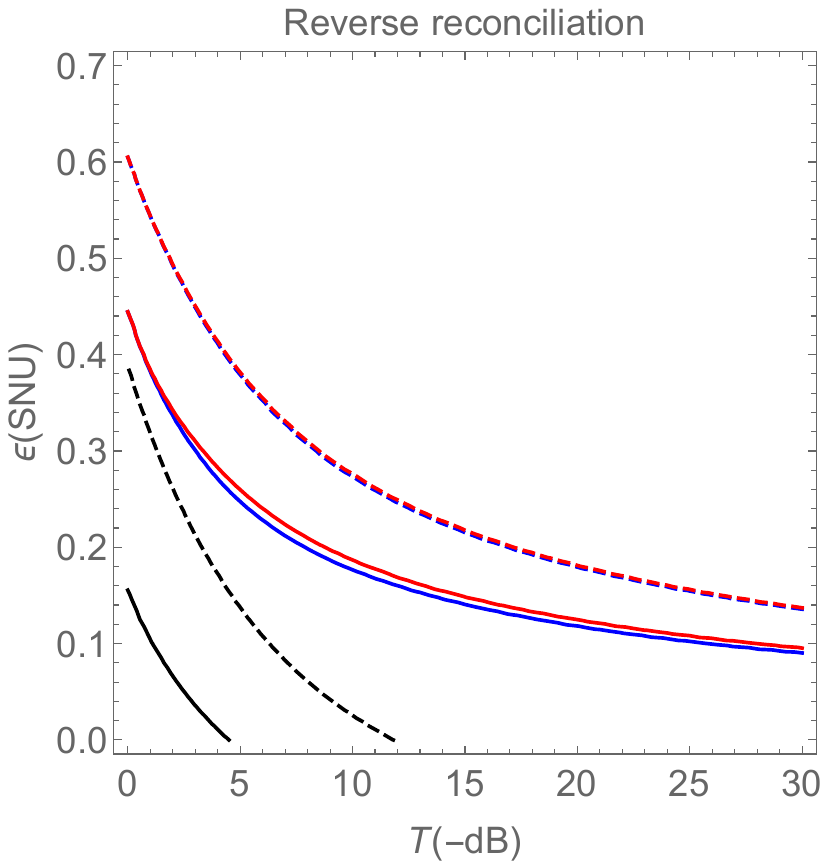}
    \end{subfigure}
    \caption{Maximal tolerable channel noise ($\epsilon$) with respect to channel attenuation $T$ (in dB scale) for highly squeezed $V=0.1$ SNU (dashed) and moderately squeezed $V = 0.5$ (solid) source state with optimised modulation variance and reconciliation efficiency $\beta = 0.95$, in the presence of 0.5 SNU AS noise, when it is trusted (blue) and when it is untrusted (black). The red plot shows the tolerance to channel noise when 10 SNU trusted AS noise is added. Adding trusted AS noise showed no difference in the tolerance for RR when the source state is highly squeezed (0.1 SNU), hence the blue dashed plot and the red dashed plot overlap, while adding trusted AS noise drastically improved the tolerance in DR when the source state is moderately squeezed.}
    \label{fig:Tol_asy}
\end{figure}

\begin{figure}
    \centering
    \begin{subfigure}[b]{0.45\linewidth}
        \includegraphics[width=\linewidth]{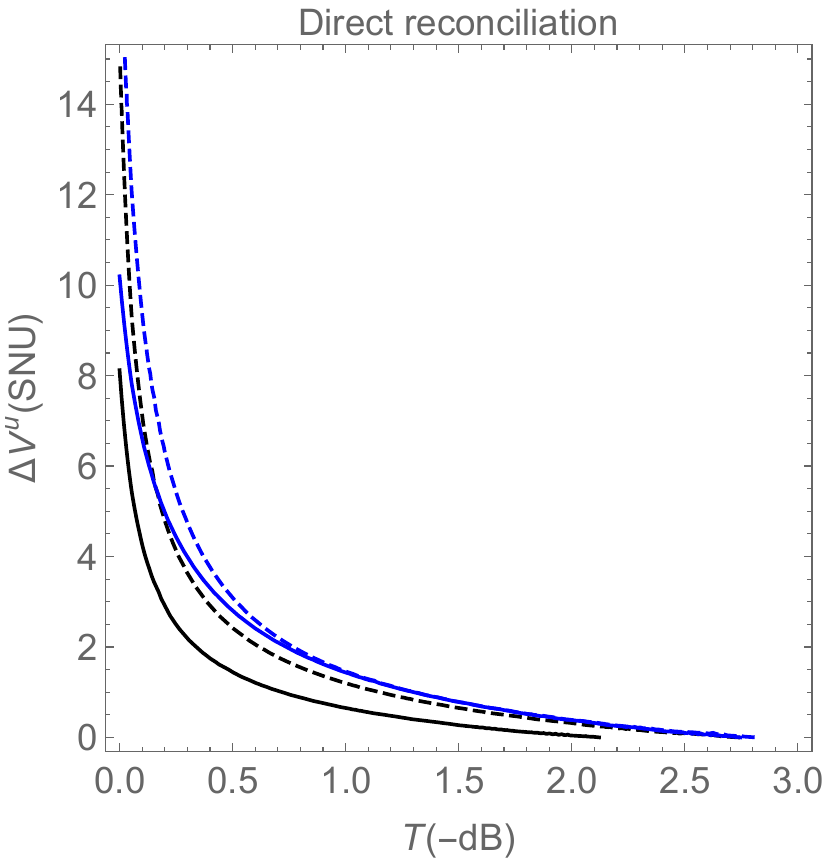}
    \end{subfigure}
    \hfill
    \begin{subfigure}[b]{0.45\linewidth}
        \includegraphics[width=\linewidth]{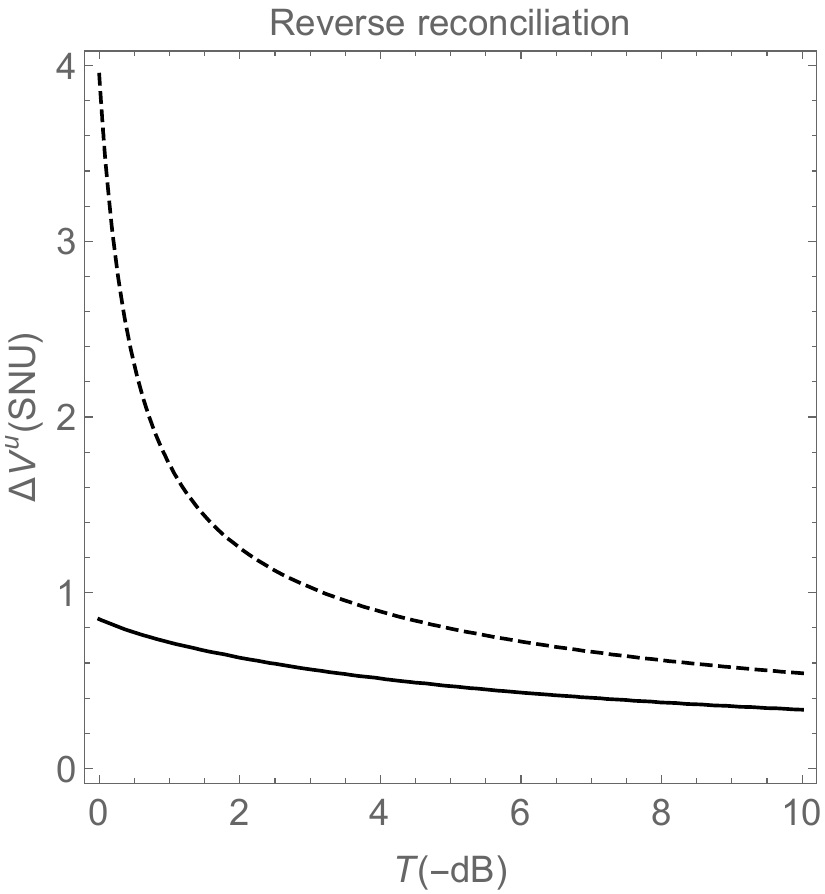}
    \end{subfigure}
    \caption{Tolerable untrusted AS noise ($\Delta V_u$) with respect to channel attenuation $T$, for highly squeezed ($V=0.1$ SNU, dashed) and moderately squeezed ($V=0.5$ SNU, solid) states, with optimized modulation variance ($\text{Max}[V_x]=10$ SNU). Shown with (blue) and without (black) trusted AS noise ($\Delta V_t=10$ SNU). Negligible impact of trusted noise in RR. Channel noise: $\epsilon=0.01$ SNU, reconciliation efficiency: $\beta=0.95$.}
    \label{fig:Tol_AS}
\end{figure}

In the presence of channel noise, no simple analytical expressions can be obtained for the Holevo bound, as we rely on the 3-mode purification described in Sec.\ref{sec:protocol}. The presence of trusted noise can improve tolerance to untrusted channel noise, as can be seen from the plots in Fig. \ref{fig:Tol_asy}, where we model the performance of the protocol assuming realistic error correction efficiency of $\beta=0.95$ \cite{mani2021multiedge} and optimizing the modulation variance. The improvement by the trusted AS noise is more evident in DR than in RR, being a case of the "fighting noise with noise" method \cite{PhysRevLett.102.130501,e18010020}. Interestingly, a moderately squeezed source ($V=0.5$ SNU) in the presence of optimal trusted AS noise has a higher tolerance to channel noise than the protocol with a higher degree of pure signal squeezing of the source state for DR. On the other hand, when the AS noise cannot be trusted, the tolerance of the protocol to channel noise is significantly reduced, as can be seen from the plots in Fig. \ref{fig:Tol_AS}, as the presence of untrusted AS noise increases the Holevo bound. This emphasises the need to control and to estimate noise in the AS quadrature, which would otherwise lead to underestimation of the Holevo bound.

\subsection{Finite-size key rate}\label{sec:finite}

A necessary requirement to verify the security of QKD is to estimate the channel parameters from which the Holevo bound can be found. Channel parameters can only be estimated within a certain confidence interval, which is inversely propositional to the number of states sent to Bob. Typically part of the received data is used for parameter estimation, and the rest is used for the key generation \cite{PhysRevA.90.062310}. It has been also shown that all the measurements can be used for both parameter estimation and key generation  by performing error correction before parameter estimation \cite{leverrier2015composable}. For completeness, we consider both cases, when part of the measurements are disclosed for parameter estimation (see \ref{sec: parameter estimation}), and when all the measurements are used for parameter estimation and key generation. 

As established in the previous section, AS noise is detrimental to the security of the squeezed-state protocol when it is untrusted. The adverse effect of AS noise is further amplified in the practical finite-size regime of limited data ensembles, as the presence of AS noise requires one to make the estimation of the channel noise independently for both the quadratures, utilising more available resources for parameter estimation, unless all the measurement outcomes are used both for the key and for the estimation.

In order not to underestimate the information available to the eavesdropper, we use the lower bound for the confidence interval of channel transmittance $T^{low}$ and the upper bound for channel noise $V_{\epsilon}^{up}$. We consider $\mathbf{E}(T^{low})=T-6.5\Upsilon$ and $\mathbf{E}(V_\epsilon^{up})=V_\epsilon+6.5s$ corresponding to the estimation error probability of $10^{-10}$, $\sigma$ and $s$ are the standard deviations of the channel transmittance and noise, respectively. In order to effectively use the available resources, we extend the parameter estimation from \cite{PhysRevA.81.062343,PhysRevA.90.062310} to use the measurements of both quadratures to potentially make a better estimation of the channel parameters (refer to \ref{sec: parameter estimation} for details). We obtain the following finite-size key rate:    
\begin{equation}\label{eq:key_untrusted antisqueeze}
    K_{finite}=\frac{n_k}{N}[K_{\infty}(T^{low},\hat{V}_{\epsilon_{x}}^{up},\hat{V}_{\epsilon_p}^{up})-\Delta(n)]
\end{equation}
In this context, $\Delta(n)$ denotes the correction term for privacy amplification\cite{PhysRevA.81.062343}. When all the measurements are used for key generation and parameter estimation, the finite-size key rate becomes:
\begin{equation}\label{eq:key_all}
    K_{finite}=K_{\infty}(T^{low},\hat{V}_{\epsilon_{x}}^{up},\hat{V}_{\epsilon_p}^{up})-\Delta(N)
\end{equation}
As using AS quadrature for the parameter estimation is detrimental to achieving large distances, the optimal finite-size key rate for the homodyne measurement approach is given by (\ref{eq:key_untrusted antisqueeze}), with $n_k=m$ ($m$ is the number of measurements of the squeezed quadrature). 

\begin{figure}
    \centering
    \begin{subfigure}{0.45\linewidth}
        \includegraphics[width=\linewidth]{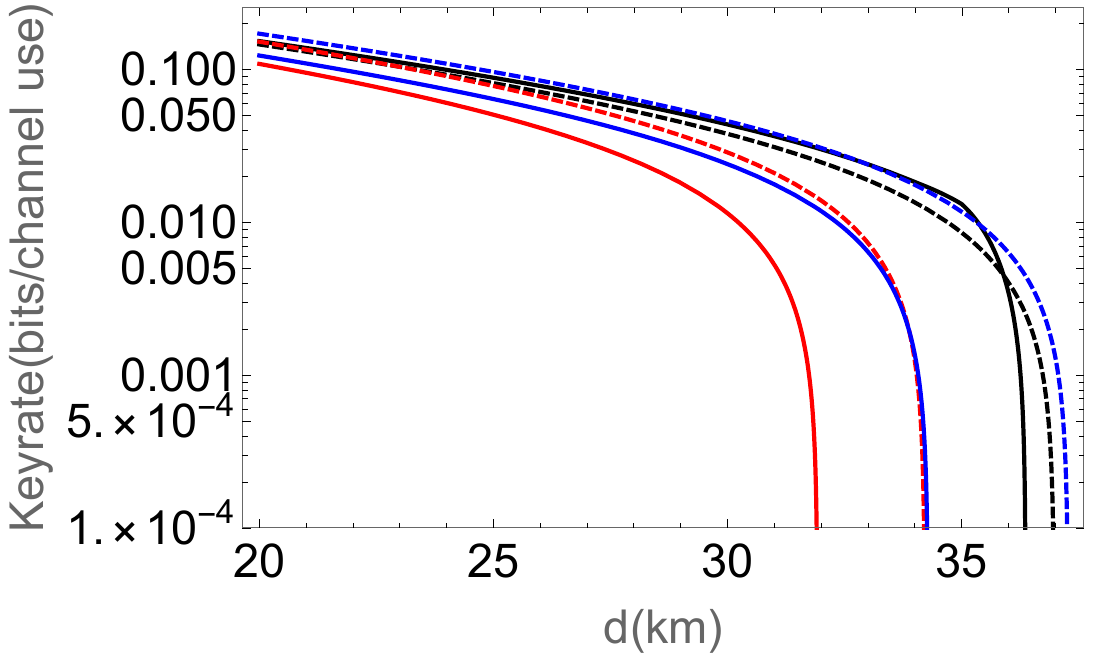}
        \caption{$\Delta V_t=0$, $\Delta V_u=0$, $N=10^6$}
        \label{fig:key_a}
    \end{subfigure}
    \begin{subfigure}{0.45\linewidth}
        \includegraphics[width=\linewidth]{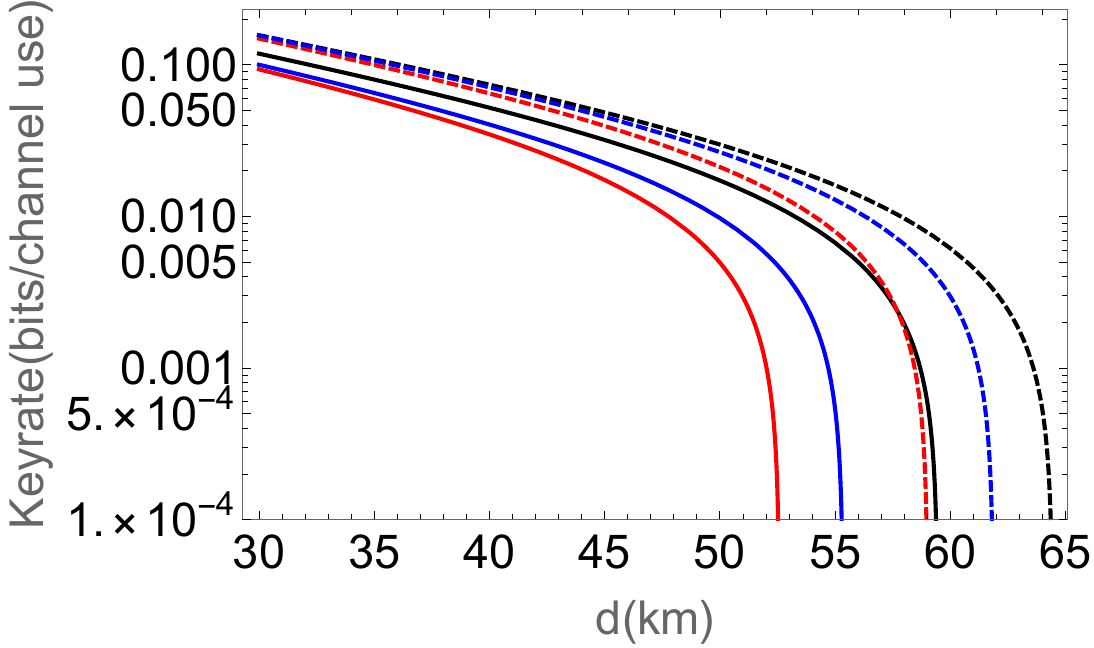}
        \caption{$\Delta V_t=0$, $\Delta V_u=0$, $N=10^7$}
        \label{fig:key_b}
    \end{subfigure}
    
    \begin{subfigure}{0.45\linewidth}
        \includegraphics[width=\linewidth]{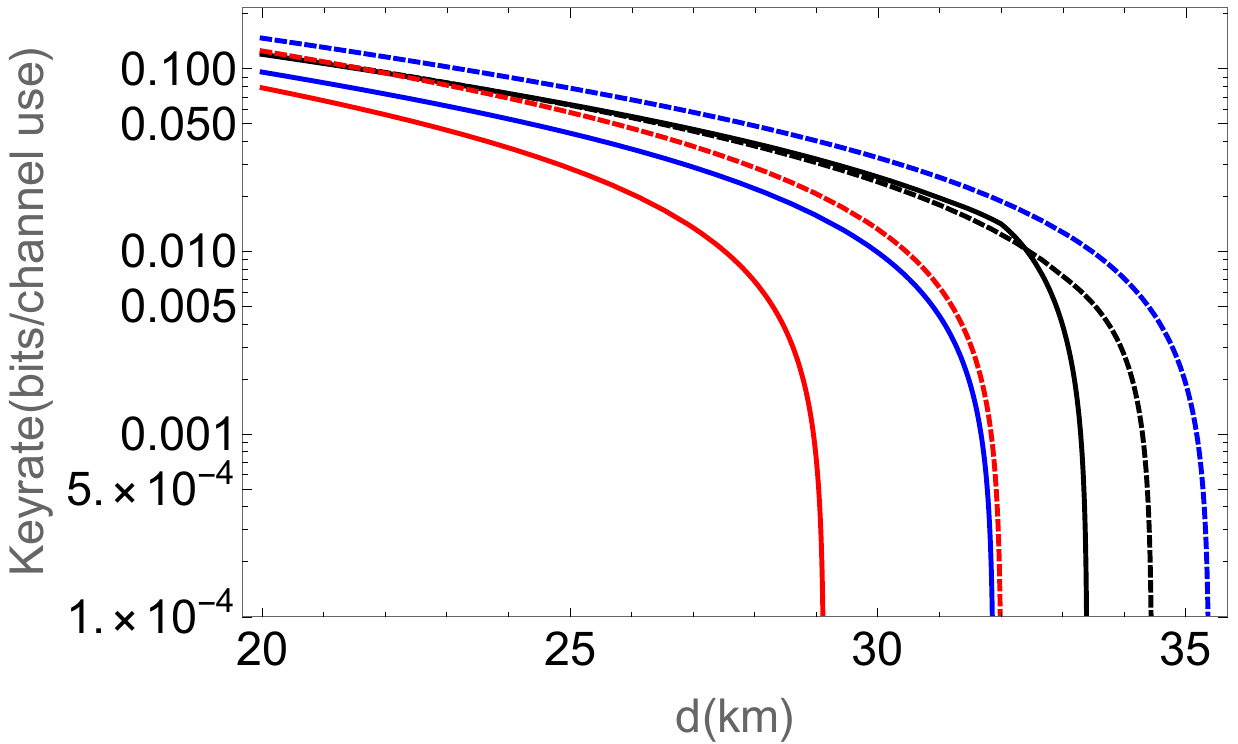}
        \caption{$\Delta V_t=0$, $\Delta V_u=0.05$, $N=10^6$}
        \label{fig:key_c}
    \end{subfigure}
    \begin{subfigure}{0.45\linewidth}
        \includegraphics[width=\linewidth]{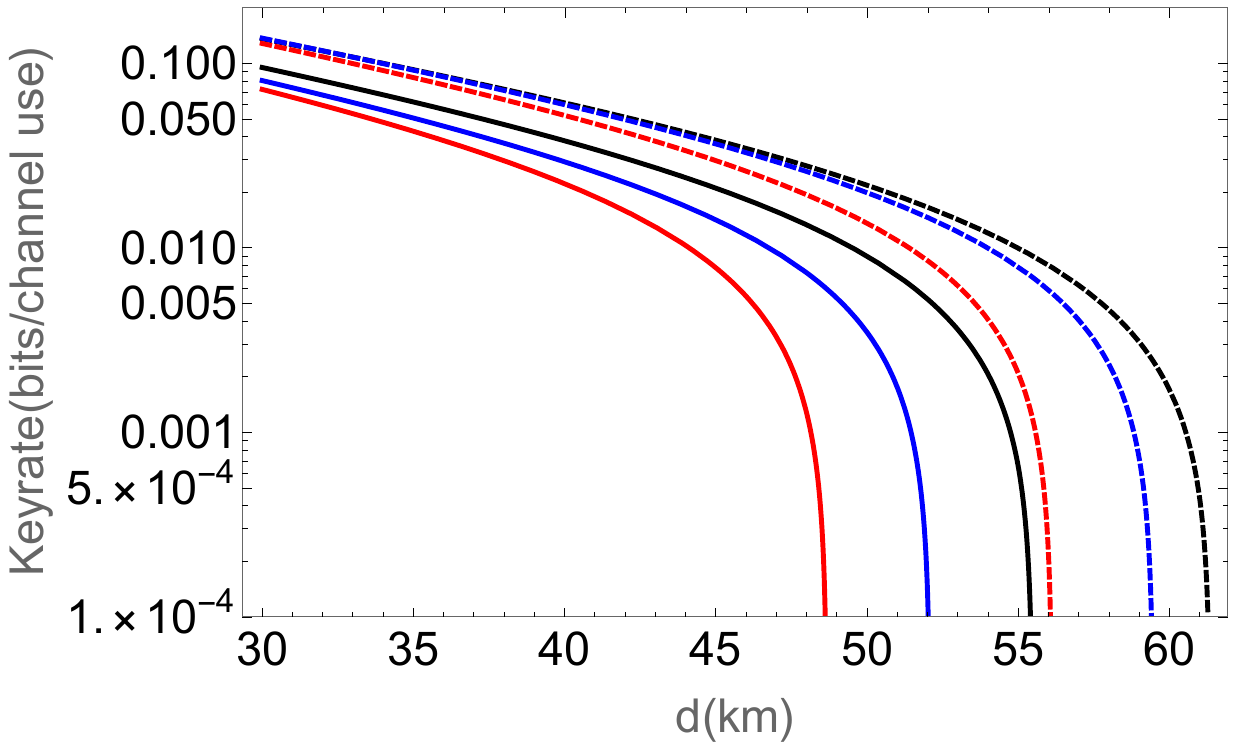}
        \caption{$\Delta V_t=0$, $\Delta V_u=0.05$, $N=10^7$}
        \label{fig:key_d}
    \end{subfigure}
    
    \begin{subfigure}{0.45\linewidth}
        \includegraphics[width=\linewidth]{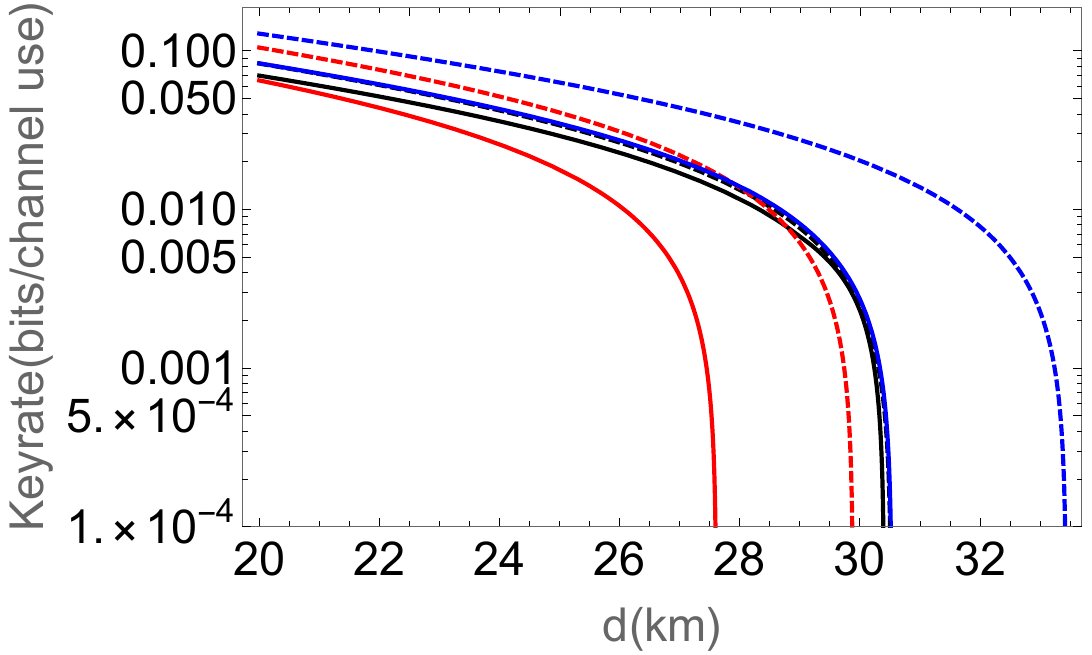}
        \caption{$\Delta V_t=5$, $\Delta V_u=0$, $N=10^6$}
        \label{fig:key_e}
    \end{subfigure}
    \begin{subfigure}{0.45\linewidth}
        \includegraphics[width=\linewidth]{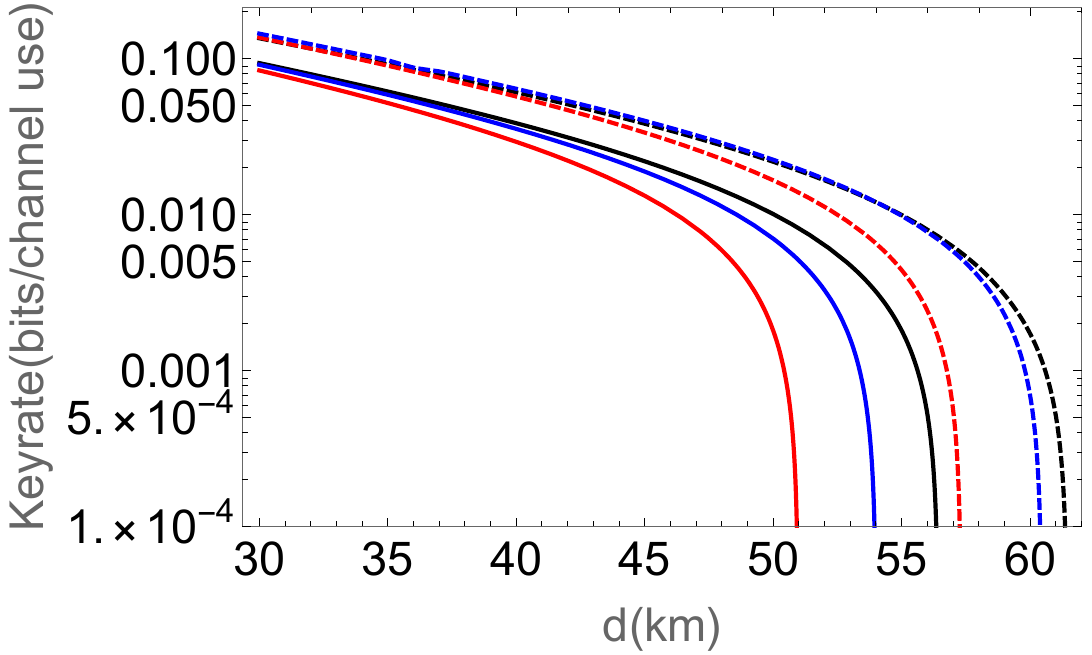}
        \caption{$\Delta V_t=5$, $\Delta V_u=0$, $N=10^7$}
        \label{fig:key_f}
    \end{subfigure}
    
    \caption{Finite-size key rate vs. distance when Bob performs biased homodyne detection (black), imbalanced heterodyne detection (blue) and when Bob performs imbalanced heterodyne detection and discloses AS measurement results (red), all in the RR scenario. The dashed plot represents a highly squeezed source state ($V=0.1$ SNU) and the solid plot represents a moderately squeezed source state ($V=0.5$ SNU). Modulation variances $V_x$ and $V_p$, transmittance of beamsplitter for heterodyne detection $t_{het}$, ratios of available resources used for key generation $r_k$, and parameter estimation $r_x$ and $r_p$ are optimized. The reconciliation efficiency is $\beta = 0.95$, the channel noise is $\epsilon = 0.05$ SNU, and the channel attenuation is 0.2 dB/km.}
    \label{fig:finite_key}
\end{figure}

\begin{figure}\centering
{\includegraphics[width=.39\linewidth]{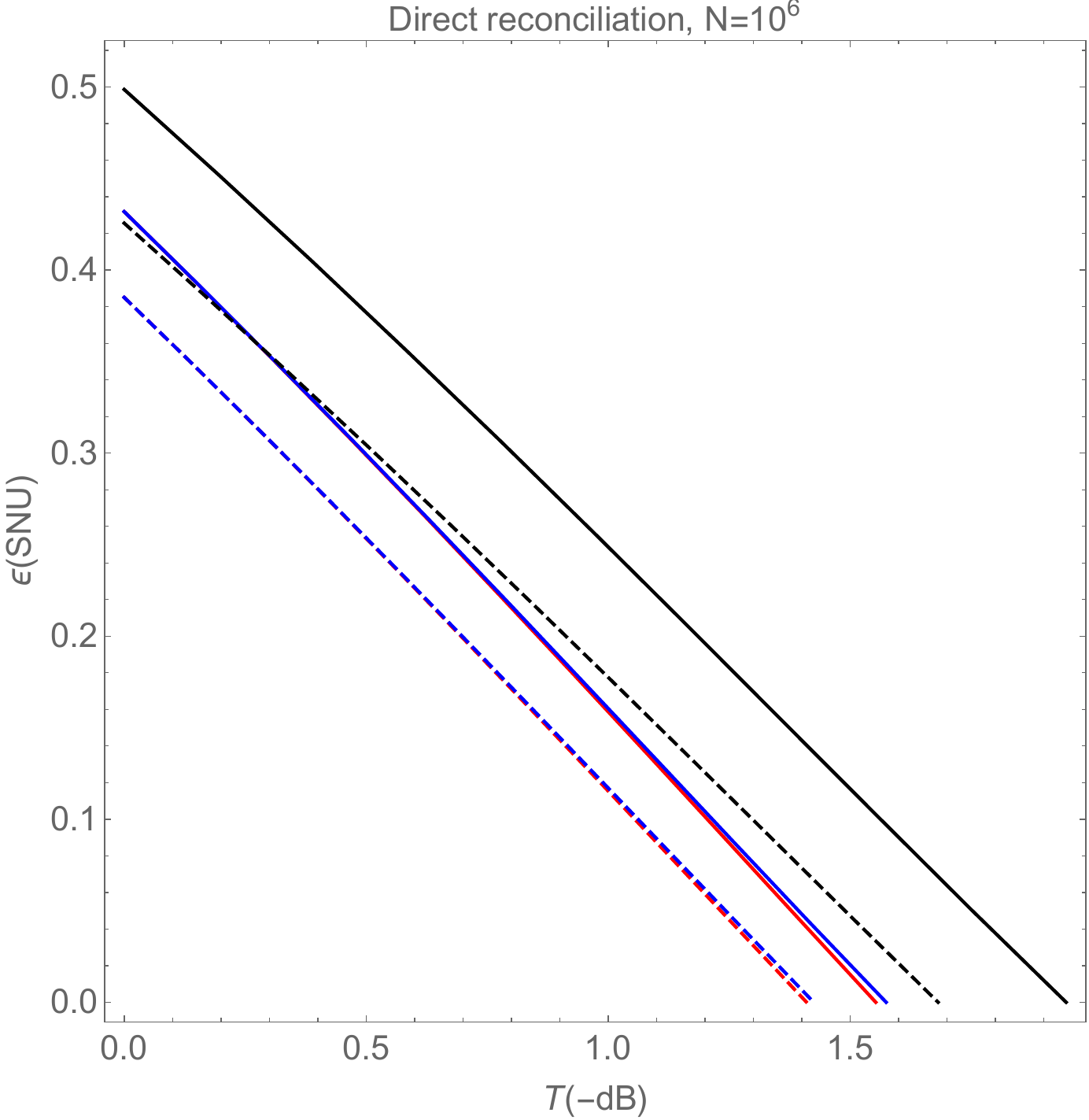}}
{\includegraphics[width=.39\linewidth]{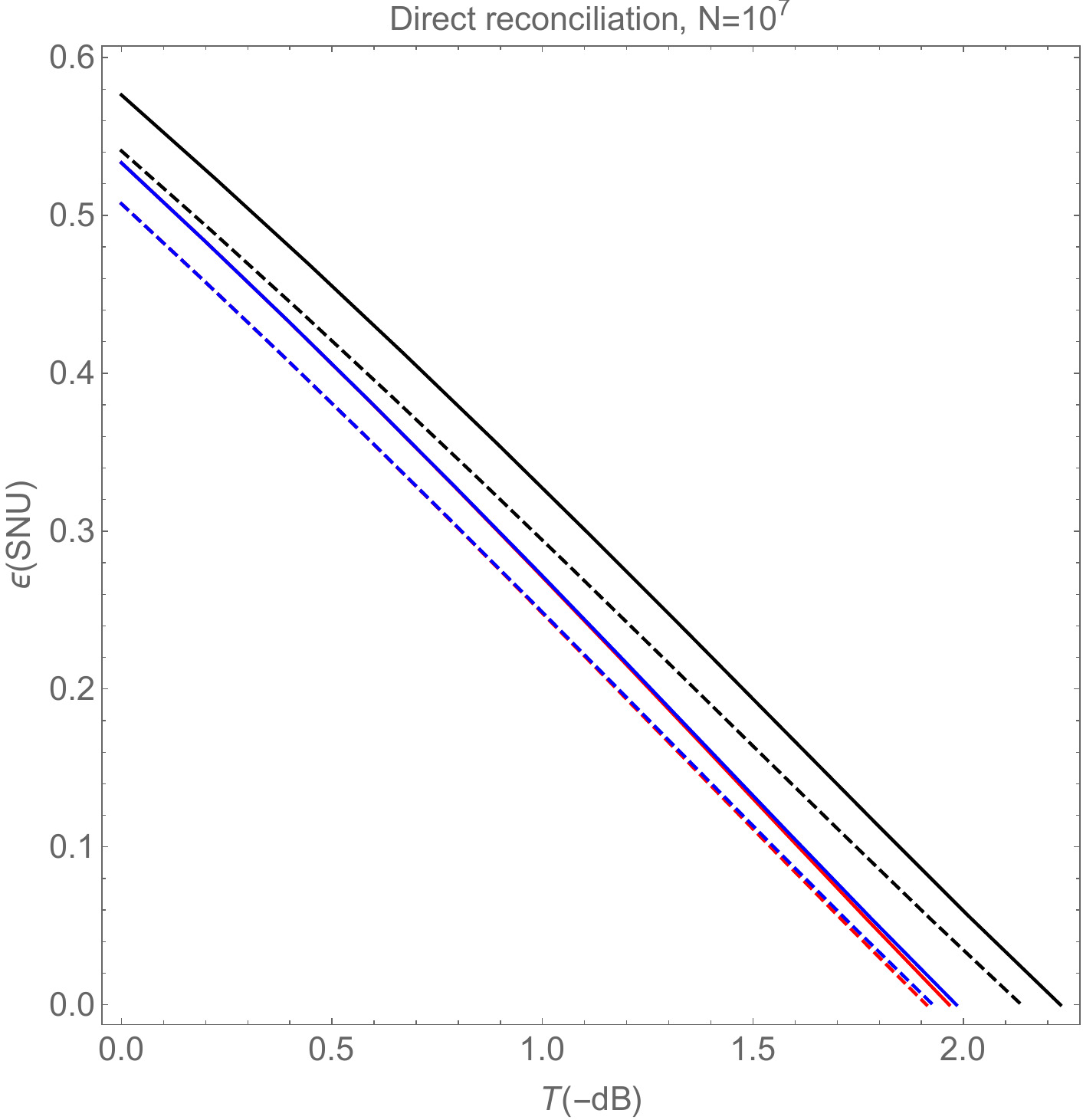}}\hfill
{\includegraphics[width=.39\linewidth]{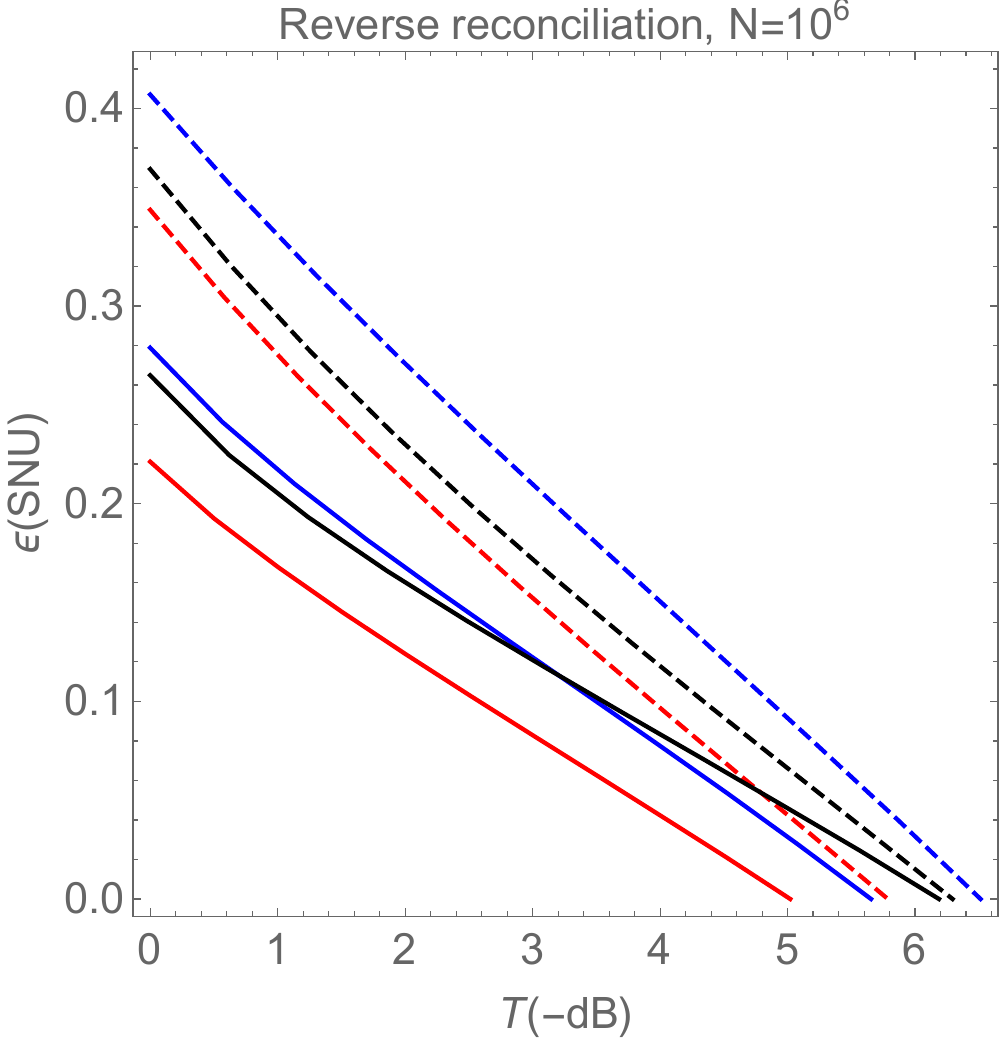}}
{\includegraphics[width=.39\linewidth]{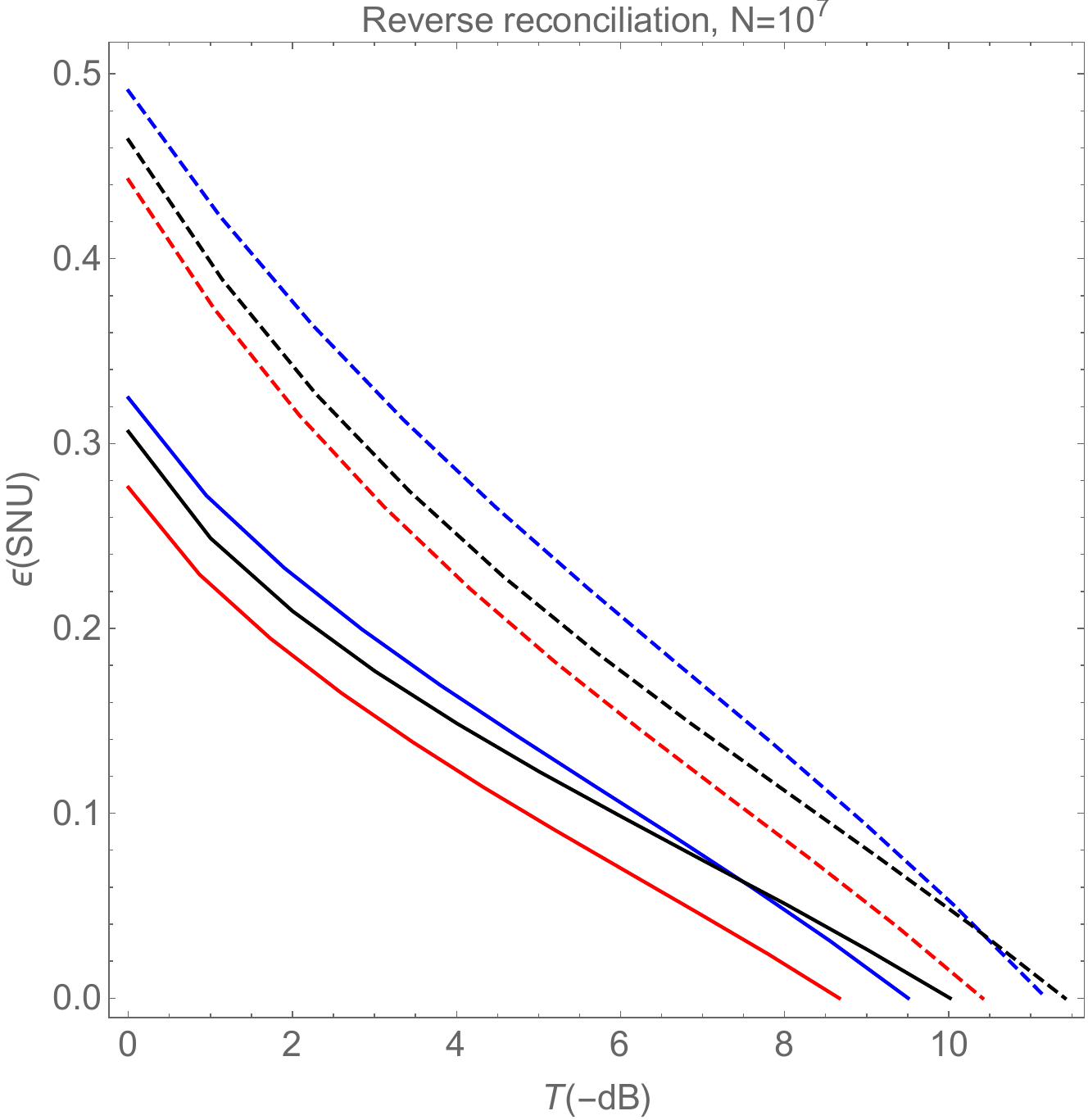}}

\caption{Tolerance to channel noise for biased homodyne (black) and imbalanced heterodyne measurement by Bob, when AS quadrature measurement is publicly announced (red) and when it is not announced (blue) in different reconciliation scenarios and at different data block sizes. The modulation variance of the squeezed and AS quadrature $V_x$ and $V_p$, ratio of squeezed quadrature measurement for biased homodyne detection and the beamsplitter transmittance $t_{het}$ of imbalanced heterodyne measurement are optimised (see Fig. \ref{fig:ratios} for optimum ratio and transmittance). Reconciliation efficiency $\beta=0.95$ and the untrusted AS noise $\Delta V_u=0.2$ SNU. }
\label{fig:tol_fin}
\end{figure}

In Fig. \ref{fig:finite_key}, our analysis reveals distinct trends based on the degree of squeezing in the source state. Imbalanced heterodyne measurements demonstrate superior performance with small block sizes ($N=10^6$) for highly squeezed states (dashed plots), while biased homodyne detection becomes more favorable with larger block sizes ($N=10^7$), see Fig. \ref{fig:finite_key}\subref{fig:key_b}, \subref{fig:key_d} and \subref{fig:key_f}. Notably, biased homodyne measurements consistently outperform other strategies for moderately squeezed states. In scenarios marked by elevated trusted noise and reduced block lengths (Fig. \ref{fig:finite_key}\subref{fig:key_e}), imbalanced heterodyne measurement proves significantly more efficacious than biased homodyne measurement. Furthermore, we present key rates when the AS quadrature measurement in imbalanced heterodyne measurements is disclosed publicly (red plots in Fig. \ref{fig:finite_key} and \ref{fig:tol_fin}), showcasing its potential contributions to parameter estimation and the selection of optimal error correction codes.\\
Fig. \ref{fig:tol_fin} adds complexity to the narrative, revealing that the efficacy of imbalanced heterodyne measurements is contingent on the presence of channel noise. Specifically, imbalanced heterodyne measurement emerges as consistently superior in scenarios characterized by a substantial amount of channel noise ($\epsilon > 0.1$ SNU), while biased homodyne measurement excels in low-channel-noise conditions ($\epsilon < 0.1$ SNU). This is because, in the presence of substantial noise ($\epsilon> 0.1 SNU$), more measurements of both quadratures are required to reduce the variance of the noise estimator, as it is proportional to the noise itself. For heterodyne measurements, this is not an issue since the number of measurements for both quadratures is always equal. In contrast, for homodyne measurements, Bob must perform additional measurements of the AS quadrature to estimate the AS noise. These AS measurements do not contribute to the key and are solely used for estimating the AS noise. The interplay of measurement strategies with channel noise highlights the nuanced considerations necessary for optimizing squeezed-state CV quantum communication systems.

\begin{figure}\centering
{\includegraphics[width=.39\linewidth]{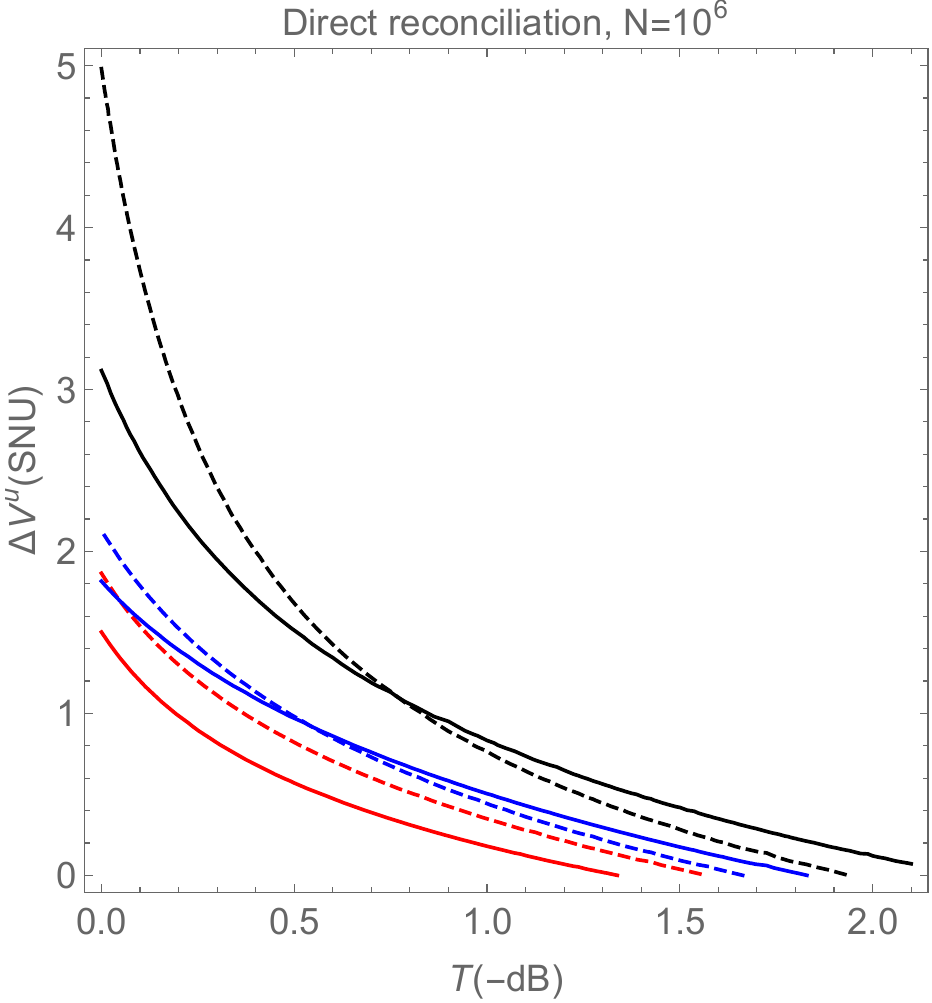}}
{\includegraphics[width=.39\linewidth]{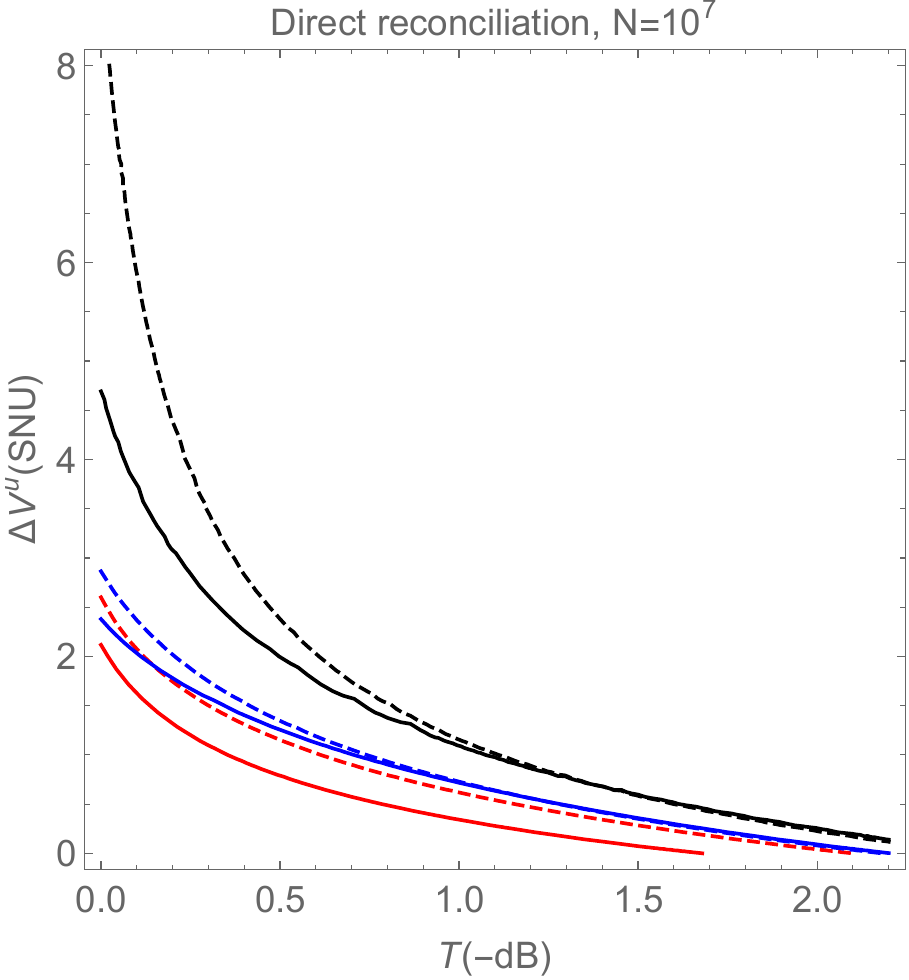}}\hfill
{\includegraphics[width=.39\linewidth]{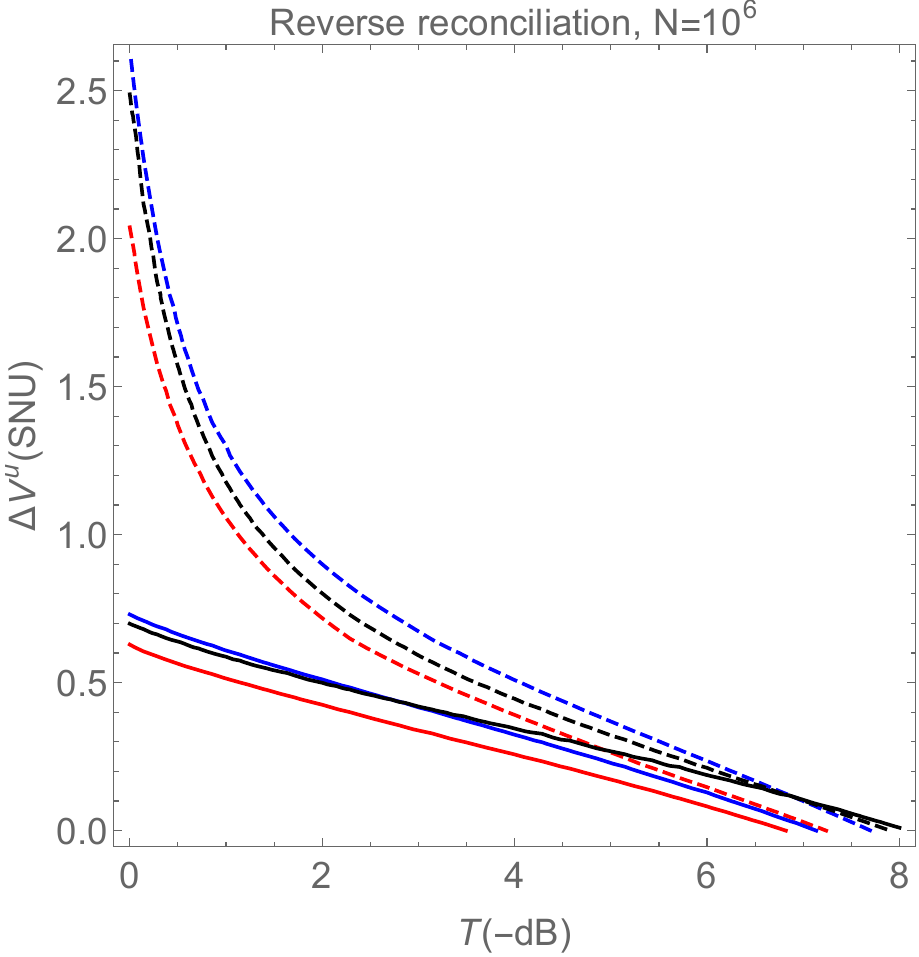}}
{\includegraphics[width=.39\linewidth]{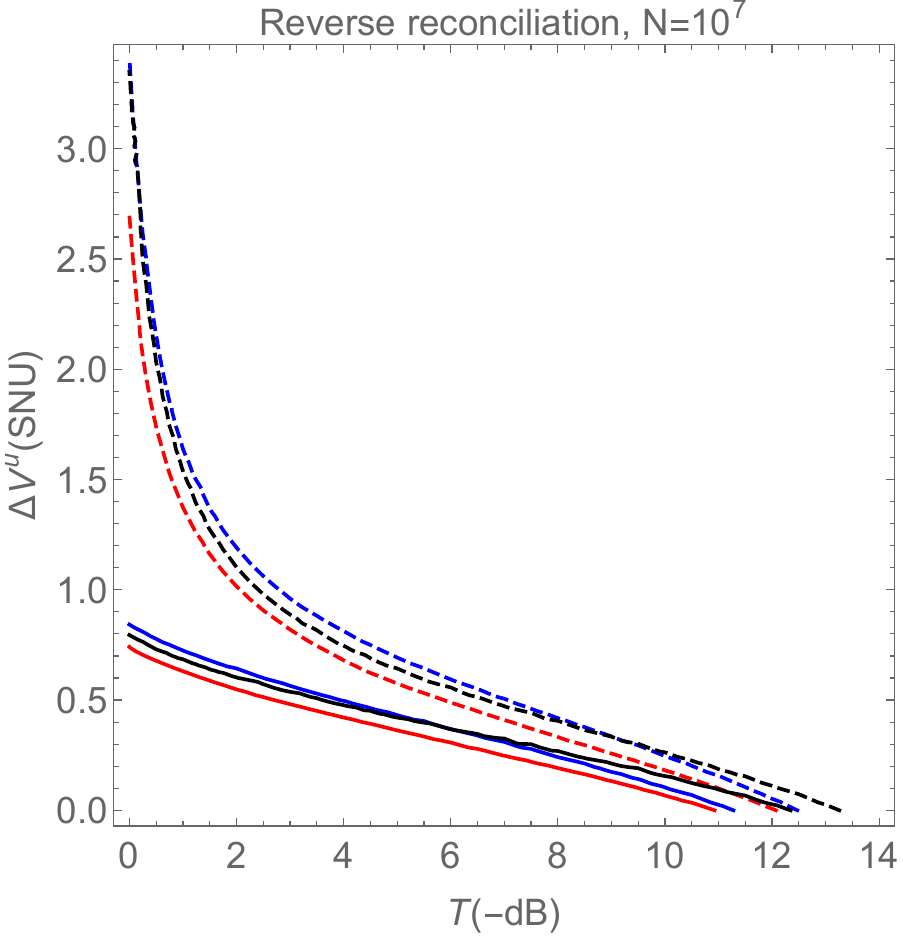}}

\caption{Tolerance to untrusted AS noise $\Delta V_u$ for biased homodyne (black) and imbalanced heterodyne measurement by Bob, when AS quadrature measurement is publicly announced (red) and when it is not announced (blue) in different reconciliation scenarios and at different data block sizes. The modulation variance of the squeezed and AS quadrature $V_x$ and $V_p$, ratio of squeezed quadrature measurement for biased homodyne detection and the beamsplitter transmittance $t_{het}$ of imbalanced heterodyne measurement are optimised. Reconciliation efficiency $\beta=0.95$ and the channel noise $\epsilon=0.01$ SNU. }
\label{fig:tol_fin_AS}
\end{figure}

Turning our attention to untrusted AS noise (Fig. \ref{fig:tol_fin_AS}), we observe a lack of significant differences in tolerance between biased homodyne and imbalanced heterodyne measurements for RR. For DR, the biased homodyne measurement strategy is always superior. The absence of a one-size-fits-all solution requires careful analysis to select the optimal measurement strategy in specific experimental conditions.


\section{Anti-squeezing noise in fluctuating channels}\label{sec:fluctuating channels}

\begin{figure}
\centering
    \includegraphics[width=.75\linewidth]{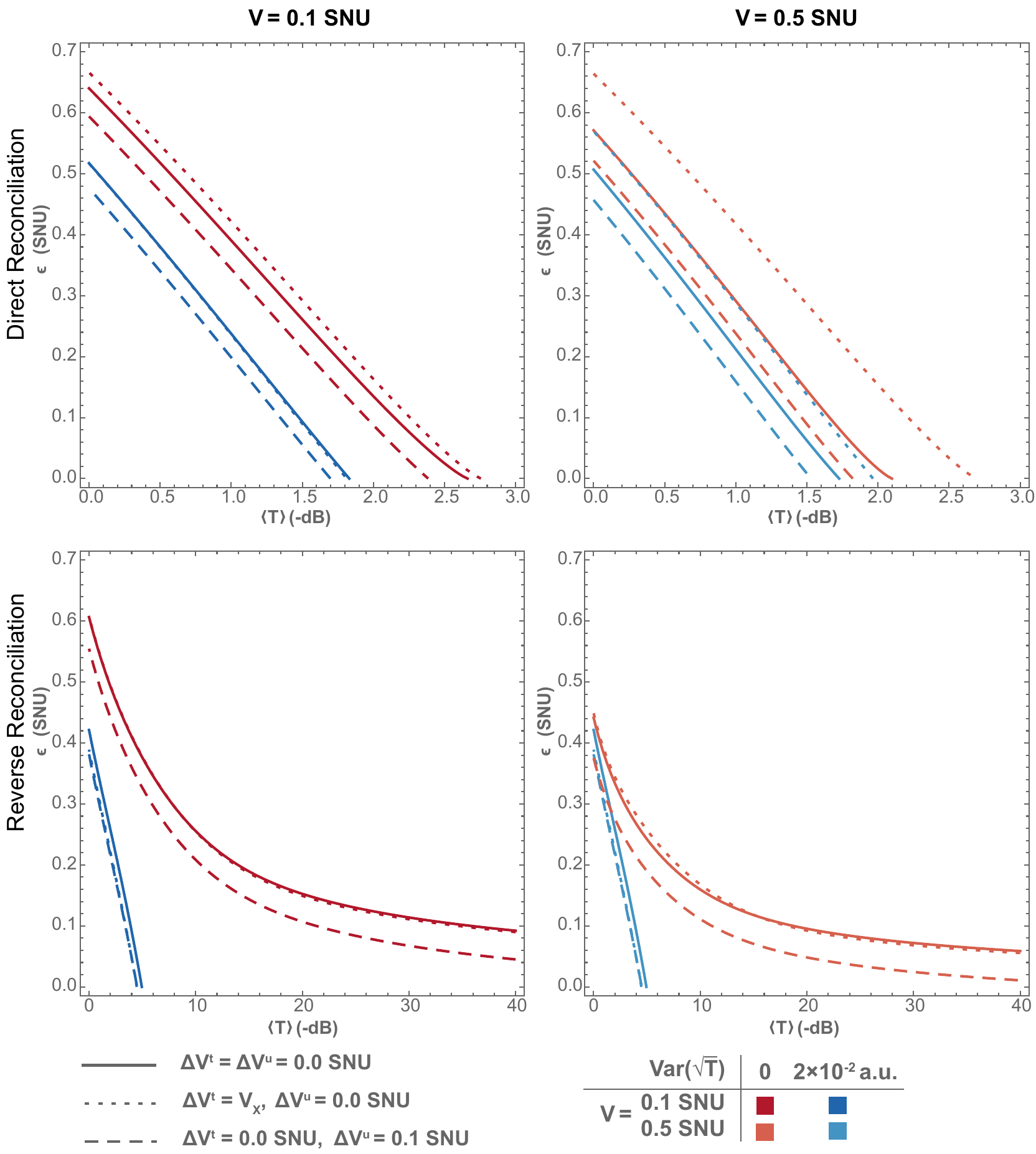}
\caption{Maximal tolerable excess noise ($\epsilon$ in SNU) at the input of the channel with respect to mean channel attenuation $\langle T\rangle$ (in dB scale) for highly squeezed $V=0.1$ SNU (left column) and moderately squeezed $V=0.5$ SNU (right column) source state, in the absence of AS noise (solid), with trusted AS noise $\Delta V_t=V_x$ (short dash), and untrusted AS noise $\Delta V_u=0.1$ SNU (long dash). Red lines correspond to the absence of transmittance fluctuations $Var(\sqrt{T})=0$, and blue to $Var(\sqrt{T})=2\times 10^{-2}$. Modulation variance $V_x$ is optimized for reconciliation efficiency of $\beta=95\%$, AS noise and respective fading.}
\label{fig:tolerance}
\end{figure}

\begin{figure}
    \centering
    \includegraphics[width=.95\linewidth]{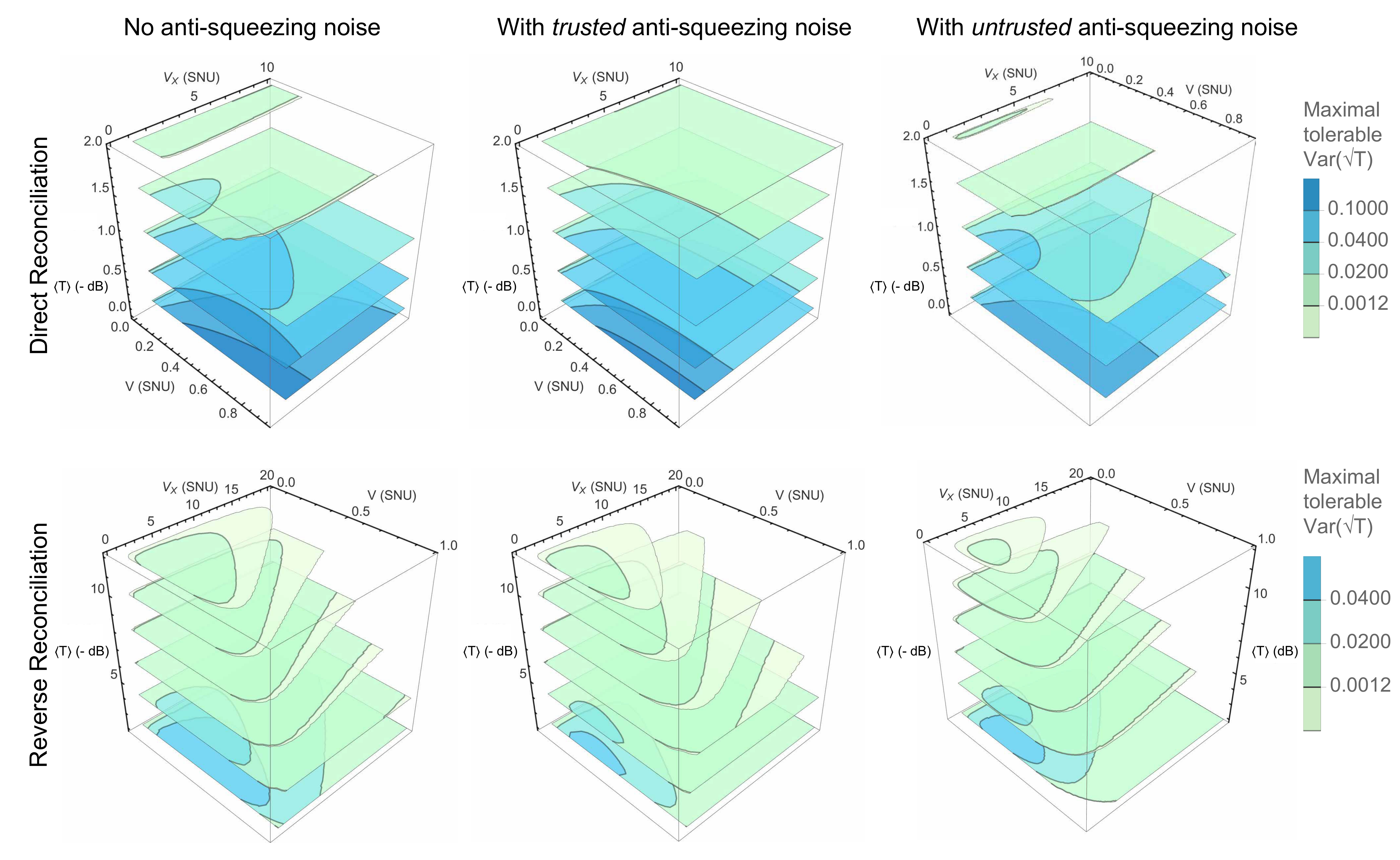}
    \caption{Maximal tolerable level of transmittance fluctuations $Var(\sqrt{T})$ against squeezing $V$, in DR (top row) and RR (bottom row) without AS noise (left column), with trusted AS noise $\Delta V_t=V_x$ (middle column), and untrusted AS $\Delta V_u=0.1$ SNU (right column). Channel excess noise $\epsilon=0.1$ SNU, and reconciliation efficiency $\beta=95\%$}
    \label{fig:fading-AS-noise}
\end{figure}

While in the previous sections we assumed the channel transmittance to be stable, which typically corresponds to fiber-type channels, in the free-space realizations it is usually not the case and the atmospheric channel transmittance fluctuates due to the turbulence effects, we further refer to such transmittance fluctuations as to the channel fading. Although the changes of transmittance are slower than the clock rate of the protocol, such attenuation fluctuations during data accumulation can significantly impede achievable secure key rate \cite{usenko2012entanglement, Derkach_2020}. The overall state shared between trusted parties after the fading channel is generally non-Gaussian and is described by the Wigner function,

which is the weighed convex sum of Wigner functions  that characterize shared state when Bob's subsystem was transmitted through a channel with fixed transmittance $T_i$ that occurs with probability $\tau_i$, with $\sum^M_{i=1}\tau_i=1$ \cite{dong2010continuous}. While the first moments remain zero, the second moments $\gamma_{AB}$ of the resulting state are hence the convex combination with transmittance $T_i$ and transmittance coefficient $\sqrt{T_i}$ of all $i\in[1,M]$. Extremality of Gaussian states \cite{wolf2006extremality} and subsequent optimality of Gaussian collective attacks \cite{PhysRevLett.97.190502,PhysRevLett.97.190503} allows us to gaussify the overall mixed state, i.e. base further security analysis on the covariance matrix $\gamma_{AB}$, which is now evidently a function of $\langle T\rangle$ and $\langle \sqrt{T}\rangle$, without underestimating the eavesdropper.
In the current section we consider the homodyne-based protocol as the best performing in the asymptotic regime, and allowing to reach longer distances in finite-size regime. 
Bob's subsystem after the fading channel is described by
    \begin{equation}
    \gamma_B=\left[ 
        \begin{array}{cc}
        \langle T\rangle(V +V_x - 1)+ 1 & 0 \\
        0 & \langle T\rangle(1/V+V_p+\Delta V - 1)+ 1
        \end{array}
        \right], 
    \end{equation}

and correlated to Alice's mode as: 

    \begin{equation}
    \varsigma_{AB}=\left[
    \begin{array}{cc}
       \langle \sqrt{T}\rangle\frac{V_2-V_1}{2\sqrt{2}}  & 0 \\
        0 &  \langle \sqrt{T}\rangle\frac{\frac{1}{V_2}-\frac{1}{V_1}}{2\sqrt{2}} 
    \end{array}
    \right]    
    \end{equation}

The mutual information (in $x$-quadrature) without AS  noise $\Delta V$ is therefore:
    \begin{equation}
        I_{AB}=\frac{1}{2}\log_2\left[\frac{1 + \langle T\rangle(V+\epsilon-1+V_x)}{1-\langle \sqrt{T}\rangle^2V_x+\langle T\rangle(V+\epsilon-1+V_x)}\right]. 
        \label{eq:iab_fading}
    \end{equation}
A fading channel can be equivalently represented as a channel with fixed transmittance $\langle\sqrt{T}\rangle^2$ and fading-related variance-dependent excess noise $\varepsilon_x^f=Var(\sqrt{T})(V+V_{x} - 1)$ and $\varepsilon_p^f=Var(\sqrt{T})(1/V+V_{p} + \Delta V- 1)$ in $x$- and $p$-quadratures respectively, contrary to symmetrical fading-related noise in the coherent-state protocol \cite{usenko2012entanglement}. The noise related to fading is to be assumed untrusted, as the transmittance fluctuations can be fully controlled by Eve. This represents another strongly negative impact of AS noise in the fluctuating channels, even if the noisy source is initially trusted. Such noise necessitates the characterization of transmittance coefficient variance $Var(\sqrt{T})=\langle T\rangle-\langle \sqrt{T}\rangle^2$ and consequent adaptation of modulation $V_{x(p)}$ and squeezing variance $V$ for optimal secure key rate. We compare all noises involved in the protocol and highlight the differences between them, as well as methods for their estimation and their influence on the security in Tab. \ref{tab:noises}.  
For weak atmospheric turbulence one can expect $Var(\sqrt{T})\leq 0.01$, and $Var(\sqrt{T})> 0.04$ for strong turbulence \cite{Derkach_2020}. Mutual information (\ref{eq:iab_fading}) reduces to (\ref{eq:IAB}) when fluctuations of transmittance are negligible, i.e. $M=1$ and $\tau_{i=1}=1$ or $Var(\sqrt{T})=0$.
The upper bound on Eve's accessible information $\chi_{AE}$ or $\chi_{BE}$ is evaluated based on $\gamma_{AB}(\langle T\rangle, Var(\sqrt{T}))$ as well. For further details regarding the modelling and security analysis of atmospheric channels in CV QKD see \cite{usenko2012entanglement}. \par
The effect of transmittance fluctuations and AS noise on the security of the squeezed-state CV QKD protocol is shown in Fig. \ref{fig:tolerance}. The value of transmittance coefficient variance is set to $Var(\sqrt{T})=0.02$ which corresponds to weak-to-moderate turbulence regime \cite{vasylyev2016atmospheric}, where beam shape deformation, broadening and wandering are expected, corresponding to daytime operation of a short free-space atmospheric link. 

\begin{table}[]
\centering
\resizebox{\textwidth}{!}{\begin{tabular}{|l|l|l|l|l|}
\hline
Noise             & \textbf{Excess} $\epsilon$                                            & \textbf{Fluctuation} $\varepsilon^f$                                                                                  & \textbf{Anti-squeezing} $\Delta V$                                                              & \textbf{Detector noise} $\nu_{el}$                                      \\ \hline
Origin            & \begin{tabular}[c]{@{}l@{}}Untrusted \\ channel\end{tabular} & Gaussification                                                                                               & \begin{tabular}[c]{@{}l@{}}Cavity loss, \\ preparation \\ imperfections\end{tabular}   & Electronic noise                                               \\ \hline
\begin{tabular}[c]{@{}l@{}}Estimation \\ method\end{tabular}& Data disclosure                                             & \begin{tabular}[c]{@{}l@{}}Separate measurement\\  and/or \\ directly from data\end{tabular} & \begin{tabular}[c]{@{}l@{}}Data disclosure \\ and/or\\ during calibration\end{tabular} & \begin{tabular}[c]{@{}l@{}}Separate\\ measurement\end{tabular} \\ \hline
Influence         & See Fig. \ref{fig:tol_fin} and \ref{fig:finite_key}                                            & See Fig. \ref{fig:fading-AS-noise}  and \ref{fig:tolerance}                                                                                                & See Fig. \ref{fig:tol_fin_AS}                                                                             & See Fig. \ref{fig:Tol_asy}                                               \\ \hline
Trust             & No                                                           & No                                                                                                           & Up to users                                                                            & Yes                                                            \\ \hline
\end{tabular}}
\caption{Comparison of different noises involved in the squeezed-state CV QKD protocol. }\label{tab:noises}
\end{table}

Regardless of the choice of reconciliation side, level of squeezing in the signal quadrature, and assumptions on the AS noise, transmittance fluctuations have a detrimental effect on the tolerance to excess noise $\epsilon$ and loss $\langle T \rangle$ of the CV QKD protocol. In the absence of fading, the protocol performances are slightly improved by trusted AS noise presence (with a negligible effect in some regimes for strong AS and RR), while untrusted AS noise leads to performance degradation. In the presence of already weak fading, the positive effect of strong trusted AS recedes and it can be helpful for DR protocols using moderately squeezed signal states. For RR protocol, however, AS is generally limiting the range of secure parameters. The sensitivity of the protocol established over fading channels underlines the need for characterization and optimization of the transmitter, as shown in Fig. \ref{fig:fading-AS-noise}. We show the maximal tolerable fading dependence on the level of squeezing $V$ and modulation variance $V_x$ at different values of mean channel loss $\langle T\rangle$. Trusting AS enables a significantly wider range of tolerable atmospheric conditions when DR protocol is used, and allows for more flexible choice of optimal combination of squeezing and modulation variance. Untrusted AS, as expected, necessitates for squeezing to be optimized (preferably to high values) and limits $V_x$ to few SNU. With RR protocol the range of optimal squeezing $V$ steers towards lower values.  Therefore, even if trusted, proper characterization, control of AS noise and consequent parameter optimization is crucial for implementations in fading (particularly free-space) channels.

\section{Summary and conclusions}

In this work, we perform a detailed investigation into the role of the excess noise in squeezed states in CV QKD, studying security against collective attacks in the asymptotic as well as finite-size regimes, and considering free-space channel fading. In the widely adopted trust assumption on the noisy squeezing, we show the robustness of the protocol against squeezing excess noise in the asymptotic regime of infinitely many data points, first for the purely attenuating channels, where strong trusted squeezing noise allows to obtain compact analytical expressions for the Holevo bound, as well as in the presence of channel noise. Furthermore, we largely confirm the stability of protocol against trusted noisy squeezing in the finite-size regime, where we consider two main strategies for the parameter estimation:
\begin{enumerate}
    \item Biased homodyne detection, where Bob alternates between measuring the squeezed and AS quadratures, and
    \item Imbalanced heterodyne detection, where Bob performs double homodyne detection with a beam splitter transmission skewed towards measuring the squeezed quadrature.
\end{enumerate}
We find that the optimal method depends on the levels of squeezing, block size, and the amount of excess noise. Specifically, for moderately squeezed states $V=0.5$ SNU with low excess noise $\epsilon < 0.1$ SNU, biased homodyne measurements enable key generation over the greatest distances. For all other scenarios, imbalanced heterodyne measurement proves to be more effective in achieving longer secure communication distances. 
In the case, when the noisy squeezing is untrusted, it introduces a security bound on the protocol realization, which is further enforced by the finite-size effects. In the case of untrusted noise, the purity of the signal squeezed states becomes crucial already in the asymptotic regime, especially in the reverse reconciliation scenario, aimed at long-distance realizations of CV QKD.
Finally, we consider the fading channels with fluctuating transmittance, which is typical for free-space atmospheric channels, and show that even trusted squeezing excess noise transforms to the untrusted channel noise due to fading. This emphasizes the importance of the state purity for squeezed-state CV QKD realizations in free-space channels already with the trusted sources. \par
Our findings present the methods for protocol optimizations, including measurement and estimation techniques, targeting the practicality and efficiency of the protocol realization. The obtained results pave the way towards optimized experimental implementation of the squeezed-state QKD protocol. Our theory shows several opportunities for experimental improvements. Firstly, increasing the signal state purity can translate into a higher performance of the QKD protocol. Secondly, stability of the squeezed light generation is essential for a relivable protocol operation, as security assessment is contingent on precise characterization of the signal state throughout whole key generation process. Lastly, any phase misalignment at the receiver side can diminish the benefits of the squeezed state. Albeit, the latter can be mitigated by employing machine learning techniques \cite{chin2021machine}. We believe that this work will enable secure quantum communication based on squeezed-state of light

in point-to-point protocols as well as in QKD-based communication networks, where great diversity of implementation conditions and combination of various protocols and link types is anticipated.

\section*{Acknowledgements}
A. O. and V. C. U. acknowledge the project 8C22002 (CVStar) and R.F. acknoledges the project 8C22001 (SPARQL) of MEYS of Czech Republic, which have received funding from the European Union’s Horizon 2020 research and innovation framework programme under grant agreement No. 731473 and 101017733. V.C.U. also acknolwedges the project No. 21-44815L of the Czech Science Foundation. A. O. acknowledges the project IGA-PrF-2024-008 of Palacky University Olomouc. I. D. acknowledges the project 22-28254O of the Czech Science Foundation. V. U. acknowledges the European Union’s Horizon Europe research and innovation programme under the project "Quantum Security Networks Partnership" (QSNP, grant agreement No. 101114043). R.F. and V.U. acknowledge the project CZ.02.01.01/00/22\textunderscore008/0004649 (QUEENTEC) of MEYS of Czech
Republic.
\clearpage
\appendix

\section{Noisy squeezing due to optical losses}\label{sec:pure-loss}

Squeezed states, typically generated using optical parametric oscillators \cite{lvovsky2015squeezed}, are unavoidably noisy already due to the optical losses in the cavity, phase noise, parasitic nonlinearities, absorptions, and inefficient collection of light at the cavity output. We model these effects in Gaussian approximation by propagating an effective pure squeezed state through an attenuation to reach the considered realistic squeezed state with an excess noise. A single-mode pure squeezed state, described by the squeezed and AS variances $V_S,1/V_S$, undergoing losses can be modeled as coupling to a vacuum mode on a beamsplitter with transmittance $\eta$. The resulting state after the beamsplitter interaction \cite{weedbrook2012gaussian} is described by the quadrature variances $\eta V_S+1-\eta,\eta/V_S+1-\eta$ and contains the fraction of vacuum noise (in shot-noise units, SNU), scaled by the beamsplitter reflectance $1-\eta$. The Gaussian purity of the state then becomes $\mu=\left[1+(\eta-\eta^2)(1/V+V-2)\right]^{-1/2}$ and is minimized at $\eta=1/2$. When detected, without precise knowledge of coupling loss $\eta$ and initial squeezing $V_S$, one cannot pinpoint the exact distribution of added noise ($\epsilon_x$ and $\epsilon_p$) in each quadrature. In the context of CV QKD, when the actual $V_S$ and $\eta$ are not known, it can be assumed that the resulting state contains a pure squeezed state with variances $V,1/V$ and some excess noise $\epsilon_x,\epsilon_p$ in either of the quadratures. The allocation of the noise once it is trusted (not controlled by Eve) does not influence the security of CV QKD. However, when the noise in the squeezed signal states is not trusted (controlled by Eve), its allocation to particular quadratures has direct security ramifications and we must adopt the most conservative stance. The latter typically implies admitting the largest quantitative amount of noise, i.e. treating all the noise as contained in the AS quadrature, see Appendix \textbf{A} for further details. 

\begin{figure}
    \centering
    \includegraphics[width=0.79\textwidth]{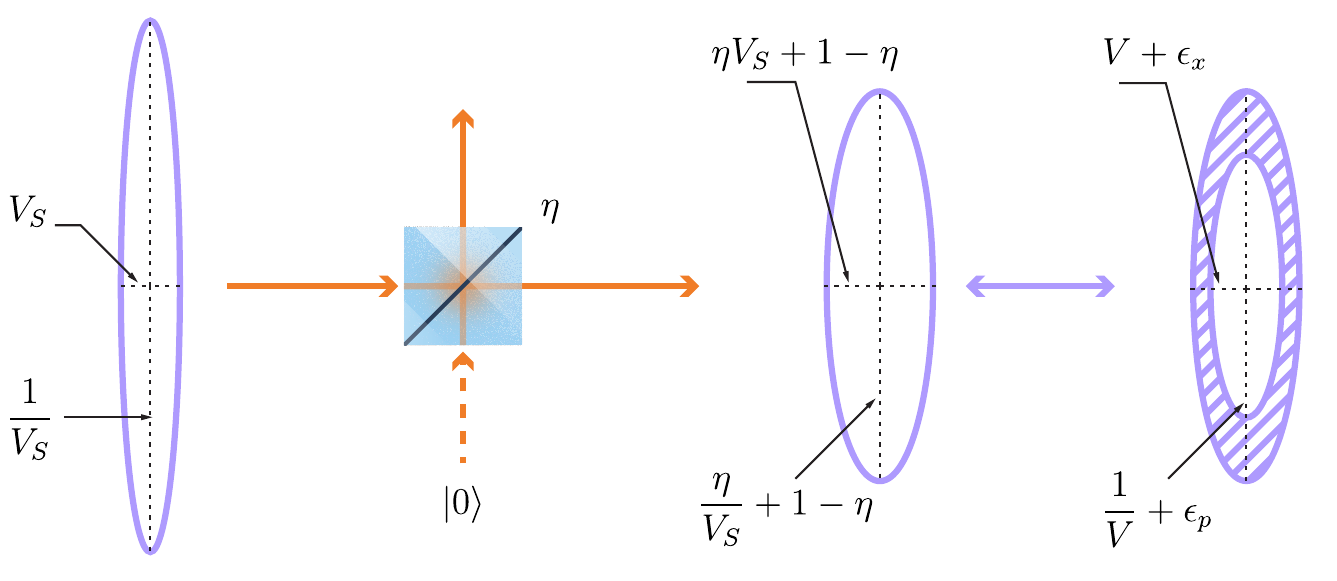}
    \caption{Evolution of a pure squeezed state with quadrature variances $V_S,1/V_S$ after attenuation $\eta$. The vacuum noise is added in both quadratures, however without the knowledge of $V_S$ and $\eta$ the output state can be viewed as a pure state with respective variances $V,1/V$ and with excess noise $\epsilon_x$ and $\epsilon_p$ in respective quadratures. The attribution of this noise to either of the quadratures, once it is untrusted, has different effect on the performance of CV QKD protocols.}
    \label{fig:pure-loss}
\end{figure}

Note, that the loss model in Fig. \ref{fig:pure-loss} can also be explicitly considered in QKD as a side channel in the sending station \cite{derkach2016preventing}, providing an eavesdropper with side information on the signal state, but not enabling control on the vacuum noise added to the squeezed state. In the current work, however, we depart from this pure loss model and adopt a more general approach, where an arbitrary amount of Gaussian noise $\Delta V$ is added to the AS quadrature (leading to significantly lower purity $\mu$ of a generated state), allowing us to maintain the most conservative security perspective. \par

\section{Squeezed-state noise allocation}\label{sec: noisy squeezed state}
In our CV QKD analysis we made a deliberate choice by incorporating all the noise, that transformed our initially pure squeezed state into a noisy one, within the AS quadrature. One might contend that the noise could equally be distributed across the squeezed quadrature. This decision holds no consequences when the noise is trusted and does not augment Eve's knowledge of the shared state between the trusted parties. However, in cases when the noise, responsible for impurities in the pure squeezed state, cannot be trusted, we are compelled to consider the most pessimistic option. This implies assigning all the noise to the AS quadrature, a demonstration of which is presented in this section.

To evaluate the potential information leakage arising from an impure squeezed state, let's consider a scenario where the exclusive source of information available to Eve is the impure squeezed state, characterized by a squeezed quadrature variance $\tilde{V}_s  = V + \epsilon_x < 1$ and an AS quadrature variance $\tilde{V}_{as}  = 1/V + \epsilon_p > 1/\tilde{V}_s$. The noise is distributed arbitrarily along the squeezed and AS quadratures, where $0 < \epsilon_x < \tilde{V}_s - 1/\tilde{V}_{as}$ and $\epsilon_p =\tilde{V}_{as} - 1/(\tilde{V}_s - \epsilon_x)$.

In an equivalent entanglement-based protocol for a P\&M scheme, where Alice modulates the squeezed quadrature with a Gaussian distribution having modulation variance $V_x$, the protocol involves two squeezed states with x quadrature variances $V_1 = V + V_x - \sqrt{(V  + V_x)V_x}$ and $V_2 = V + V_x + \sqrt{(V  + V_x)V_x}$. These states are operated on a balanced beamsplitter, resulting in the following covariance matrix:
\begin{equation}
   \gamma_{AB}=\left[ \begin{array}{cccc}
        \frac{V_1+V_2}{2} &0 &\frac{V_2-V_1}{2} & 0\\
       0 &\frac{1}{2}\left(\frac{1}{V_2}+\frac{1}{V_1}\right) &0 &\frac{1}{2}\left(\frac{1}{V_2}-\frac{1}{V_1}\right) \\
       \frac{V_2-V_1}{2} & 0 & \frac{V_1+V_2}{2}+\epsilon_x & 0\\
        0& \frac{1}{2}\left(\frac{1}{V_2}-\frac{1}{V_1}\right)&0 & \frac{1}{2}\left(\frac{1}{V_2}+\frac{1}{V_1}\right)+\epsilon_p
    \end{array}\right]
\end{equation}

\begin{figure}
  \includegraphics[width=\linewidth]{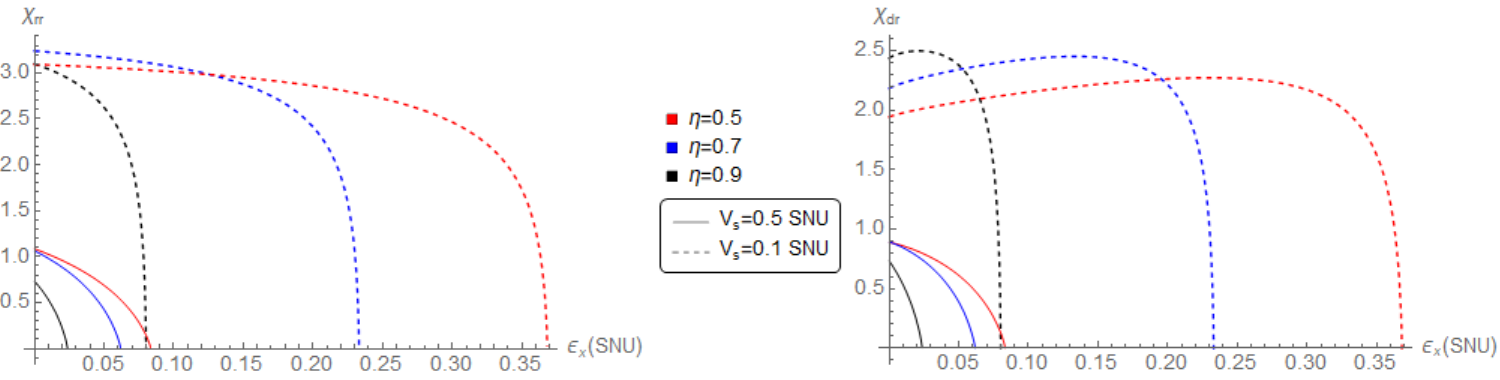}
   \caption{Holevo bound vs excess noise in the squeezed quadrature $\epsilon_x$ for RR (left) and DR (right), for an initially pure squeezed state with squeezed quadrature $V_s$, attenuated by loss $\eta$. Modulation variance $V_x=10$ SNU.}
   \label{fig:HolevoVsnoise}
\end{figure}

The conditional covariance matrix of Alice upon Bob's squeezed quadrature measurement is $\gamma_{A|B}=Diag[(2 V_1 V_2 + \epsilon_x (V_1 + V_2))/(2 \epsilon_x + V_1 + V_2),(V_1 + V_2)/(2 V_1 V_2)]$ and conditional covariance matrix of Bob upon Alice's squeezed quadrature measurement is $\gamma_{B|A}=Diag[ (2 V_1 V_2)/(V_1 + V_2)+\epsilon_x,1/2 (1/V_1 + 1/V_2)+\epsilon_p]$. The Holevo bound, which provides the upper bound on Eve's knowledge of Alice's or Bob's measurement outcome, for DR or RR protocols, respectively, is obtained from the symplectic eigenvalues of the covariance matrices $\gamma_{AB}$, $\gamma_{A|B}$ and $\gamma_{B|A}$, as explained in Sec.\ref{sec:protocol}. Although the symplectic eigenvalues of $\gamma_{A|B}$ and $\gamma_{B|A}$ are easy to calculate (they are $\pm \sqrt{Det[\gamma_{A|B}]}$ and $\pm \sqrt{Det[\gamma_{B|A}]}$ respectively), the symplectic eigenvalues of $\gamma_{AB}$ do not have a simple form in general. However, we are able to obtain a simple symplectic eigenvalue expression for extreme cases, i.e., $\epsilon^{max}_x=\tilde{V}_s-1/\tilde{V}_{as}$ and $\epsilon_p=0$ or $\epsilon_x=0$ and $\epsilon^{max}_p=\tilde{V}_{as}-1/\tilde{V}_s$. When all the noise is considered to be in the squeezed quadrature, the symplectic eigenvalues of $\gamma_{AB}$ after using the P\&M equivalence relations are $\{\pm 1,\pm (\sqrt{[\epsilon^{max}_x+V]/V}-1)/2\}$. When the noise is considered in AS quadrature the symplectic eigen values are $\{\pm 1, \pm (\sqrt{1+\epsilon^{max}_p[V+V_x]}-1)/2\}$. The Holevo bounds are:
\begin{align}
    \chi_{RR}(V,V_x,\epsilon^{max}_x,0) &= G\left(\frac{\sqrt{[\epsilon^{max}_x+V]/V}-1}{2}\right) \nonumber \\
    &\quad - G\left(\frac{\sqrt{[(\epsilon^{max}_x + V) (V + V_x)]/[V (\epsilon^{max}_x + V + V_x)]}-1}{2}\right) \\
    \chi_{RR}(V,V_x,0,\epsilon^{max}_p) &= G\left(\frac{\sqrt{1 + \epsilon^{max}_p (V + V_x)}-1}{2}\right) \\
    \chi_{DR}(V,V_x,\epsilon^{max}_x,0) &= 0 \\
    \chi_{DR}(V,V_x,0,\epsilon^{max}_p) &= G\left(\frac{\sqrt{1 + \epsilon^{max}_p (V + V_x)}-1}{2}\right) \nonumber \\
    &\quad - G\left(\frac{\sqrt{1 + \epsilon^{max}_p V}-1}{2}\right)
\end{align}
 As $V > 0$, $V_x > 0$, and $\epsilon^{max}_p > \epsilon^{max}_x$ (where $\epsilon^{max}_p=(\tilde{V}{as}/\tilde{V}_s ) \epsilon^{max}_x$), the comparison of terms involving $G$ between $\chi_{RR}(V,V_x,\epsilon^{max}_x,0)$ and $\chi_{RR}(V,V_x,0,\epsilon^{max}_p)$, the larger arguments in the latter case result in greater values due to the monotonic nature of $G$. Additionally, it is worth noting that when all the noise is in the squeezed quadrature, $S(AB)=S(B|A)$, and when all the noise is in the AS quadrature, $S(A|B)=0$. \\
To gain insights into the impact of arbitrary noise distributions, we perform numerical investigations focusing on impure squeezed state obtained due to attenuation of pure squeezed state Fig. \ref{fig:pure-loss}. Fig. \ref{fig:HolevoVsnoise} displays the Holevo bound versus excess squeezed quadrature variance $\epsilon_x$ for various transmittance values and degrees of squeezing, with the excess noise in the (AS) quadrature being $\epsilon_p=1+\eta(1/V_s-1)-1/(1+\eta[V_s-1]-\epsilon_x)$. The plots provide insights into the consequences of employing an impure squeezed state. One notable observation is that, under worst-case considerations, increased impurity does not necessarily translate to reduced security. This phenomenon stems from the fact that a higher impurity in a noisy squeezed state does not inherently result in higher AS noise. To illustrate, consider the attenuation of a highly squeezed state ($V_s$ = 0.1 SNU) with beamsplitter transmittance values $\eta$ = 0.5 and $\eta$ = 0.7. Despite the state with higher losses being less pure, the maximum AS excess noise for $\eta$ = 0.5 is less than that for $\eta$ = 0.7.\\

\begin{figure}
    \centering
    \begin{subfigure}[b]{0.32\linewidth}
        \includegraphics[width=\linewidth]{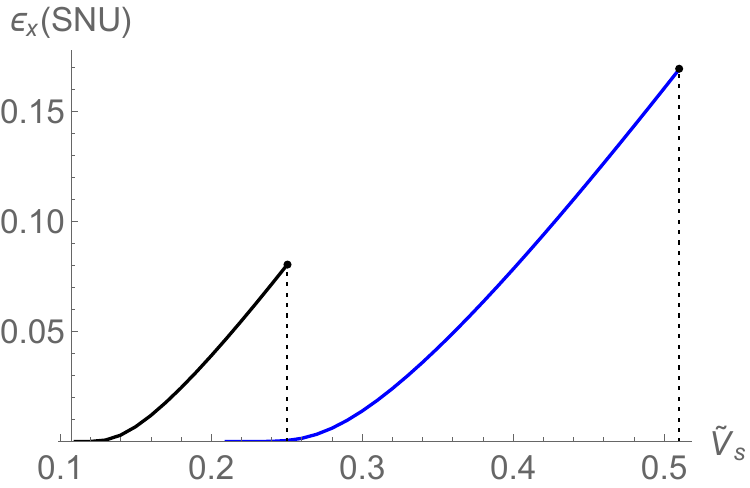}
        \caption{}
        \label{fig:opteps}
    \end{subfigure}\hfill
    \begin{subfigure}[b]{0.32\linewidth}
        \includegraphics[width=\linewidth]{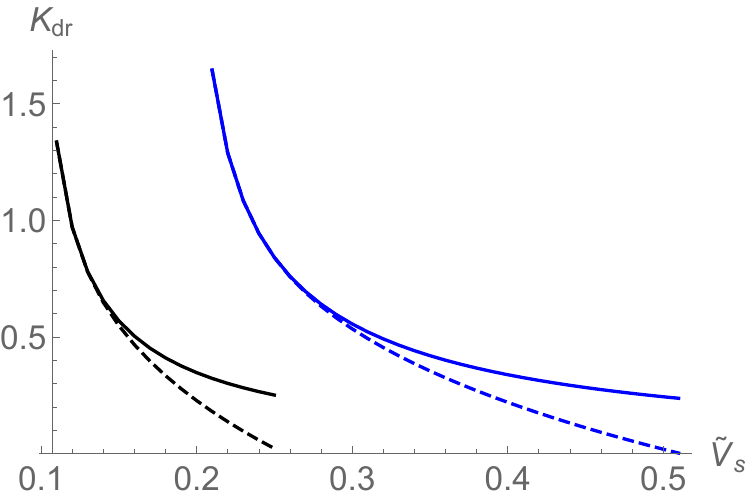}
        \caption{}
        \label{fig:keynoise}
    \end{subfigure}\hfill
    \begin{subfigure}[b]{0.32\linewidth}
        \includegraphics[width=\linewidth]{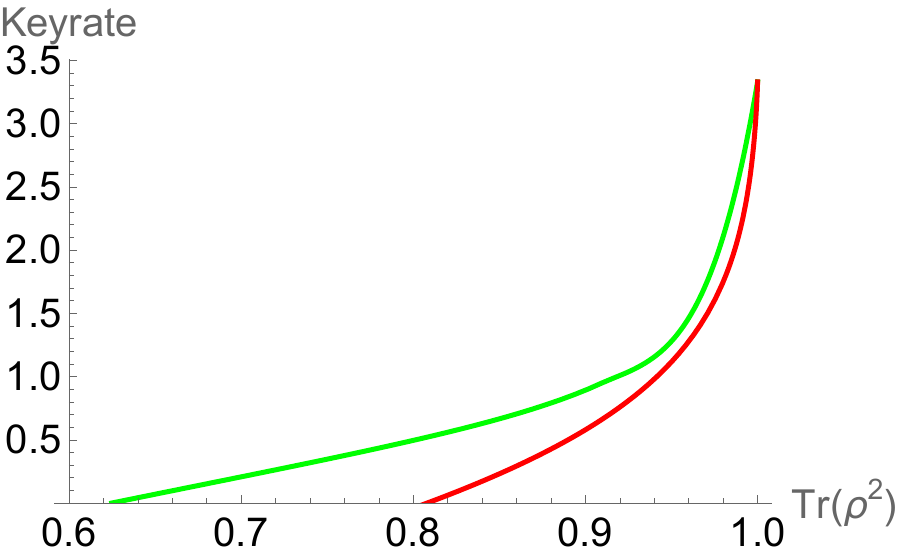} 
        \caption{}
        \label{fig:purity}
    \end{subfigure}
    \caption{(a) shows optimal $\epsilon_x$ to obtain minimal key rate Eq. \ref{eq:drnoise}, for different noisy squeezed state with squeezed quadrature variance $\tilde{V}_s$, and AS quadrature variance $\tilde{V}_{AS}=10$ SNU (black) and $\tilde{V}_{AS}=5$ SNU (blue). (b) shows key rates for DR, when optimal noise distribution from (a) is used (dashed) and when all the noise is considered to be in AS quadrature (solid). (c) shows the key rates with a pessimistic noise distribution for RR (red) and DR (green), varying the purity of the squeezed state.}
    \label{fig:noisedistribution}
\end{figure}

In the context of DR, accounting for all the noise in the AS quadrature may not necessarily lead to a worse outcome. To delve deeper into this, we assess the key rate given by the equation:
\begin{equation}\label{eq:drnoise}
     K_{dr} = \frac{1}{2} \log_2\left(1 + \frac{V_x}{\tilde{V}_s}\right) - \chi_{DR}(V, V_x, \epsilon_x, \epsilon_p) 
\end{equation}
This evaluation is performed for noisy squeezed states with AS quadrature variance $\tilde{V}_{\text{as}} = 10$ (black plot from Fig. \ref{fig:noisedistribution}) and $\tilde{V}_{\text{AS}} = 5$ (blue plot from Fig. \ref{fig:noisedistribution}), considering squeezed quadrature variances $0.1 \leq \tilde{V}_s \leq \tilde{V}_{s}^{\text{max}}$ and $0.2 \leq \tilde{V}_s \leq \tilde{V}_{s}^{\text{max}}$, respectively. Here, $\tilde{V}_{s}^{\text{max}}$ represents the maximum $\tilde{V}_s$ for which a positive key rate can be achieved.

Fig. \ref{fig:opteps} illustrates the squeezed quadrature noise $\epsilon_x$ for which the key rate is minimized, considering the rest of the noise in the AS quadrature $\epsilon_p$. From the figure, we infer that for lower noise levels, the pessimistic approach is to account for all the noise in the AS quadrature, as the noise increases, the pessimistic distribution of noise also evolves.

Fig. \ref{fig:keynoise} showcases the key rates when the most pessimistic distribution is used and when all noise is considered to be in the AS quadrature. Despite the pessimistic distribution of noise changing with the noisy squeezed state for DR, it consistently holds true that $\epsilon_p > \epsilon_x$. Hence, in this paper, we investigate the role of AS noise $\Delta V = \epsilon_p - \epsilon_x$, distinguishing the noise as symmetric noise ($\epsilon = \epsilon_x$) and additional AS noise ($\Delta V = \epsilon_p - \epsilon_x$).

 Fig. \ref{fig:purity} illustrates the impact of the squeezed state's purity on the protocol's security. It should be noted that, as previously explained and demonstrated, the security of the protocol hinges on the noise distribution causing the impurity. The plots display the key rates for the worst-case noise distribution corresponding to that purity. 

\section{Parameter estimation}\label{sec: parameter estimation}
 We start by defining $\widehat{C_{AB}}:=\frac{1}{m+n}(\sum^m_{i=1}M^x_iB^x_i+\sum^n_{j=1}M^p_jB^p_j)$. Here $m$ and $n$ are the number of measurements of $x$ and $p$ quadratures, respectively, used for parameter estimation, $M^x,M^p,B^x$ and $B^p$ are the realizations of Alice's random number generator corresponding to quadrature variables $(x_m,p_m)$ and realizations of Bob's quadrature variables $(x_b,p_b)$, respectively. We can define estimator of transmittance: $\widehat{T}=\frac{1}{(r_x V_x +r_p V_p)^2 }(\widehat{C_{AB}})^2$, since
\begin{equation}
    E[\widehat{C_{AB}}]=\frac{1}{m+n}(m E[x_m x_b] + n E[p_m p_b])
  \end{equation}
  \begin{equation}
    E[\widehat{C_{AB}}]=\sqrt{T}(\underbrace{m/(m+n)}_{r_x} V_x + \underbrace{n/(m+n)}_{r_p} V_p)
\end{equation}
On the other hand, we have
\begin{equation}
    \begin{split}
         Var[\widehat{C_{AB}}]&=\frac{1}{(m+n)^2}\left(\sum^m_{i=1}Var[M^x_iB^x_i]+\sum^n_{j=1}Var[M^p_jB^p_j]\right)\\  &=\frac{1}{(m+n)^2}(m Var[x_m x_b]+n Var[p_m p_b])\\ &=\frac{1}{(m+n)^2}[m (T 2 V_x^2+V_x V_{Nx})+n (T 2 V_p^2+V_p V_{Np})]\\ 
    \end{split}
\end{equation}
where $V_{Nx}=1+T(V+\epsilon-1)$ and $V_{Np}=1+T(1/V+\epsilon+\Delta V_u-1)$. As shown in \cite{PhysRevA.81.062343}, $Var[\widehat{T}]$ is a function of $E[\widehat{C_{AB}}]$ and $Var[\widehat{C_{AB}}]$:
 \begin{equation}
 \begin{split}
     Var[\hat{T}]&= \frac{2Var[\widehat{C_{AB}}](Var[\widehat{C_{AB}}]+2(E[\widehat{C_{AB}}])^2)}{(r_x V_x+r_p V_p)^4}\\&=\frac{2Var[\widehat{C_{AB}}]2(E[\widehat{C_{AB}}])^2}{(r_x V_x+r_p V_p)^4}+\frac{2(Var[\widehat{C_{AB}}])^2}{(r_x V_x+r_p V_p)^4}
      \end{split}
 \end{equation}
 Notice that $Var[\widehat{C_{AB}}]$ is of the order of $1/(m+n)^2$. For large number of pulses, the second term is significantly small, and the variance of the estimator $\hat{T}$ can be approximated to:
 \begin{equation}
    Var[\hat{T}]\approx \frac{4 T [n V_p (V_{Np} + 2 T V_p) + m V_x (V_{Nx} + 2 T V_x)]}{(n V_p + m V_x)^2}=:\Upsilon^2
 \end{equation}
 
 The variance of the estimator for the channel noise of squeezed quadrature defined as $\hat{V}_{\epsilon_x}=T\epsilon:=\frac{1}{m}\sum^m_{i=1}(B^x_i-\sqrt{\hat{T}}M^x_i)^2+\hat{T}(1-V)-1$ can be approximated to:
 \begin{equation}
    Var[\hat{V}_{\epsilon_x}]\approx\frac{2}{m}[1+T(V+\epsilon-1)]^2+(1-V)^2\Upsilon^2=:s_{x}^2
\end{equation}
  The variance of the estimator for the channel noise of AS quadrature $\hat{V}_{\epsilon_{p}}=T(\epsilon+\Delta_V^u):=\frac{1}{n}\sum^n_{i=1}(B^p_i-\sqrt{\hat{T}}M^p_i)^2+\hat{T}(1-1/V)-1$ can be approximated to:
 \begin{equation}
    Var[\hat{V}_{\epsilon_p}]\approx\frac{2}{n}[1+T(1/V+\epsilon+\Delta V_u-1)]^2+(1-1/V)^2\Upsilon^2=:s_{p}^2
\end{equation}
Note that when the source radiates the coherent state and the channel noise is same for both the quadrature ($\Delta V_u$) the variance of the estimator $\hat{V}_{\epsilon}$ can be approximated to $Var[V_{\epsilon}]\approx 2(1+V_{\epsilon})^2/(m+n)$, by using the channel noise estimator $\hat{V_\epsilon}:=\frac{1}{m+n}(\sum^m_{i=1}(B^x_i-\sqrt{\hat{T}}M^x_i)^2+\sum^n_{i=1}(B^p_i-\sqrt{\hat{T}}M^p_i)^2)-1$ .\\
To account for the trusted AS noise when using AS quadrature for parameter estimation, we incorporate the trusted noise $\Delta V_t$ transforming the expressions for standard deviation to 
\begin{equation}
    \Upsilon^2:=\frac{4 T [n [ V_p-\Delta V_t] (V_{Np}+\Delta V_t + 2 T[ V_p-\Delta V_t]) + m V_x (V_{Nx} + 2 T V_x)]}{[n ( V_p-\Delta V_t) + m V_x]^2}
\end{equation}
\begin{equation}
    s^2_p:=\frac{2}{n}[1+T(1/V+\Delta V_t+\epsilon+\Delta V_u-1)]^2+(1-1/V-\Delta V_t)^2\Upsilon^2
\end{equation}
For the imbalanced heterodyne measurement, the estimators defined  above are no longer valid, since the overall estimated transmittance values for the squeezed and AS quadrature are not equal. We further extend the parameter estimation to include an imbalanced heterodyne measurement. We start by defining $\widehat{C'_{AB}}:=\frac{1}{m+n}(\sum^m_{i=1}M^x_iB^x_i/\sqrt{t_{het}}+\sum^n_{j=1}M^p_jB^p_j/\sqrt{1-t_{het}})$,
where $t_{het}$ is the transmittance of the beamsplitter of imbalanced heterodyne measurement. Following the same procedure as above, we find the variance of the estimator $\hat{T}$ can be approximated as
\begin{equation}
    Var[\hat{T}]\approx \frac{4 T [n V_p (V'_{Np} + 2 T V_p) + m V_x (V'_{Nx} + 2 T V_x)]}{(n V_p + m V_x)^2}=:\Upsilon^2
 \end{equation}
 Where $V'_{Nx}=1/t_{het}+T(V+\epsilon-1)$ and $V'_{Np}=1/(1-t_{het})+T(1/V+\epsilon+\Delta V_u-1)$.
\\ \\ The estimator of the channel noise for imbalanced heterodyne detection is $\hat{V}_{\epsilon_x}:=\frac{1}{m}\sum^m_{i=1}(B^x_i/t_{het}-\sqrt{\hat{T}}M^x_i)^2+\hat{T}(1-V)-1$ for $x$-quadrature, and $\hat{V}_{\epsilon_p}:=\frac{1}{m}\sum^m_{i=1}(B^p_i/[1-t_{het}]-\sqrt{\hat{T}}M^p_i)^2+\hat{T}(1-1/V)-1$ for $p$-quadrature. The variance of which can be approximated as:
\begin{equation}
    Var[\hat{V}_{\epsilon_x}]\approx\frac{2}{m}[V'_{Nx}]^2+(1-V)^2 Var[\hat{T}]
\end{equation}
for $x$-quadrature and 
\begin{equation}
    Var[\hat{V}_{\epsilon_p}]\approx\frac{2}{m}[V'_{Np}]^2+(1-1/V)^2 Var[\hat{T}]
\end{equation}
for $p$-quadrature.

\begin{figure}
    \centering
    \begin{subfigure}[b]{0.3\linewidth}
        \includegraphics[width=\linewidth, height=5cm]{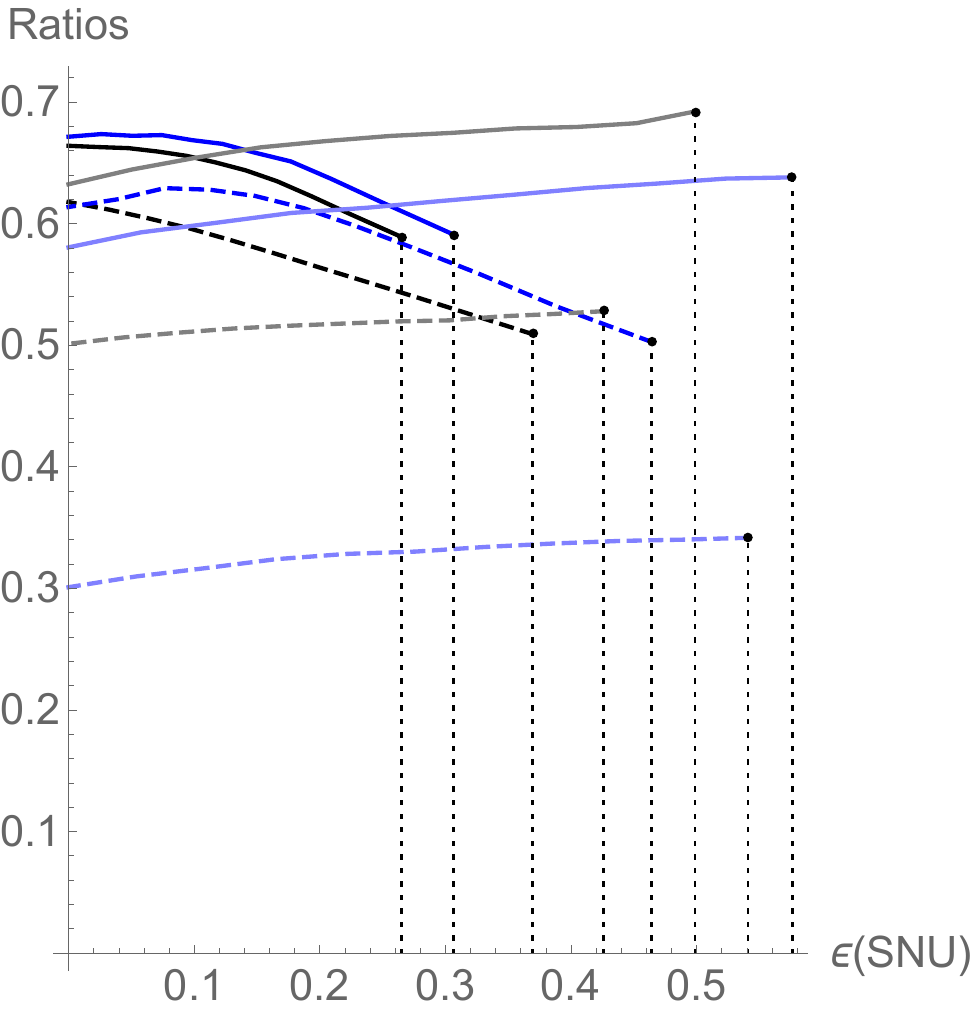}
        \caption{}
        \label{fig:ratio_a}
    \end{subfigure}
    \begin{subfigure}[b]{0.3\linewidth}
        \includegraphics[width=\linewidth, height=5cm]{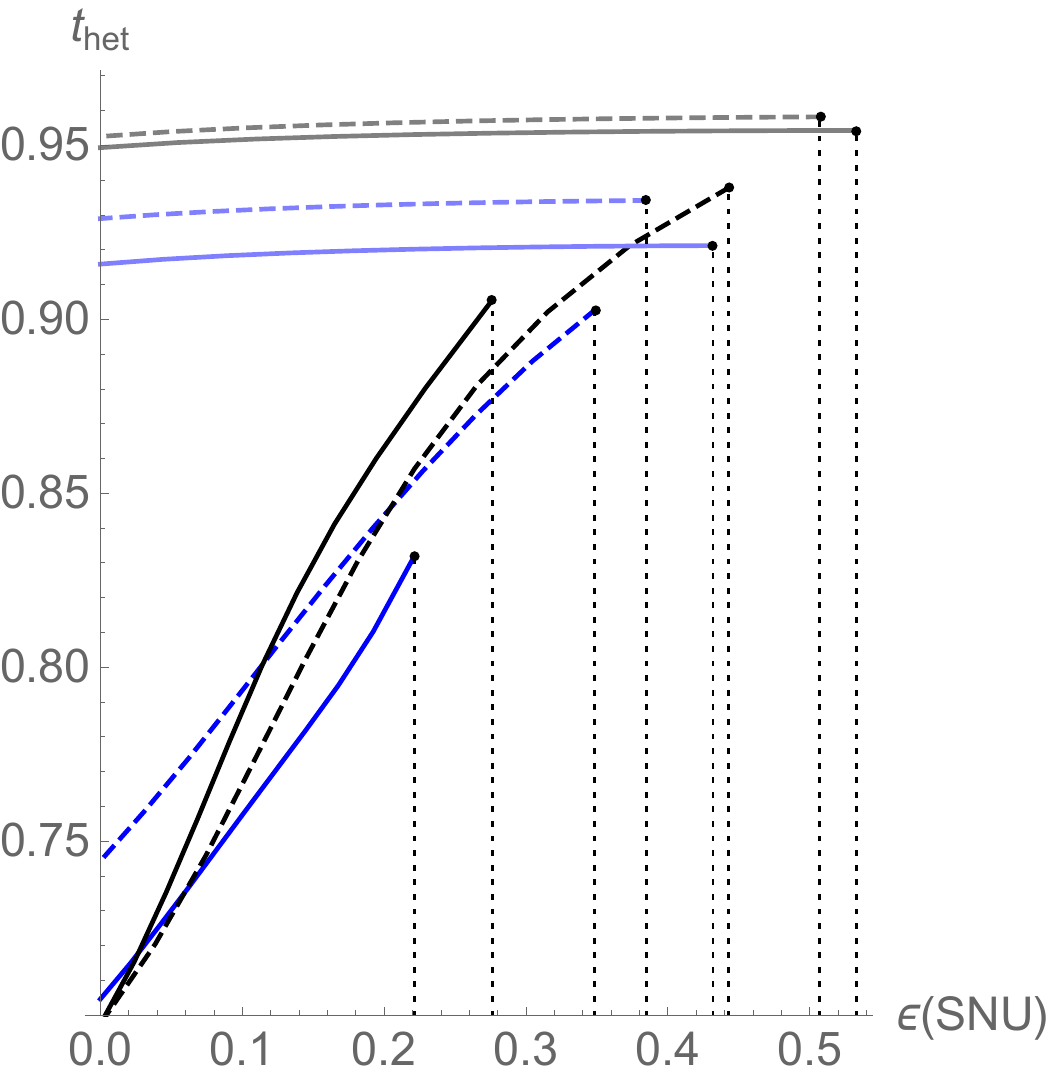}
        \caption{}
        \label{fig:ratio_c}
    \end{subfigure}
    \begin{subfigure}[b]{0.3\linewidth}
        \includegraphics[width=\linewidth, height=5cm]{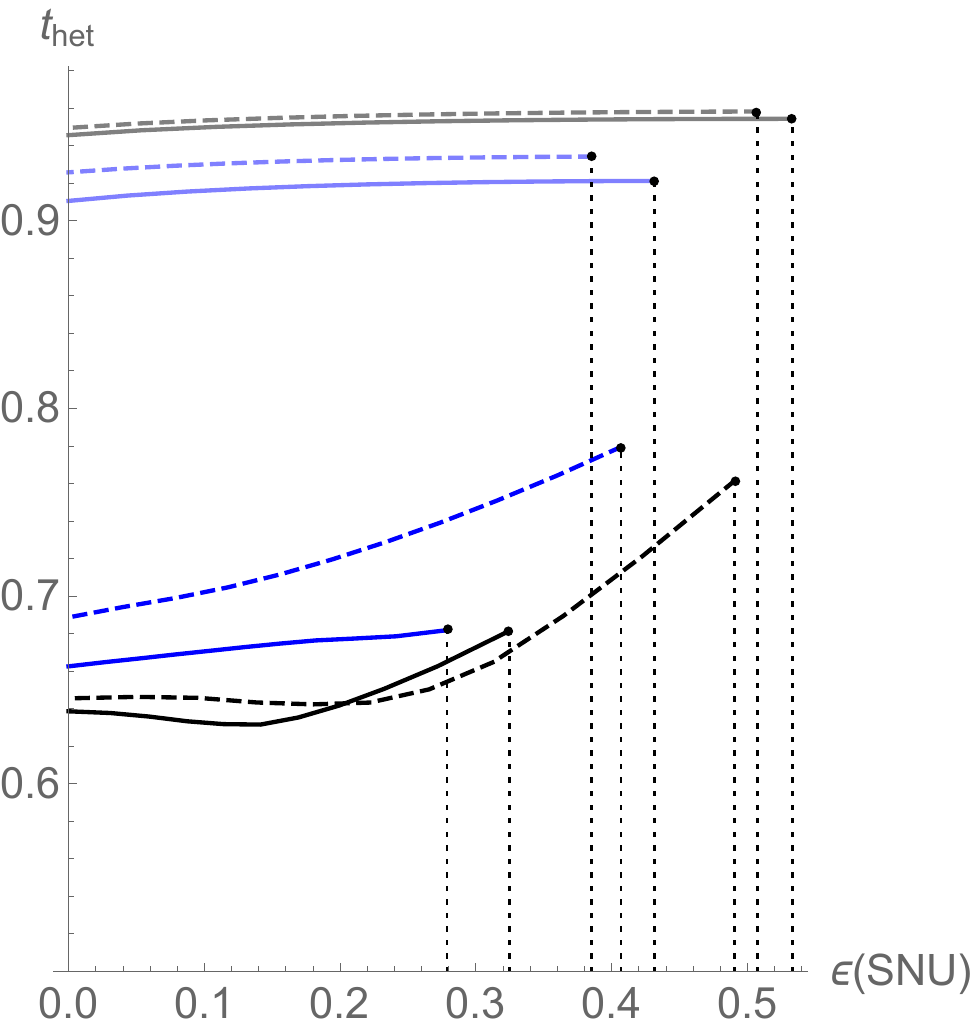}
        \caption{}
        \label{fig:ratio_d}
    \end{subfigure}
    \caption{Optimal ratios of squeezed quadrature measurement for biased homodyne measurement and optimal beamsplitter transmittance $t_{het}$ for the imbalanced heterodyne measurement (used in Fig. \ref{fig:tol_fin}). Black plots represent a block size of $10^7$, blue plots represent a block size of $10^6$, solid plots represent a moderately squeezed source state $V=0.5$ SNU, dashed plots represent a highly squeezed source state $V=0.1$ SNU. Light colors correspond to DR, and dark colors correspond to RR. The vertical lines show the maximum tolerable channel noise.}
    \label{fig:ratios}
\end{figure}
\section{}\label{Sec:disclose}
\begin{figure}[H]\centering
\subfloat[Direct reconciliation]{\includegraphics[width=.45\linewidth,height=5cm]{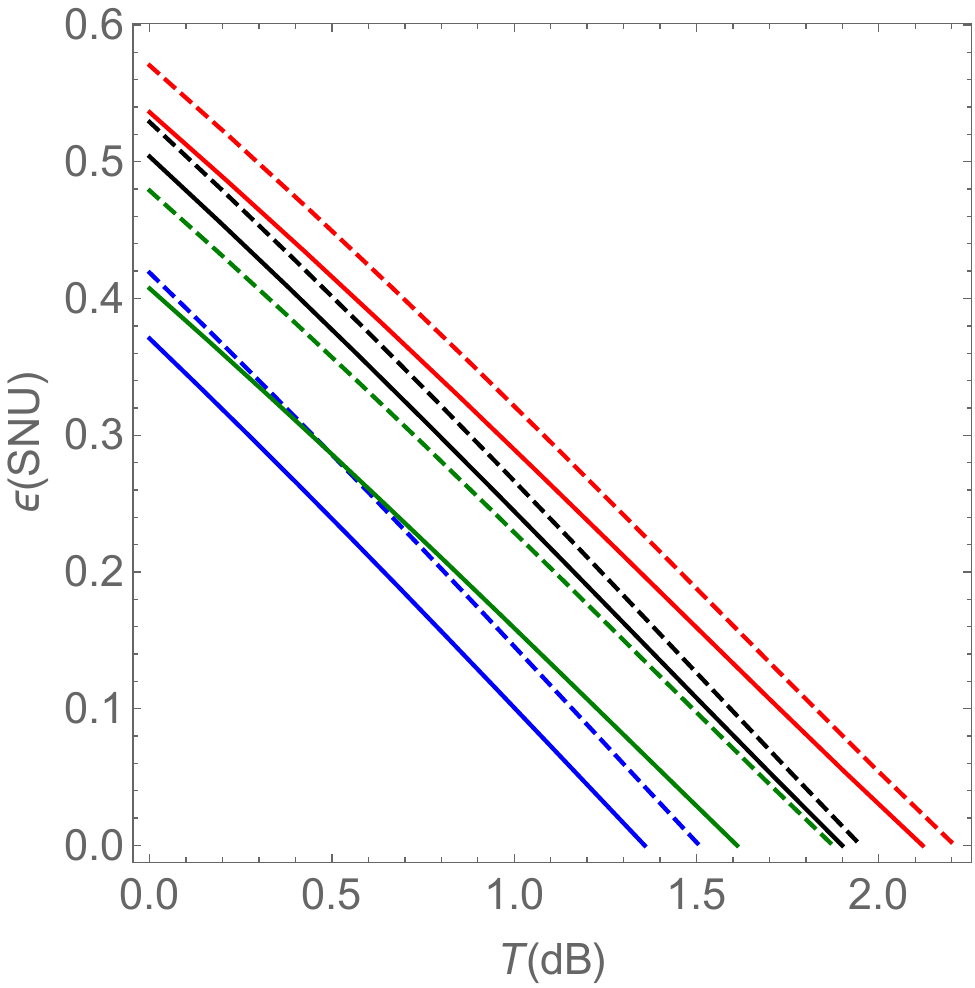}}\hfill
\subfloat[Reverse reconciliation]{\includegraphics[width=.45\linewidth,height=5cm]{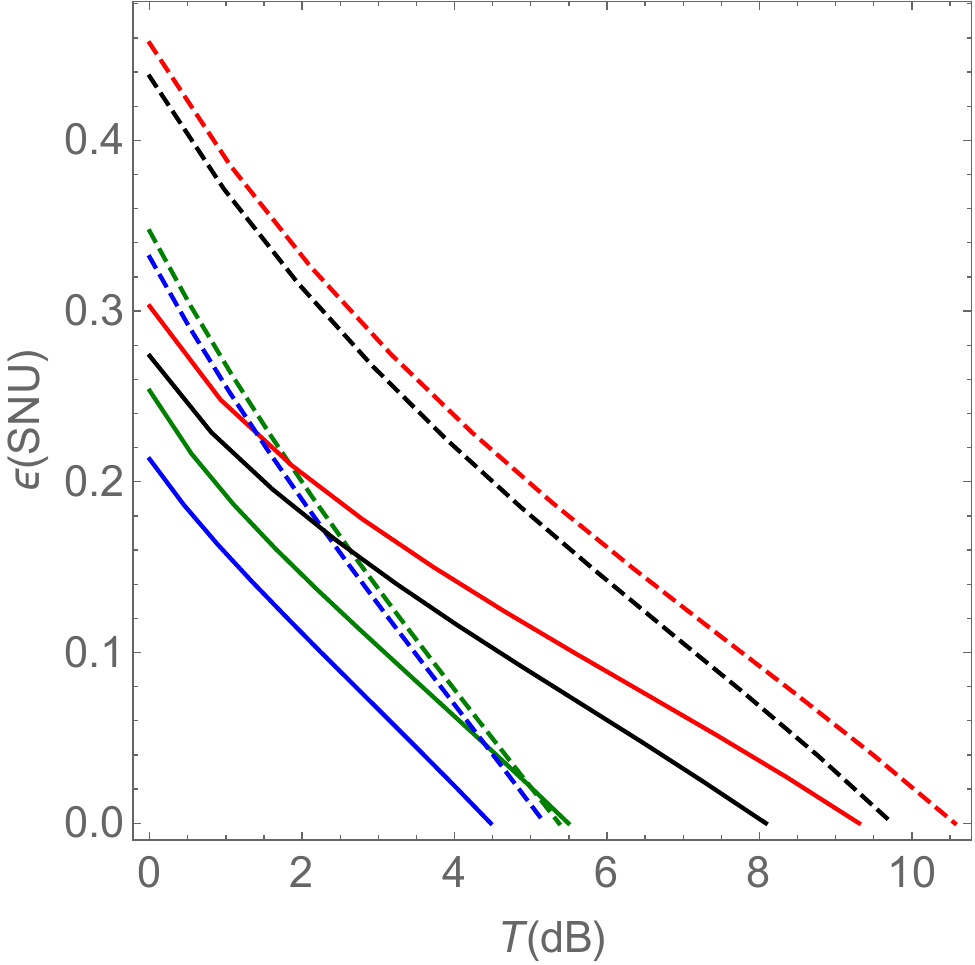}}
\caption{Tolerance to channel noise for biased homodyne with block size $N=10^7$ (red) and $N=10^6$ (green), and imbalanced heterodyne measurement with block size $N=10^7$ (black) and $N=10^6$ (blue), For moderately squeezed source state (solid) and highly squeezed source state (dashed). The modulation variance of the squeezed and AS quadrature $V_x$ and $V_p$, ratio of squeezed quadrature measurement for biased homodyne detection and the beamsplitter transmittance $t_{het}$ of imbalanced heterodyne measurement are optimised (see Fig. \ref{fig:ratio_dis} for optimum ratio and transmittance). Reconciliation efficiency $\beta=0.95$ }
\label{fig:tol_fin_dis}
\end{figure}
In the main body of the paper, we illustrated the performance of various scenarios in which no measurement results are disclosed for parameter estimation. Here, we show the performance of the scenarios when an optimal amount of measurement results is disclosed to facilitate parameter estimation. 
 Our analysis, depicted in Fig. \ref{fig:tol_fin_dis}, reveals that biased homodyne measurement consistently outperforms imbalance heterodyne measurement, particularly in terms of tolerable symmetric excess noise.

\begin{figure}
    \centering
    \begin{subfigure}[b]{0.45\linewidth}
        \includegraphics[width=\linewidth]{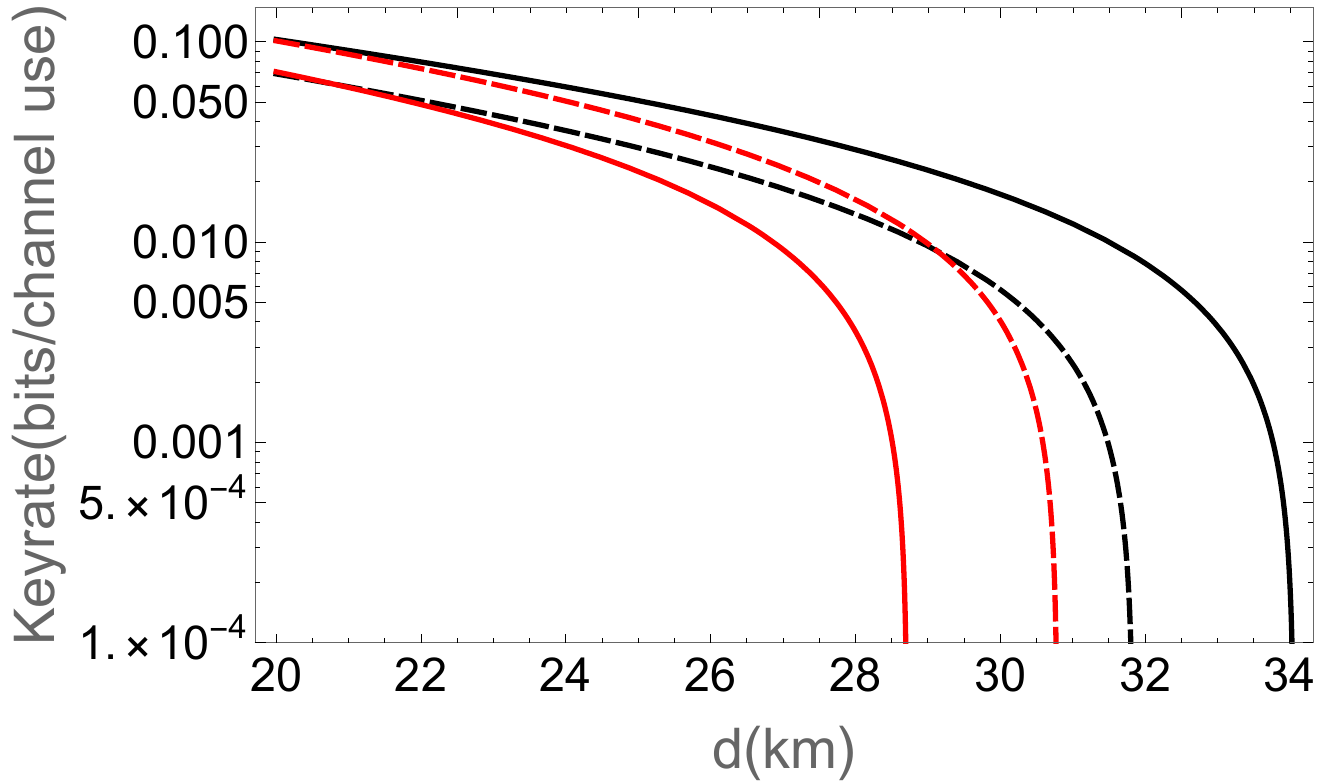}
        \caption{$\Delta V_t=0$, $\Delta V_u=0$, $N=10^6$}
        \label{fig:key_a2}
    \end{subfigure}
    \begin{subfigure}[b]{0.45\linewidth}
        \includegraphics[width=\linewidth]{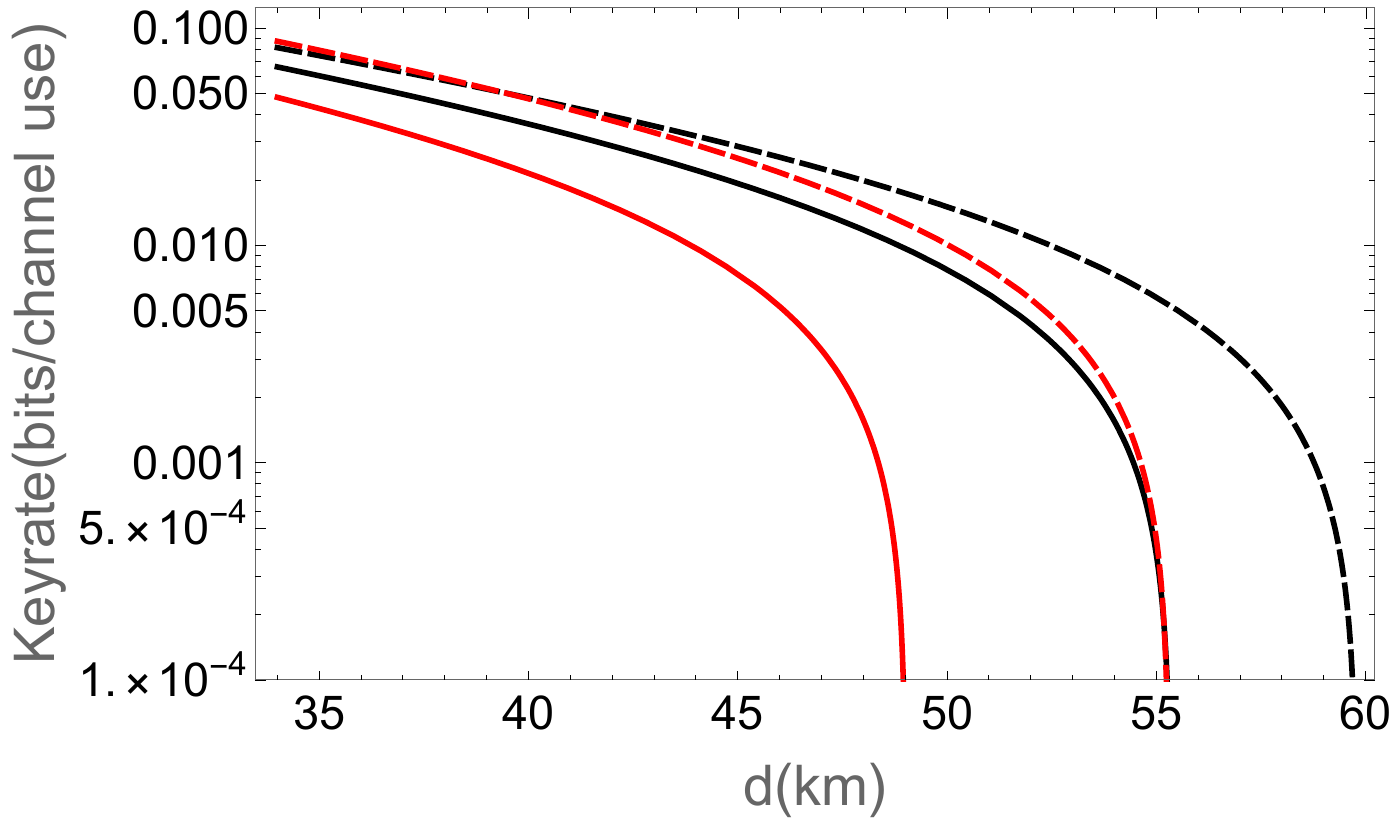}
        \caption{$\Delta V_t=0$, $\Delta V_u=0$, $N=10^7$}
        \label{fig:key_b2}
    \end{subfigure}
    
    \begin{subfigure}[b]{0.45\linewidth}
        \includegraphics[width=\linewidth]{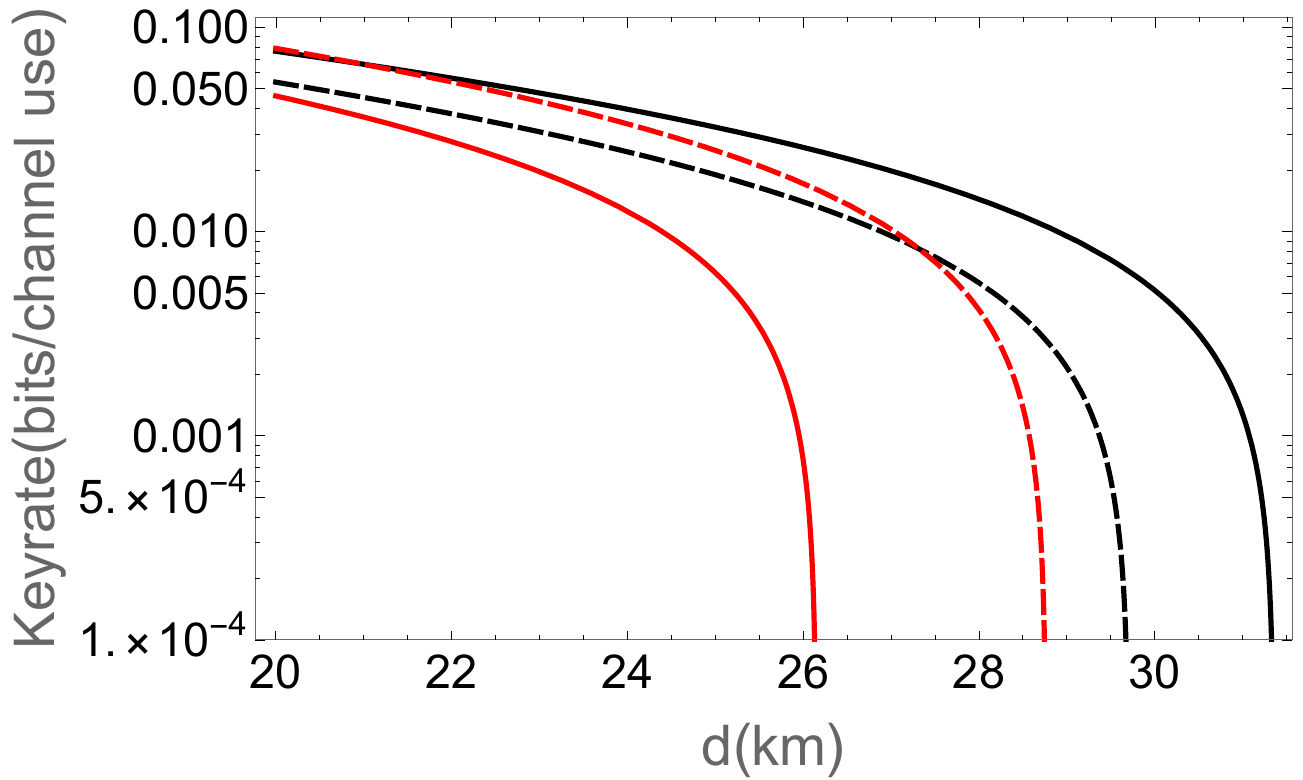}
        \caption{$\Delta V_t=0$, $\Delta V_u=0.05$, $N=10^6$}
        \label{fig:key_c2}
    \end{subfigure}
    \begin{subfigure}[b]{0.45\linewidth}
        \includegraphics[width=\linewidth]{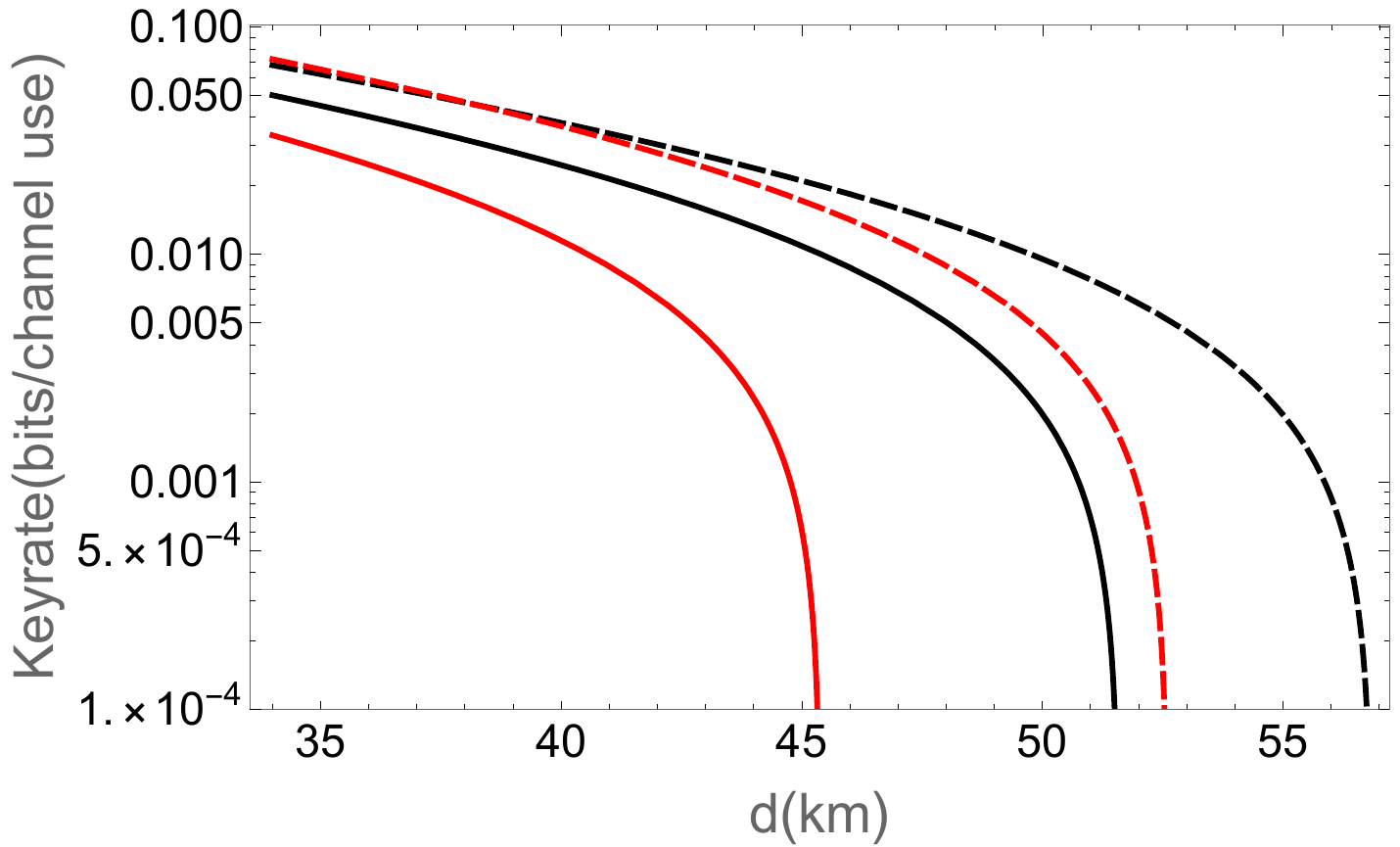}
        \caption{$\Delta V_t=0$, $\Delta V_u=0.05$, $N=10^7$}
        \label{fig:key_d2}
    \end{subfigure}
    
    \begin{subfigure}[b]{0.45\linewidth}
        \includegraphics[width=\linewidth]{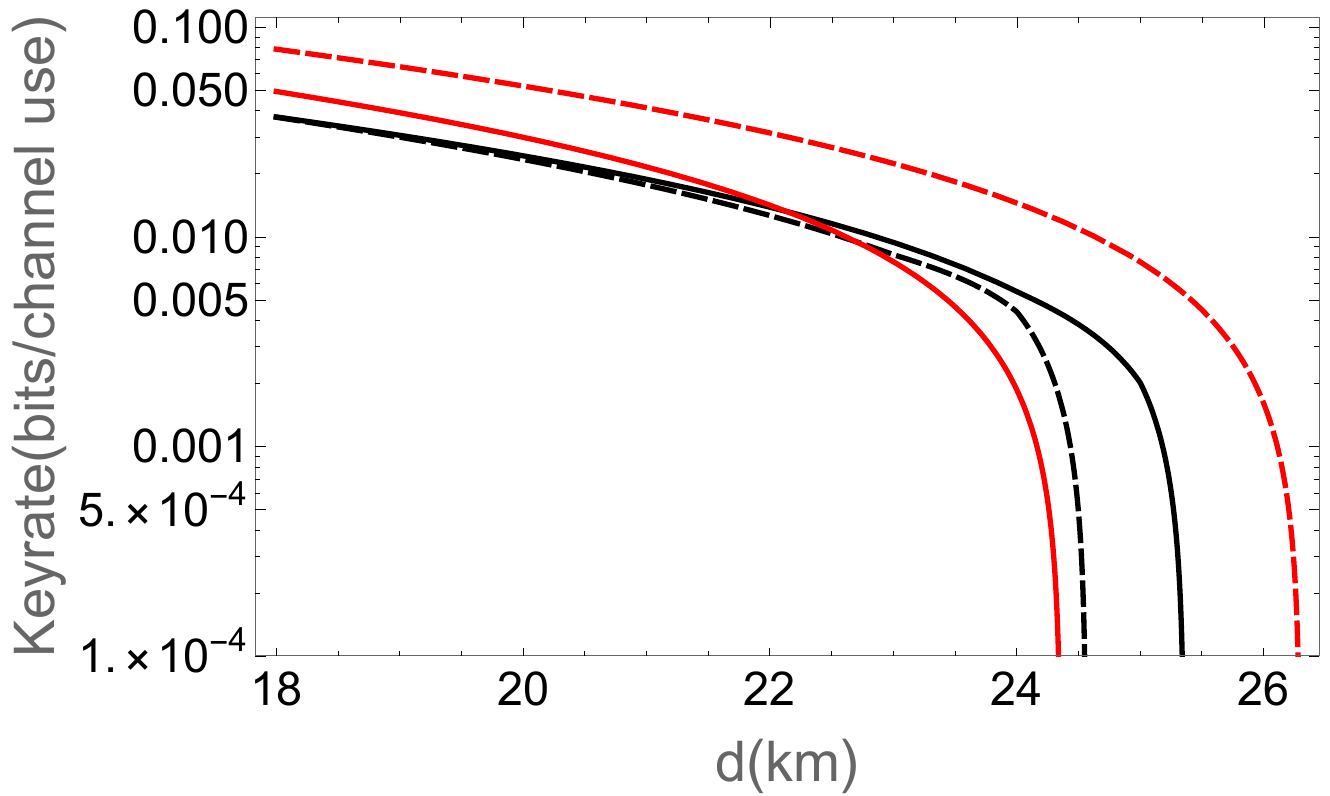}
        \caption{$\Delta V_t=5$, $\Delta V_u=0$, $N=10^6$}
        \label{fig:key_e2}
    \end{subfigure}
    \begin{subfigure}[b]{0.45\linewidth}
        \includegraphics[width=\linewidth]{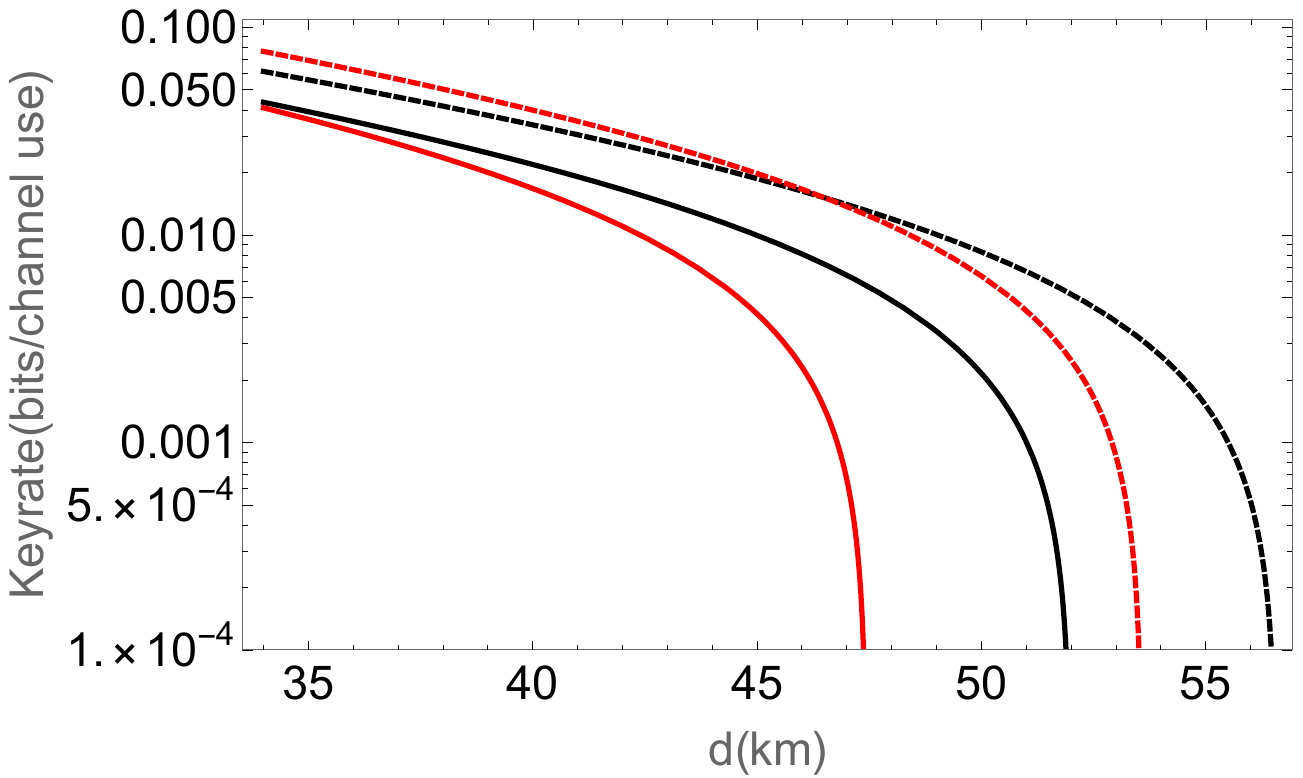}
        \caption{$\Delta V_t=5$, $\Delta V_u=0$, $N=10^7$}
        \label{fig:key_f2}
    \end{subfigure}
    
    \caption{Finite-size key rate vs. distance when Bob performs biased homodyne detection (black) and imbalanced heterodyne detection (red), when the parameter estimation requires disclosure of the measurement outcome. The dashed plot represents a highly squeezed source state ($V=0.1$ SNU) and the solid plot represents a moderately squeezed source state ($V=0.5$ SNU). Modulation variances $V_x$ and $V_p$, transmittance of beamsplitter for heterodyne detection $t_{het}$, ratios of available resources used for key generation $r_k$, and parameter estimation $r_x$ and $r_p$ are optimized. The reconciliation efficiency is $\beta = 0.95$, the channel noise is $\epsilon = 0.05$ SNU, and the channel attenuation is 0.2 dB/km.}
    \label{fig:finite_key_rate}
\end{figure}

Exceptionally, imbalance heterodyne measurement surpasses biased homodyne measurement in cases involving elevated antisqueezed noise, a small block size, and substantial squeezing (Fig. \ref{fig:finite_key_rate}).

\begin{figure}
    \centering
    \begin{subfigure}[b]{0.23\linewidth}
        \includegraphics[width=\linewidth]{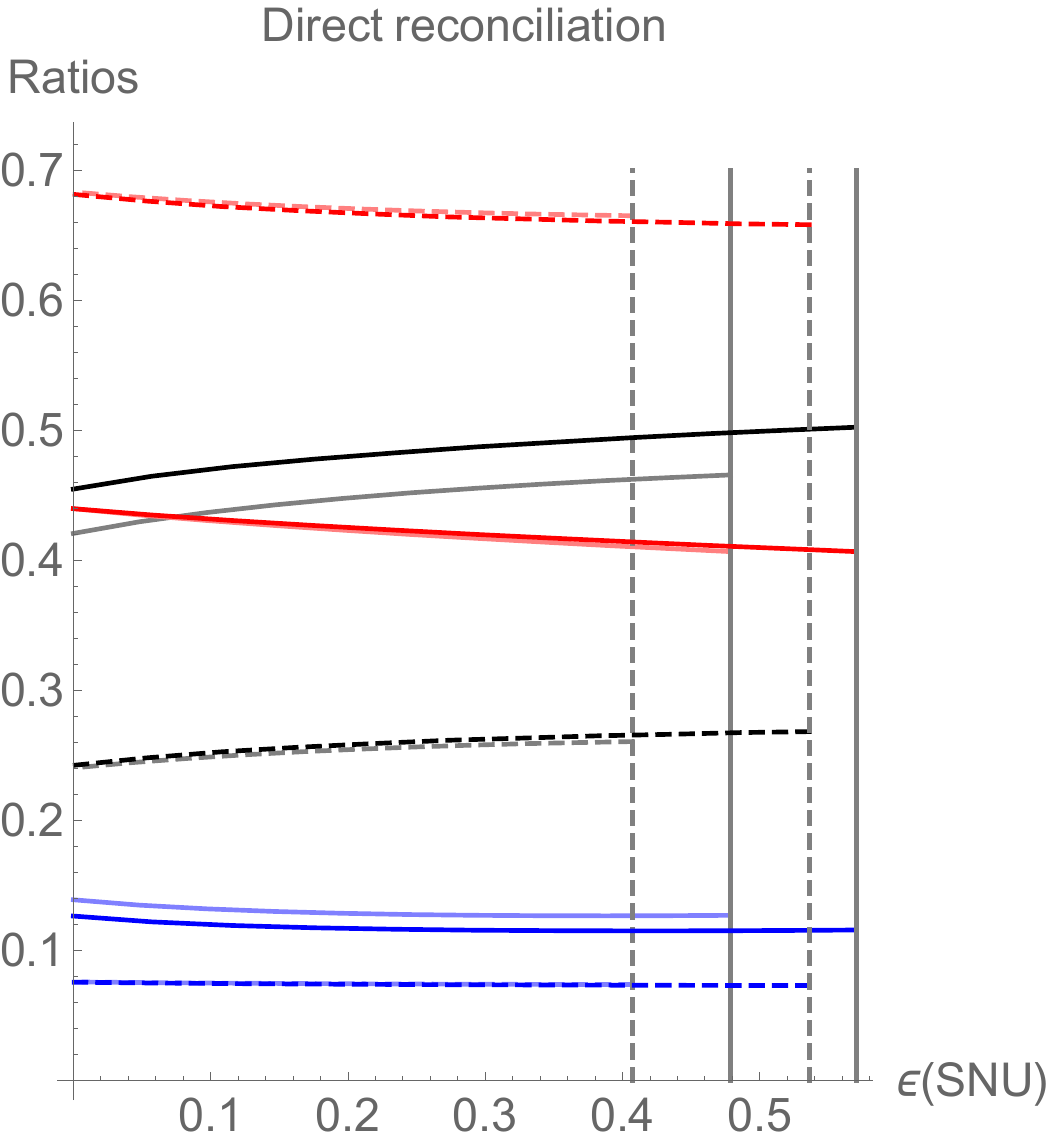}
        \caption{}
        \label{fig:r_a}
    \end{subfigure}
    \begin{subfigure}[b]{0.23\linewidth}
        \includegraphics[width=\linewidth]{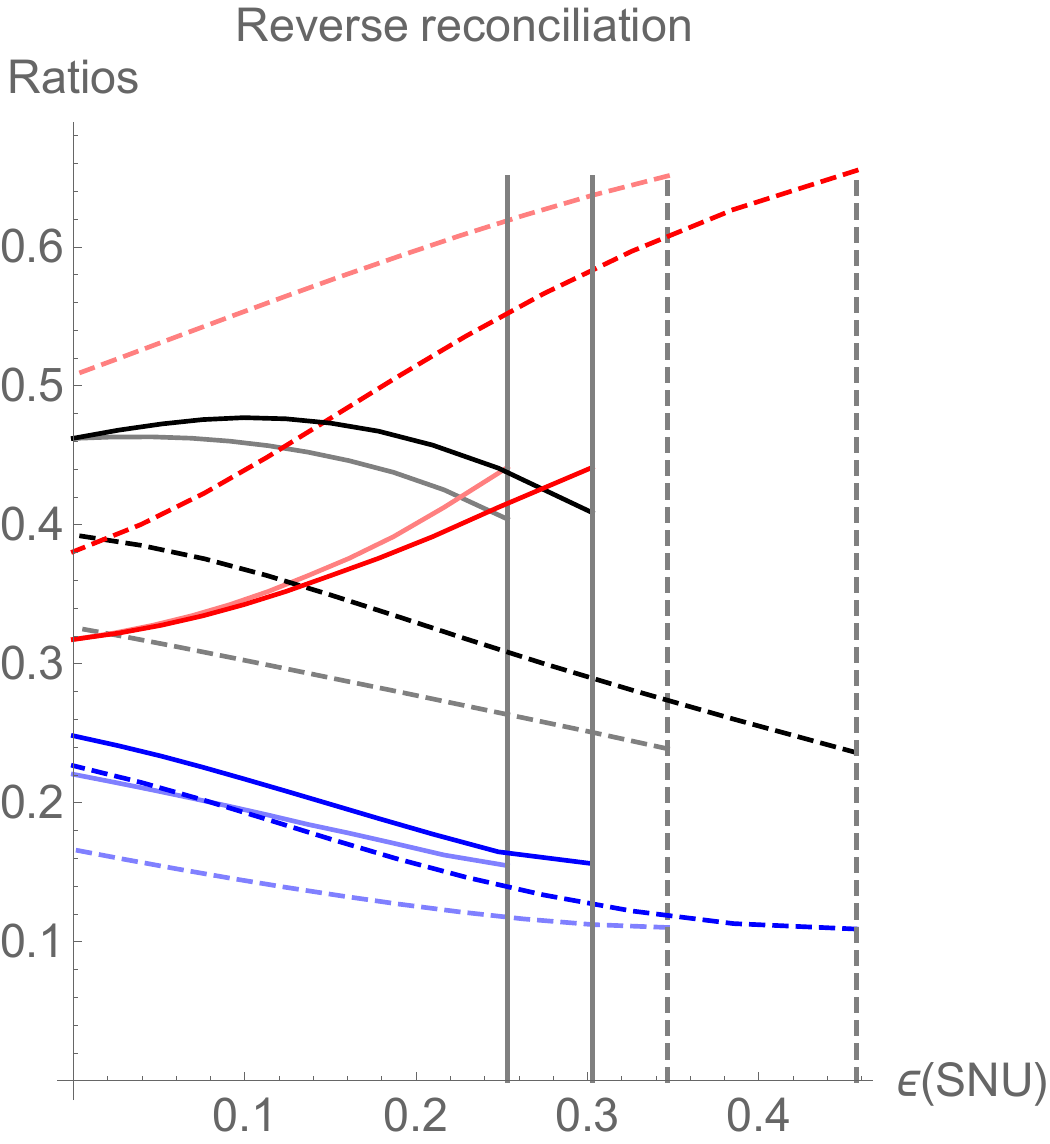}
        \caption{}
        \label{fig:r_b}
    \end{subfigure}
    \begin{subfigure}[b]{0.23\linewidth}
        \includegraphics[width=\linewidth]{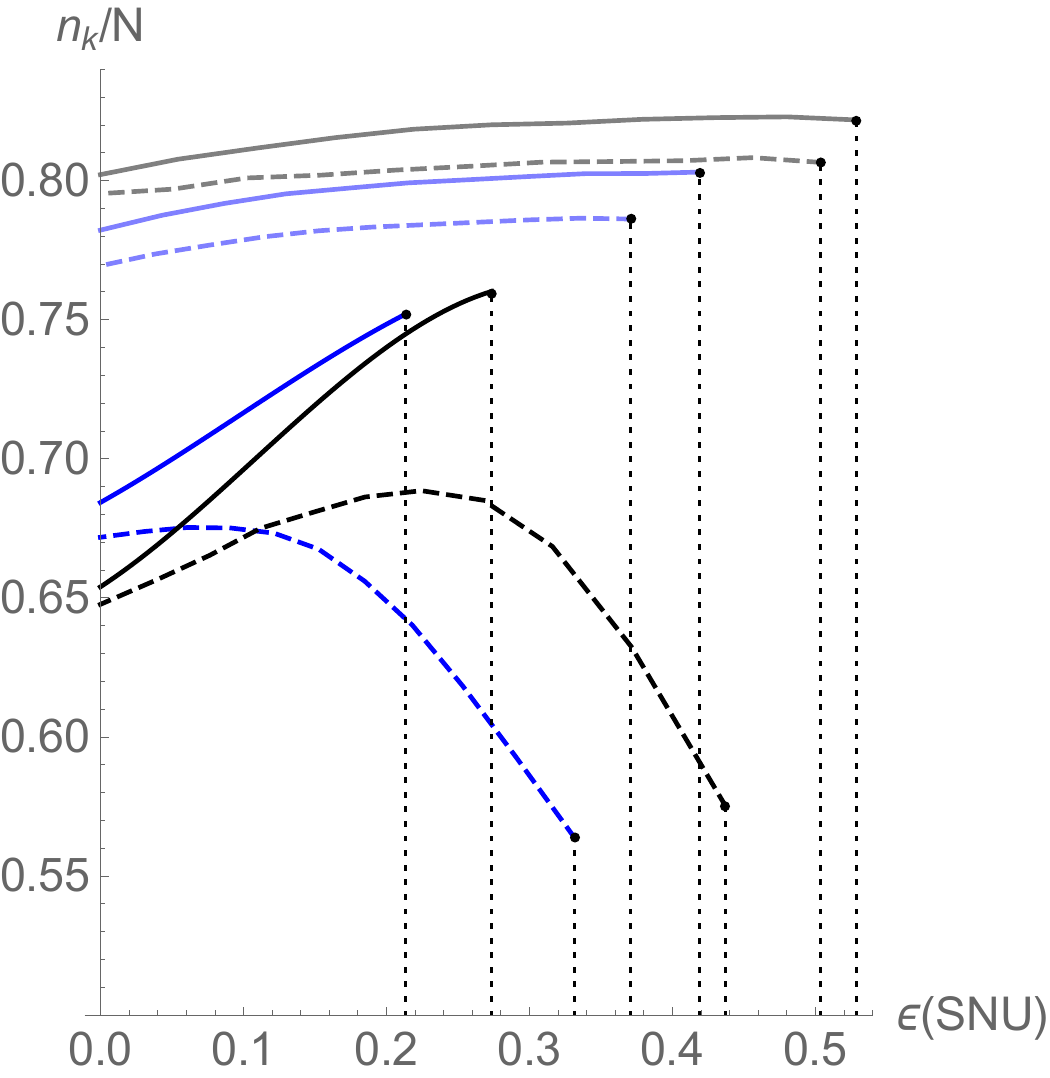}
        \caption{}
        \label{fig:r_c}
    \end{subfigure}
    \begin{subfigure}[b]{0.23\linewidth}
        \includegraphics[width=\linewidth]{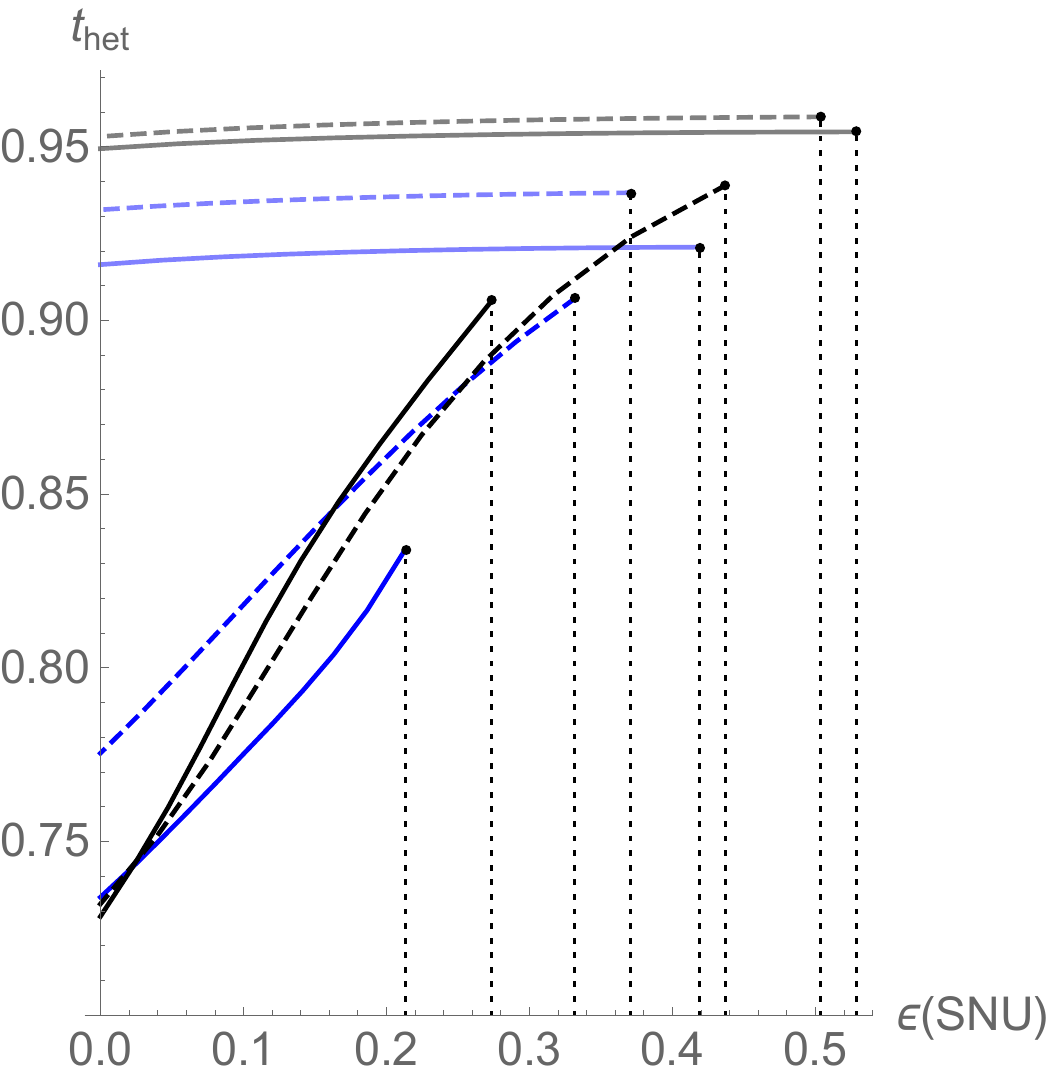}
        \caption{}
        \label{fig:r_d}
    \end{subfigure}

    \caption{(a) and (b) show the optimal ratios (used in Fig. \ref{fig:tol_fin_dis}) for biased homodyne measurement used for key generation $n_k/N$ (black), channel noise estimation of AS quadrature $n/N$ (red), and channel noise estimation of squeezed quadrature $m/N$ (blue), for moderately squeezed $V=0.5$ SNU (solid) and highly squeezed $V=0.1$ SNU (dashed) source to achieve maximum distance in the presence of 0.2 SNU AS noise. Darker shades of a color represent a block size of $10^7$, and lighter shades represent a block size of $10^6$. The gray vertical line shows the maximum tolerable channel noise for the moderately squeezed (dashed) and highly squeezed (solid) source state. (c) and (d) show the optimal transmittance of the beamsplitter of imbalanced heterodyne measurement $t_{het}$, and ratio $n_k/N$ of available states used for key generation (used in Fig. \ref{fig:tol_fin_dis}). Black plots represent a block size of $10^7$, blue plots represent a block size of $10^6$, solid plots represent a moderately squeezed source state $V=0.5$ SNU, dashed plots represent a highly squeezed source state $V=0.1$ SNU, light colors correspond to DR, and dark colors correspond to RR. The vertical lines show the maximum tolerable channel noise.}
    \label{fig:ratio_dis}
\end{figure}

Notably, the optimal ratios of block sizes for parameter estimation and key generation remain relatively constant in the context of DR (Fig. \ref{fig:ratio_dis}). However, for RR, these optimal ratios exhibit variations corresponding to the level of excess noise.

\clearpage

\bibliographystyle{unsrt}
\bibliography{biblo.bib}

\begin{thebibliography}{10}

\bibitem{Gisin2002}
Nicolas Gisin, Gr{\'e}goire Ribordy, Wolfgang Tittel, and Hugo Zbinden.
\newblock Quantum cryptography.
\newblock {\em Reviews of Modern Physics}, 74(1):145, 2002.

\bibitem{Scarani2009}
Valerio Scarani, Helle Bechmann-Pasquinucci, Nicolas~J. Cerf, Miloslav Du{\v{s}}ek, Norbert L{\"{u}}tkenhaus, and Momtchil Peev.
\newblock The security of practical quantum key distribution.
\newblock {\em Reviews of Modern Physics}, 81(3):1301--1350, sep 2009.

\bibitem{Diamanti2016}
Eleni Diamanti, Hoi-Kwong Lo, Bing Qi, and Zhiliang Yuan.
\newblock Practical challenges in quantum key distribution.
\newblock {\em npj Quantum Information}, 2:16025, 2016.

\bibitem{xu2020secure}
Feihu Xu, Xiongfeng Ma, Qiang Zhang, Hoi-Kwong Lo, and Jian-Wei Pan.
\newblock Secure quantum key distribution with realistic devices.
\newblock {\em Reviews of Modern Physics}, 92(2):025002, 2020.

\bibitem{pirandolaadvances}
Stefano Pirandola, Ulrik~L Andersen, Leonardo Banchi, Mario Berta, Darius Bunandar, Roger Colbeck, Dirk Englund, Tobias Gehring, Cosmo Lupo, Carlo Ottaviani, et~al.
\newblock Advances in quantum cryptography.
\newblock {\em Advances in Optics and Photonics}, 12(4):1012--1236, 2020.

\bibitem{Braunstein2005}
Samuel~L Braunstein and Peter Van~Loock.
\newblock Quantum information with continuous variables.
\newblock {\em Reviews of Modern Physics}, 77(2):513, 2005.

\bibitem{GG02}
Fr{\'e}d{\'e}ric Grosshans and Philippe Grangier.
\newblock Continuous variable quantum cryptography using coherent states.
\newblock {\em Physical review letters}, 88(5):057902, 2002.

\bibitem{grosshans2003quantum}
Fr{\'e}d{\'e}ric Grosshans, Gilles Van~Assche, J{\'e}r{\^o}me Wenger, Rosa Brouri, Nicolas~J Cerf, and Philippe Grangier.
\newblock Quantum key distribution using gaussian-modulated coherent states.
\newblock {\em Nature}, 421(6920):238--241, 2003.

\bibitem{jouguet2013experimental}
Paul Jouguet, S{\'e}bastien Kunz-Jacques, Anthony Leverrier, Philippe Grangier, and Eleni Diamanti.
\newblock Experimental demonstration of long-distance continuous-variable quantum key distribution.
\newblock {\em Nature photonics}, 7(5):378--381, 2013.

\bibitem{PhysRevA.81.062343}
Anthony Leverrier, Fr\'ed\'eric Grosshans, and Philippe Grangier.
\newblock Finite-size analysis of a continuous-variable quantum key distribution.
\newblock {\em Phys. Rev. A}, 81:062343, Jun 2010.

\bibitem{PhysRevLett.118.200501}
Anthony Leverrier.
\newblock Security of continuous-variable quantum key distribution via a gaussian de finetti reduction.
\newblock {\em Phys. Rev. Lett.}, 118:200501, May 2017.

\bibitem{PRXQuantum.2.040334}
Wen-Bo Liu, Chen-Long Li, Yuan-Mei Xie, Chen-Xun Weng, Jie Gu, Xiao-Yu Cao, Yu-Shuo Lu, Bing-Hong Li, Hua-Lei Yin, and Zeng-Bing Chen.
\newblock Homodyne detection quadrature phase shift keying continuous-variable quantum key distribution with high excess noise tolerance.
\newblock {\em PRX Quantum}, 2:040334, Nov 2021.

\bibitem{su2023experimental}
Zikang Su, Jintao Wang, Dajian Cai, Xiaojie Guo, Dawei Wang, and Zhaohui Li.
\newblock Experimental demonstration of phase-sensitive multimode continuous variable quantum key distribution with improved secure key rate.
\newblock {\em Photonics Research}, 11(11):1861--1869, 2023.

\bibitem{wang2020high}
Heng Wang, Yaodi Pi, Wei Huang, Yang Li, Yun Shao, Jie Yang, Jinlu Liu, Chenlin Zhang, Yichen Zhang, and Bingjie Xu.
\newblock High-speed gaussian-modulated continuous-variable quantum key distribution with a local local oscillator based on pilot-tone-assisted phase compensation.
\newblock {\em Optics express}, 28(22):32882--32893, 2020.

\bibitem{hajomer2024highspeed}
Adnan~AE Hajomer, C{\'e}dric Bruynsteen, Ivan Derkach, Nitin Jain, Axl Bomhals, Sarah Bastiaens, Ulrik~L Andersen, Xin Yin, and Tobias Gehring.
\newblock Continuous-variable quantum key distribution at 10 gbaud using an integrated photonic-electronic receiver.
\newblock {\em Optica}, 11(9):1197--1204, 2024.

\bibitem{pietri2024experimental}
Yoann Pi{\'e}tri, Luis Trigo~Vidarte, Matteo Schiavon, Laurent Vivien, Philippe Grangier, Amine Rhouni, and Eleni Diamanti.
\newblock Experimental demonstration of continuous-variable quantum key distribution with a silicon photonics integrated receiver.
\newblock {\em Optica Quantum}, 2(6):428--437, 2024.

\bibitem{zhang2020long}
Yichen Zhang, Ziyang Chen, Stefano Pirandola, Xiangyu Wang, Chao Zhou, Binjie Chu, Yijia Zhao, Bingjie Xu, Song Yu, and Hong Guo.
\newblock Long-distance continuous-variable quantum key distribution over 202.81 km of fiber.
\newblock {\em Physical review letters}, 125(1):010502, 2020.

\bibitem{hajomer2023long}
Adnan~AE Hajomer, Ivan Derkach, Nitin Jain, Hou-Man Chin, Ulrik~L Andersen, and Tobias Gehring.
\newblock Long-distance continuous-variable quantum key distribution over 100 km fiber with local local oscillator.
\newblock {\em arXiv preprint arXiv:2305.08156}, 2023.

\bibitem{chin2021machine}
Hou-Man Chin, Nitin Jain, Darko Zibar, Ulrik~L Andersen, and Tobias Gehring.
\newblock Machine learning aided carrier recovery in continuous-variable quantum key distribution.
\newblock {\em npj Quantum Information}, 7(1):20, 2021.

\bibitem{liu2022automated}
Zhi-Ping Liu, Min-Gang Zhou, Wen-Bo Liu, Chen-Long Li, Jie Gu, Hua-Lei Yin, and Zeng-Bing Chen.
\newblock Automated machine learning for secure key rate in discrete-modulated continuous-variable quantum key distribution.
\newblock {\em Optics Express}, 30(9):15024--15036, 2022.

\bibitem{PhysRevA.63.052311}
N.~J. Cerf, M.~L\'evy, and G.~Van Assche.
\newblock Quantum distribution of gaussian keys using squeezed states.
\newblock {\em Phys. Rev. A}, 63:052311, Apr 2001.

\bibitem{lvovsky2015squeezed}
Alexander~I Lvovsky.
\newblock Squeezed light.
\newblock {\em Photonics: Scientific Foundations, Technology and Applications}, 1:121--163, 2015.

\bibitem{PhysRevLett.102.130501}
Ra\'ul Garc\'{\i}a-Patr\'on and Nicolas~J. Cerf.
\newblock Continuous-variable quantum key distribution protocols over noisy channels.
\newblock {\em Phys. Rev. Lett.}, 102:130501, Mar 2009.

\bibitem{naturecom2012}
L.~Madsen, V.~Usenko, M.~Lassen, and et~al.
\newblock Continuous variable quantum key distribution with modulated entangled states.
\newblock {\em Nat Commun}, 3, Aug 2012.

\bibitem{Usenko_2011}
Vladyslav~C Usenko and Radim Filip.
\newblock Squeezed-state quantum key distribution upon imperfect reconciliation.
\newblock {\em New Journal of Physics}, 13(11):113007, nov 2011.

\bibitem{jacobsen2018complete}
Christian~S Jacobsen, Lars~S Madsen, Vladyslav~C Usenko, Radim Filip, and Ulrik~L Andersen.
\newblock Complete elimination of information leakage in continuous-variable quantum communication channels.
\newblock {\em npj Quantum Information}, 4(1):1--6, 2018.

\bibitem{usenko2012entanglement}
Vladyslav~C Usenko, Bettina Heim, Christian Peuntinger, Christoffer Wittmann, Christoph Marquardt, Gerd Leuchs, and Radim Filip.
\newblock Entanglement of gaussian states and the applicability to quantum key distribution over fading channels.
\newblock {\em New Journal of Physics}, 14(9):093048, 2012.

\bibitem{Derkach_2020}
Ivan Derkach, Vladyslav~C Usenko, and Radim Filip.
\newblock Squeezing-enhanced quantum key distribution over atmospheric channels.
\newblock {\em New Journal of Physics}, 22(5):053006, may 2020.

\bibitem{e23010055}
Ivan Derkach and Vladyslav~C. Usenko.
\newblock Applicability of squeezed- and coherent-state continuous-variable quantum key distribution over satellite links.
\newblock {\em Entropy}, 23(1), 2021.

\bibitem{Kish2024comparisonof}
Sebastian~P. Kish, Patrick~J. Gleeson, Angus Walsh, Ping~Koy Lam, and Syed~M. Assad.
\newblock Comparison of {D}iscrete {V}ariable and {C}ontinuous {V}ariable {Q}uantum {K}ey {D}istribution {P}rotocols with {P}hase {N}oise in the {T}hermal-{L}oss {C}hannel.
\newblock {\em {Quantum}}, 8:1382, June 2024.

\bibitem{PhysRevA.102.012608}
Liyun Hu, M.~Al-amri, Zeyang Liao, and M.~S. Zubairy.
\newblock Continuous-variable quantum key distribution with non-gaussian operations.
\newblock {\em Phys. Rev. A}, 102:012608, Jul 2020.

\bibitem{vahlbruch2016detection}
Henning Vahlbruch, Moritz Mehmet, Karsten Danzmann, and Roman Schnabel.
\newblock Detection of 15 db squeezed states of light and their application for the absolute calibration of photoelectric quantum efficiency.
\newblock {\em Physical review letters}, 117(11):110801, 2016.

\bibitem{mondain2019}
F.~Mondain, T.~Lunghi, A.~Zavatta, E.~Gouzien, F.~Doutre, M.~De Micheli, S.~Tanzilli, and V.~D'Auria.
\newblock Chip-based squeezing at a telecom wavelength.
\newblock {\em Photonics Research}, 7(7):A36--A39, July 2019.

\bibitem{sun2019}
Xiaocong Sun, Yajun Wang, Long Tian, Shaoping Shi, Yaohui Zheng, and Kunchi Peng.
\newblock Dependence of the squeezing and anti-squeezing factors of bright squeezed light on the seed beam power and pump beam noise.
\newblock {\em Optics Letters}, 44(7):1789--1792, April 2019.

\bibitem{kashiwazaki2020}
Takahiro Kashiwazaki, Naoto Takanashi, Taichi Yamashima, Takushi Kazama, Koji Enbutsu, Ryoichi Kasahara, Takeshi Umeki, and Akira Furusawa.
\newblock Continuous-wave 6-{{dB-squeezed}} light with 2.5-{{THz-bandwidth}} from single-mode {{PPLN}} waveguide.
\newblock {\em APL Photonics}, 5(3):036104, March 2020.

\bibitem{PSA10068233}
Karol Łukanowski, Konrad Banaszek, and Marcin Jarzyna.
\newblock Quantum limits on the capacity of multispan links with phase-sensitive amplification.
\newblock {\em Journal of Lightwave Technology}, pages 1--9, 2023.

\bibitem{gehring2015implementation}
Tobias Gehring, Vitus H{\"a}ndchen, J{\"o}rg Duhme, Fabian Furrer, Torsten Franz, Christoph Pacher, Reinhard~F Werner, and Roman Schnabel.
\newblock Implementation of continuous-variable quantum key distribution with composable and one-sided-device-independent security against coherent attacks.
\newblock {\em Nature communications}, 6(1):1--7, 2015.

\bibitem{PhysRevApplied.10.064028}
Ning Wang, Shanna Du, Wenyuan Liu, Xuyang Wang, Yongmin Li, and Kunchi Peng.
\newblock Long-distance continuous-variable quantum key distribution with entangled states.
\newblock {\em Phys. Rev. Appl.}, 10:064028, Dec 2018.

\bibitem{Holevo}
A.~S. Holevo and R.~F. Werner.
\newblock Evaluating capacities of bosonic gaussian channels.
\newblock {\em Phys. Rev. A}, 63:032312, Feb 2001.

\bibitem{PhysRevLett.97.190502}
Miguel Navascu\'es, Fr\'ed\'eric Grosshans, and Antonio Ac\'{\i}n.
\newblock Optimality of gaussian attacks in continuous-variable quantum cryptography.
\newblock {\em Phys. Rev. Lett.}, 97:190502, Nov 2006.

\bibitem{PhysRevLett.97.190503}
Ra\'ul Garc\'{\i}a-Patr\'on and Nicolas~J. Cerf.
\newblock Unconditional optimality of gaussian attacks against continuous-variable quantum key distribution.
\newblock {\em Phys. Rev. Lett.}, 97:190503, Nov 2006.

\bibitem{Serafini_2005}
A~Serafini, M~G~A Paris, F~Illuminati, and S~De Siena.
\newblock Quantifying decoherence in continuous variable systems.
\newblock {\em Journal of Optics B: Quantum and Semiclassical Optics}, 7(4):R19--R36, feb 2005.

\bibitem{PhysRevLett.94.020504}
Fr\'ed\'eric Grosshans.
\newblock Collective attacks and unconditional security in continuous variable quantum keydistribution.
\newblock {\em Phys. Rev. Lett.}, 94:020504, Jan 2005.

\bibitem{e18010020}
Vladyslav~C. Usenko and Radim Filip.
\newblock Trusted noise in continuous-variable quantum key distribution: A threat and a defense.
\newblock {\em Entropy}, 18(1), 2016.

\bibitem{mani2021multiedge}
Hossein Mani, Tobias Gehring, Philipp Grabenweger, Bernhard {\"O}mer, Christoph Pacher, and Ulrik~Lund Andersen.
\newblock Multiedge-type low-density parity-check codes for continuous-variable quantum key distribution.
\newblock {\em Physical Review A}, 103(6):062419, 2021.

\bibitem{PhysRevA.90.062310}
L\'aszl\'o Ruppert, Vladyslav~C. Usenko, and Radim Filip.
\newblock Long-distance continuous-variable quantum key distribution with efficient channel estimation.
\newblock {\em Phys. Rev. A}, 90:062310, Dec 2014.

\bibitem{leverrier2015composable}
Anthony Leverrier.
\newblock Composable security proof for continuous-variable quantum key distribution with coherent states.
\newblock {\em Physical review letters}, 114(7):070501, 2015.

\bibitem{dong2010continuous}
Ruifang Dong, Mikael Lassen, Joel Heersink, Christoph Marquardt, Radim Filip, Gerd Leuchs, and Ulrik~L Andersen.
\newblock Continuous-variable entanglement distillation of non-gaussian mixed states.
\newblock {\em Physical Review A}, 82(1):012312, 2010.

\bibitem{wolf2006extremality}
Michael~M Wolf, Geza Giedke, and J~Ignacio Cirac.
\newblock Extremality of gaussian quantum states.
\newblock {\em Physical review letters}, 96(8):080502, 2006.

\bibitem{vasylyev2016atmospheric}
D~Vasylyev, AA~Semenov, and W~Vogel.
\newblock Atmospheric quantum channels with weak and strong turbulence.
\newblock {\em Physical review letters}, 117(9):090501, 2016.

\bibitem{weedbrook2012gaussian}
Christian Weedbrook, Stefano Pirandola, Ra{\'u}l Garc{\'\i}a-Patr{\'o}n, Nicolas~J Cerf, Timothy~C Ralph, Jeffrey~H Shapiro, and Seth Lloyd.
\newblock Gaussian quantum information.
\newblock {\em Reviews of Modern Physics}, 84(2):621, 2012.

\bibitem{derkach2016preventing}
Ivan Derkach, Vladyslav~C Usenko, and Radim Filip.
\newblock Preventing side-channel effects in continuous-variable quantum key distribution.
\newblock {\em Physical Review A}, 93(3):032309, 2016.

\end{thebibliography}

\end{document}